\pdfoutput=1
\documentclass[11pt]{article}
\usepackage[pdftex]{graphicx,color} 
\usepackage{jheppub}
\usepackage{amsmath}
\usepackage{amssymb}
\usepackage{ytableau}

\usepackage{multirow}
\usepackage{mathtools}

\newcommand{\bali}{\begin{align}}
\newcommand{\eali}{\end{align}}
\newcommand{\bea}{\begin{equation}\begin{aligned}}
\newcommand{\eea}[1]{\label{#1}\end{aligned}\end{equation}}
\newcommand{\beg}{\begin{equation}\begin{gathered}}
\newcommand{\eeg}[1]{\label{#1}\end{gathered}\end{equation}}

\newcommand{\bt}[1]{{\mathord{\vcenter{\hbox{\includegraphics[scale=0.4]{{{#1}}}}}}}}
\newcommand{\btm}[1]{{\mathord{\vcenter{\hbox{\includegraphics[scale=0.3]{{{#1}}}}}}}}
\newcommand{\bts}[1]{{\mathord{\vcenter{\hbox{\includegraphics[scale=0.15]{{{#1}}}}}}}}

\graphicspath{ {./figures/} }

\newcommand{\beq}{\begin{equation}}
\newcommand{\eeq}{\end{equation}}

\newcommand{\D}{\Delta}

\newcommand{\bra}{\big \langle}
\newcommand{\ket}{\big \rangle}

\newcommand{\calO}{\mathcal{O}}

\newcommand{\Ncal}{\mathcal{N}}

\newcommand{\calS}{\mathcal{S}}
\newcommand{\calC}{\mathcal{C}}
\newcommand{\bz}{{\bf z}}

\newcommand{\bZ}{{\bf Z}}

\newcommand{\bdel}{\boldsymbol{\partial}}

\newcommand{\G}{\Gamma}

\newcommand{\barz}{{\bar z}}
\newlength\Colsep
\title{Projectors and seed conformal blocks for traceless mixed-symmetry tensors}
\author{Miguel S. Costa$ ^{\dagger,\Diamond}$,}
\author{Tobias Hansen$ ^{\dagger,\ddagger}$,}
\author{Jo\~ao Penedones$ ^{\dagger,\Diamond,\sharp}$,}
\author{Emilio Trevisani$ ^{\dagger}$}
\affiliation{$ ^{\dagger}$Centro de F\'\i sica do Porto,
Departamento de F\'\i sica e Astronomia\\
Faculdade de Ci\^encias da Universidade do Porto\\
Rua do Campo Alegre 687,
4169--007 Porto, Portugal}
\affiliation{$ ^{\Diamond}$Theory Division, Department of Physics, CERN\\CH-1211 Gen\`eve 23, Switzerland}
\affiliation{$ ^{\ddagger}$II. Institut f\"ur Theoretische Physik, Universit\"at Hamburg\\ Luruper Chaussee 149, D-22761 Hamburg, Germany }
\affiliation{$ ^{\sharp}$ Fields and Strings Laboratory, Institute of Physics, EPFL\\ CH-1015 Lausanne, Switzerland}

\preprint{CERN-TH-2016-080}


\abstract{In this paper we derive the projectors to all irreducible $SO(d)$ representations (traceless mixed-symmetry tensors)
that appear in the partial wave decomposition of a conformal correlator of four stress-tensors in $d$ dimensions.
These projectors are given in a closed form for arbitrary length $l_1$ of the first row of the Young diagram.
The appearance of Gegenbauer polynomials leads directly to recursion relations in $l_1$ for seed conformal blocks.
Further results include a differential operator that generates the projectors to traceless mixed-symmetry tensors
and the general normalization constant of the shadow operator.}

\begin{document}
\maketitle

\section{Introduction}

One missing ingredient 
for the conformal bootstrap  program
\cite{Ferrara:1973yt, Polyakov:1974gs, arXiv:0807.0004,
arXiv:0905.2211, arXiv:1009.2087, arXiv:1009.2725, arXiv:1009.5985, arXiv:1109.5176, arXiv:1203.6064, arXiv:1210.4258, arXiv:1304.1803, arXiv:1307.3111, arXiv:1307.6856, arXiv:1309.5089, arXiv:1403.4545, arXiv:1404.0489, arXiv:1406.4814, arXiv:1406.4858, arXiv:1412.4127, arXiv:1412.7541, arXiv:1502.02033, arXiv:1502.07217, arXiv:1503.02081, arXiv:1504.07997, arXiv:1507.05637, arXiv:1508.00012, arXiv:1510.03866,  arXiv:1511.04065, arXiv:1511.07552, arXiv:1511.08025, arXiv:1601.03476,
arXiv:1602.02810, arXiv:1603.03771, arXiv:1603.04436}
for correlation functions with operators with spin in $d$ dimensions is the explicit knowledge of 
seed conformal blocks exchanging mixed-symmetry tensors.
Seed conformal blocks are the conformal blocks exchanging a given irreducible representation (irrep) of $SO(d)$, while having a minimal amount of spin in the external operators, such that the conformal block is unique.
They are seeds in the sense that conformal blocks for the exchange of the same representation, but for external operators with higher spin, 
can be derived from them by acting with differential operators that generate the required extra tensor 
structures, as described in \cite{Costa:2011dw, Echeverri:2015rwa, arXiv:1508.00012}.

Much of the structure of a seed conformal block is encoded by the projector to the $SO(d)$ irrep which labels the exchanged operator. This can be seen by considering the integral expressions of the conformal blocks
in the shadow formalism \cite{Ferrara:1972uq, Dolan:2011dv,SimmonsDuffin:2012uy}. Indeed, the lack of explicit expressions for the projectors is the main reason why the seed conformal blocks are still unknown.
So far, the projectors and seed conformal blocks in $d$ dimensions are only known for the exchanged operators in the irreps
$(l_1) = \bts{young_hook0}$ \cite{Dolan:2000ut,Dolan:2003hv,Dolan:2011dv} and
$(l_1,1) = \bts{young_hook1}$ \cite{Rejon-Barrera:2015bpa}.
Expressions for the projectors to the irreps $(l_1,1,1)$ and $(l_1,2)$ were given in \cite{Geyer:2000ig,Eilers:2006kd}.
Table \ref{table:blocks_in_correlators} shows all irreps that appear in a correlator of four stress-tensors,
and for each irrep the correlator where it appears in a seed conformal block.

\begin{table}[h!]
\centering
{\renewcommand{\arraystretch}{2}
\begin{tabular}{ c  c  }
Correlator & Exchanged $SO(d)$ irreps as seed conformal blocks \\ \hline
$\langle \phi_1  \phi_2 \phi_3 \phi_4 \rangle$ & $\btm{young_hook0}$ \\  
$\langle \phi_1  J_2^\mu \phi_3 J_4^\nu \rangle$ & $\btm{young_hook1}$ \\  
$\langle \phi_1  T_2^{\mu \nu} \phi_3 T_4^{\rho \sigma} \rangle$ & $\btm{young_hook2b}$ \\  
$\langle J_1^\mu  J_2^\nu J_3^\rho J_4^\sigma \rangle$ & $\btm{young_hook2b}$, \ $\btm{young_hook2a}$\\  
$\langle J_1^\mu  T_2^{\nu \rho} J_3^\sigma T_4^{\lambda \kappa} \rangle$ & $\btm{young_hook3b}$, \ $\btm{young_hook3a}$\\  
$\langle T_1^{\mu \nu}  T_2^{\rho \sigma} T_3^{\lambda \kappa} T_4^{\tau \omega} \rangle$ & $\btm{young_hook4c}$, \ $\btm{young_hook4b}$, \ $\btm{young_hook4a}$ \\ 
\end{tabular}
}
\caption[]{Exchanged irreps in correlators of currents and stress-tensors. Each line shows only the irreps exchanged as a seed block. 
For a correlator in a given line, the irreps in the lines above can  also be exchanged, but those conformal blocks can be constructed by acting with derivatives.}
\label{table:blocks_in_correlators}
\end{table}

In section \ref{sec:projectors} the projectors for all irreps appearing in Table \ref{table:blocks_in_correlators} will be derived in a compact form.
The length of the first row of the Young diagram $l_1$ is left unspecified and only appears in the final results as a parameter of Gegenbauer polynomials and in the overall normalization. A consequence of this are universal recursion relations in $l_1$ for the seed conformal blocks. These recursion relations are shown to
hold for the seed conformal blocks of all the correlators in Table \ref{table:blocks_in_correlators} and conjectured to hold for any seed conformal block of bosonic operators.
These relations are derived in section \ref{sec:recursion}, making use of 
the integral representations of the conformal blocks in the shadow formalism, where the projector appears explicitly.
Section \ref{sec:remarks} presents final remarks.

The appendices contain several other general results related to projectors and seed conformal blocks.
In appendix \ref{sec:todorov} we derive a differential operator that generates projectors to traceless mixed-symmetry tensors
for Young diagrams of two rows. This operator is a generalization of a well known operator for traceless symmetric tensors.
Appendix \ref{ref:relating_similar_projectors} deals with a relation between projectors of different irreps that arise from certain index contractions.
In appendix \ref{sec:more_projectors} we state some of the longer explicit results for projectors.
Appendix \ref{sec:shadow_constant} computes the normalization constant that arises when a shadow transformation is performed on an operator
in any three-point function that can appear in a seed conformal block.
In appendix \ref{sec:ope_limit} the OPE limit of general seed conformal blocks in the shadow formalism is computed to facilitate comparisons to other results.
Finally, in appendix \ref{sec:spherical_harmonics} we explain the relation between projectors to traceless mixed-symmetry tensors and tensor harmonics on the sphere.
Included with this work is a Mathematica notebook containing the derived projectors.

\section{Projectors to traceless mixed-symmetry tensors}
\label{sec:projectors}

\subsection{Review of projectors to traceless symmetric tensors}

As an inspiration for the ideas ahead let us briefly review how to quickly derive the
projector to traceless symmetric tensors encoded in a simple polynomial, following \cite{Dolan:2011dv}.
The projector is defined by its symmetry
\beq
\pi_{(l)}^{a_1 \ldots a_l , b_1 \ldots b_l}
= \pi_{(l)}^{(a_1 \ldots a_l) , (b_1 \ldots b_l)}\,,
\eeq
tracelessness
\beq
\delta_{a_1 a_2} \pi_{(l)}^{a_1 \ldots a_l , b_1 \ldots b_l}
= \pi_{(l)}^{a_1 \ldots a_l , b_1 \ldots b_l}  \delta_{b_1 b_2} = 0\,,
\eeq
and idempotence
\beq
\pi_{(l)}^{a_1 \ldots a_l , b_1 \ldots b_l}
\pi_{(l) \, b_1 \ldots b_l}^{\qquad \quad c_1 \ldots c_l}
= \pi_{(l)}^{a_1 \ldots a_l , c_1 \ldots c_l}\,.
\eeq
Due to the first property the projector can be implemented as a function of a single variable by
contracting with two auxiliary vectors $z^a, \barz^b \in \mathbb{R}^d$,
\beq
\pi_{(l)} (z, \bar z) 
= z^{a_1} \ldots z^{a_l}
\pi_{(l)}{\,\!}_{a_1 \ldots a_l , b_1 \ldots b_l}
{\bar z}^{b_1} \ldots {\bar z}^{b_l}
= \left( z^2 {\bar z}^2 \right)^{\frac{l}{2}} f_l(t) \,, 
\qquad 
t = \frac{z \cdot {\bar z}}{\left( z^2 {\bar z}^2 \right)^{\frac{1}{2}}}\,.
\eeq
The tracelessness can be imposed using the differential operator
\beq
\frac{\partial}{\partial z} \cdot \frac{\partial}{\partial z}  \, \pi_{\lambda}(z,\bar z) = 0\,,
\eeq
which results in a Gegenbauer differential equation for $f_l(t)$,
\beq
(t^2 - 1) f''_{l} (t) + (d-1)t f'_{l} (t) = l (l+d-2) f_{l} (t)\,,
\eeq
constraining $f_{l} (t)$ to be proportional to the Gegenbauer polynomial $C_{l}^{\left(\tfrac{d}{2}-1\right)} (t)$.
The result including normalization is \cite{Dolan:2011dv}
\beq
\pi_{(l)} (z, \bar z) 
= c_{(l)} \left( z^2 {\bar z}^2 \right)^{\frac{l}{2}} C_l^{\left(\tfrac{d}{2}-1\right)} (t)\,,
\quad \text{with } c_{(l)}=\frac{l!}{2^l \left(\tfrac{d}{2}-1\right)_l}\,,
\label{eq:symmetric_pi_as_gegenbauer}
\eeq
where $(x)_n = \frac{\G(x+n)}{\G(x)}$ is the Pochhammer symbol.

\subsection{Projectors for Young diagrams with two rows}

Next let us consider the irreps which are labeled by Young diagrams with two rows $\lambda=(l_1,l_2)$.
The total number of boxes in the Young diagram $\lambda$ will be denoted by $|\lambda|$, so in the case at hand we have $|\lambda|=l_1+l_2$.
Again we want to construct these projectors as polynomials and then impose differential equations.
In this case the projector can be encoded in a polynomial depending on four vectors
\bea
\pi_{\lambda}(z_1,z_2,\bar z_1, \bar z_2)
=z_1^{a_1}  \ldots  z_1^{a_{l_1}}
z_2^{a_{l_1 + 1}} \ldots z_2^{a_{|\lambda|}}
\pi_{\lambda} {\,\!}_{a_1 \ldots a_{|\lambda|} , b_1 \ldots b_{|\lambda|}}
\barz_1^{b_1}  \ldots  \barz_1^{b_{l_1}}
\barz_2^{b_{l_1 + 1}} \ldots \barz_2^{b_{|\lambda|}}\,.
\eea{eq:gen_poly_def}
Our strategy is now to first implement the mixed-symmetry property and then the tracelessness.
Along the way we will always make sure to keep the construction symmetric under exchange of $z_i$ and $\barz_i$,
\beq
\pi_{\lambda}(z_1,z_2,\bar z_1, \bar z_2)
=\pi_{\lambda}(\bar z_1, \bar z_2,z_1,z_2)\,.
\eeq
This means it is enough to impose conditions on one side of the projector.

The mixed symmetry or Young symmetrization of a tensor $f$ in the symmetric representation of the irrep $(l_1,l_2)$ amounts to the following two conditions\footnote{This is explained in section 2.3 of \cite{Costa:2014rya}.}
\begin{align}
\label{eq:symmetrization_ex}
f_{a_1\ldots a_{l_1} b_1\ldots b_{l_2}} &= f_{(a_1\ldots a_{l_1})(b_1\ldots b_{l_2})}\,,
\\
-f_{(a_1\ldots a_{l_1})(b_1\ldots b_{l_2})}
&= f_{(b_1 a_2\ldots a_{l_1})(a_1 b_2\ldots b_{l_2})}
+ f_{(a_1 b_1 a_3\ldots a_{l_1})(a_2 b_2\ldots b_{l_2})}
+ \ldots + f_{(a_1\ldots a_{l_1-1}b_1 )(a_{l_1} b_2\ldots b_{l_2})} \,.
\nonumber
\end{align}
The first condition is automatically satisfied after the contraction with polarizations. 
The second one can be rephrased as
\beq
z_1^{a_1}  \ldots  z_1^{a_{l_1}} z_1^{b_1}
z_2^{b_2} \ldots z_2^{b_{l_2}} f_{(a_1\ldots a_{l_1})( b_1\ldots b_{l_2})} = 0\,.
\eeq
One can see the equivalence by acting on the latter expression with
\beq
\partial_{z_1^{a_1}}  \ldots  \partial_{z_1^{a_{l_1}}} \partial_{z_1^{b_1}}
\partial_{z_2^{b_2}} \ldots \partial_{z_2^{b_{l_2}}}\,.
\eeq
Hence, the mixed-symmetry property just means that the polynomial must vanish whenever one of the vectors $z_2$ is replaced by $z_1$, or
\beq
\left( z_1 \cdot \frac{\partial}{\partial z_2}\right)  \pi_{\lambda}(z_1,z_2,\bar z_1, \bar z_2) = 0\,.
\label{eq:Young_symmetry}
\eeq
This can be easily implemented by constructing $\pi_{\lambda}(z_1,z_2,\bar z_1, \bar z_2)$ out of structures $Q_s$ that have this property
\beq
\pi_{\lambda}(z_1,z_2,\bar z_1, \bar z_2) = c_\lambda \left( z_1^2 {\bar z}_1^2 \right)^{\frac{l_1}{2}} \sum\limits_s f_s(t) Q_s(z_1,z_2,\bar z_1, \bar z_2)\,, 
\qquad
t = \frac{z_1 \cdot \bar z_1}{(z_1^2 {\bar z}_1^2)^{\frac{1}{2}}}\,.
\label{eq:pi_ansatz_2rows}
\eeq
The structures $Q_s$  have weight $l_2$ in $z_2$ and ${\bar z}_2$, and zero weight in $z_1$ and ${\bar z}_1$.
We are already familiar with such transverse structures from the construction of conformally invariant correlators in embedding space.
These structures can be built out of transverse building blocks.
In this case one can use
\begin{align}
\begin{split}
\label{eq:building_blocks}
H(z_2,\barz_2) &= z_2^a H_{ab} \barz_2^b =  z_2^a \left( \delta_{ab} - \frac{\barz_{1a} z_{1b}}{z_1 \cdot \barz_1} \right) \barz_2^b\,,\\
V(z_2) &= z_2^a V_{a} = z_2^a \left( \delta_{ab} - \frac{z_{1a} z_{1b}}{z_1^2} \right) \frac{\barz_1^b}{\sqrt{t \barz_1^2}}\,, \\
\bar{V}(\barz_2) &=\bar{V}_{b} \barz_2^b = \frac{z_1^a}{\sqrt{t z_1^2}} \left( \delta_{ab} - \frac{\barz_{1a} \barz_{1b}}{\barz_1^2} \right)  \barz_2^b\,, \\
T(z_2,z_2) &= z_2^{a_1} z_2^{a_2} T_{ a_1 a_2} = z_2^{a_1} z_2^{a_2} \left( \delta_{a_1 a_2} - \frac{z_{1a_1} z_{1a_2}}{z_1^2} \right)\,, \\
\bar{T}(\barz_2,\barz_2) &=   \bar{T}_{b_1 b_2} \barz_2^{b_1} \barz_2^{b_2} = \left( \delta_{b_1 b_2} - \frac{\barz_{1 b_1} \barz_{1b_2}}{\barz_1^2} \right) \barz_2^{b_1} \barz_2^{b_2}\,.
\end{split}
\end{align}
This construction ensures that the number of undetermined functions $f_s(t)$ is as small as possible.
Note that the dependence of the building blocks on $t$ is
\bea
H(z_2,\barz_2) &= \calO(1) + \calO(t^{-1})\,,\\
V(z_2), \bar{V}(\barz_2) &= \calO(t^{-\frac{1}{2}}) + \calO(t^{\frac{1}{2}})\,,\\
T(z_2,z_2), \bar{T}(\barz_2,\barz_2) &= \calO(1)\,.
\eea{eq:bb_t_dep}
The individual terms in the sum of \eqref{eq:pi_ansatz_2rows} can have negative powers of $t$, however
any such terms must cancel in the sum. The building blocks were chosen such that the  $f_s(t)$
are polynomials of minimal degree, i.e.\ they do not have overall factors of $t$. This can be demonstrated by considering the example $\lambda=(l_1,1)$,
and using that $\pi_{\lambda}(z_1,z_2,\bar z_1, \bar z_2)$ must be a polynomial in products of $z_1,z_2,\bar z_1, \bar z_2$, that is 
\begin{align}
&\pi_{(l_1,1)}(z_1,z_2,\bar z_1, \bar z_2)
\propto \left( z_1^2 {\bar z}_1^2 \right)^{\frac{l_1}{2}}
\left(f_1(t) H(z_2,\barz_2) + f_2(t) V(z_2) \bar{V}(\barz_2)\right)
\nonumber\\
&=
\left( z_1^2 {\bar z}_1^2 \right)^{\frac{l_1}{2}}
\left( z_2 \cdot {\bar z}_2 \right) f_1(t) 
\label{eq:f_poly_order_ex}\\
&-\left( z_1^2 {\bar z}_1^2 \right)^{\frac{l_1-1}{2}}
\left( z_1 \cdot {\bar z}_2 \right) \left( z_2 \cdot {\bar z}_1 \right) \left( \frac{f_1(t) - f_2(t)}{t} \right)
\nonumber\\
&+\left( z_1^2 {\bar z}_1^2 \right)^{\frac{l_1-2}{2}}
\Big( (z_1 \cdot z_2) (\barz_1 \cdot \barz_2)(z_1 \cdot \barz_1)
-  (z_1 \cdot z_2) (z_1 \cdot \barz_2) \barz_1^2
-  (\barz_1 \cdot z_2) (\barz_1 \cdot \barz_2) z_1^2 \Big) f_2(t) \,.
\nonumber
\end{align}
From the second, third and fourth line of this equation we conclude that
\beq
 f_1(t)=\sum\limits_{i=0}^{l_1}a_i t^i \,,\qquad
\frac{f_1(t) - f_2(t)}{t}=\sum\limits_{i=0}^{l_1-1}b_i t^i\,, \qquad
f_2(t)=\sum\limits_{i=0}^{l_1-2}c_i t^i\,,
\eeq
respectively.
Combining these three conditions yields for this case
\beq
f_1(t)=\sum\limits_{i=0}^{l_1}a_i t^i \,, \qquad
f_2(t)=\sum\limits_{i=0}^{l_1-2}c_i t^i \,, \qquad
a_0 = c_0\,.
\eeq
Furthermore, both functions have to be even to avoid a square root in $z_1^2 {\bar z}_1^2$.

These polynomials are determined by solving a coupled system of second order differential equations arising from the tracelessness conditions
\bea
\frac{\partial}{\partial z_1}\! \cdot\! \frac{\partial}{\partial z_1}  \pi_{\lambda}(z_1,z_2,\bar z_1, \bar z_2) =
\frac{\partial}{\partial z_1} \!\cdot \!\frac{\partial}{\partial z_2}  \pi_{\lambda}(z_1,z_2,\bar z_1, \bar z_2) =
\frac{\partial}{\partial z_2} \!\cdot \!\frac{\partial}{\partial z_2}  \pi_{\lambda}(z_1,z_2,\bar z_1, \bar z_2) = 0 \,.
\eea{eq:laplac}
After discussing a first example, we will describe the algorithm that is used to solve these equations in Section \ref{sec:algorithm}.
Then the structures of the individual families of projectors will be presented.
Finally, the overall normalization constants $c_\lambda$ appearing in \eqref{eq:pi_ansatz_2rows} will be computed in Section \ref{sec:normalization}.

\subsubsection{Birdtrack notation}

To construct Young symmetrized structures $Q_i$ it is convenient to use birdtrack notation, with lines denoting index contractions.
Using multiple copies of the same vector results in a group of symmetric indices, which is denoted by a white bar
\beq
 {\mathord{\vcenter{\hbox{\scalebox{0.8}{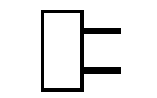}}}}} \ = z_{a_1} z_{a_2} \,,
\eeq
while a black bar denotes antisymmetrization
\beq
 {\mathord{\vcenter{\hbox{\scalebox{0.8}{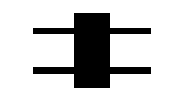}}}}} \ = \frac{1}{2}\left( \delta_{a_1 b_1} \delta_{a_2 b_2} - \delta_{a_1 b_2}\delta_{a_2 b_1}\right).
\eeq
The building blocks transverse to $z_1^a$ and to  $\barz_1^b$ that were defined in \eqref{eq:building_blocks} will be
denoted by the following symbols
\bea
H_{ab} &= \, {\mathord{\vcenter{\hbox{\scalebox{0.8}{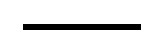}}}}}\ , \quad
&\delta_{ab}  &= {\mathord{\vcenter{\hbox{\scalebox{0.8}{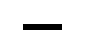}}}}} \ ,\\
V_{a} &= \, {\mathord{\vcenter{\hbox{\scalebox{0.8}{\begin{picture}(0,0)%
\includegraphics{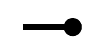}%
\end{picture}%
\setlength{\unitlength}{4144sp}%
\begingroup\makeatletter\ifx\SetFigFont\undefined%
\gdef\SetFigFont#1#2#3#4#5{%
  \reset@font\fontsize{#1}{#2pt}%
  \fontfamily{#3}\fontseries{#4}\fontshape{#5}%
  \selectfont}%
\fi\endgroup%
\begin{picture}(408,228)(526,-175)
\put(541,-106){\makebox(0,0)[b]{\smash{{\SetFigFont{12}{14.4}{\familydefault}{\mddefault}{\updefault}{\color[rgb]{0,0,0}$a$}%
}}}}
\end{picture}%
}}}}}\ , \quad
&\bar{V}_{b}  &= {\mathord{\vcenter{\hbox{\scalebox{0.8}{\begin{picture}(0,0)%
\includegraphics{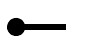}%
\end{picture}%
\setlength{\unitlength}{4144sp}%
\begingroup\makeatletter\ifx\SetFigFont\undefined%
\gdef\SetFigFont#1#2#3#4#5{%
  \reset@font\fontsize{#1}{#2pt}%
  \fontfamily{#3}\fontseries{#4}\fontshape{#5}%
  \selectfont}%
\fi\endgroup%
\begin{picture}(408,228)(958,-175)
\put(1351,-106){\makebox(0,0)[b]{\smash{{\SetFigFont{12}{14.4}{\familydefault}{\mddefault}{\updefault}{\color[rgb]{0,0,0}$b$}%
}}}}
\end{picture}%
}}}}} \ ,\\
T_{a_1 a_2} &= \, {\mathord{\vcenter{\hbox{\scalebox{0.8}{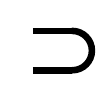}}}}}\ , \quad
&\bar{T}_{b_1 b_2} &= {\mathord{\vcenter{\hbox{\scalebox{0.8}{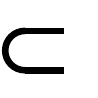}}}}} \ .
\eea{eq:birdtracks}
The $\delta_{ab}$ was defined to indicate that short lines connecting $\bt{zz_pdf}$ and $\bt{antisym_pdf}$
do not stand for $H_{ab}$. The notation should be clear after the first examples which are given both in birdtrack and explicit notation.

\subsubsection{Projectors to the irreps $(l_1,1)$}

The projectors to $SO(d)$ irreps with Young diagrams of shape $\bts{young_hook1}$ were already derived in \cite{Rejon-Barrera:2015bpa}. We include them here for completeness.
The structures are
\bea
Q_1 &= \ {\mathord{\vcenter{\hbox{\scalebox{0.8}{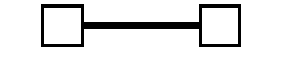}}}}} &&=  H(z_2,\barz_2)\,,\\
Q_2 &= \ {\mathord{\vcenter{\hbox{\scalebox{0.8}{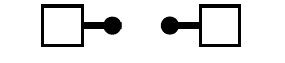}}}}} &&=  V(z_2) \bar{V}(\barz_2)\,.
\eea{eq:structures_l1}
Imposing the tracelessness conditions \eqref{eq:laplac} results in many differential equations for the functions $f_1(t)$ and $f_2(t)$. For example the
term proportional to $(z_1 \cdot z_2) (z_1 \cdot \barz_2)$ in the condition 
$\frac{\partial}{\partial z_1} \cdot \frac{\partial}{\partial z_1}  \pi_{\lambda}(z_1,z_2,\bar z_1, \bar z_2)=0$ 
is
\beq
-(l_1 - 2)(l_1+d)  f_2(t) + (d + 3) t f'_2(t) + (t^2-1) f''_2(t) = 0\,.
\eeq
This is the Gegenbauer differential equation, solved by $f_2(t) = - \partial_t^2 C_{l_1}^{\left(\tfrac{d}{2}-1\right)} (t) \propto C_{{l_1}-2}^{\left(\tfrac{d}{2}+1\right)}(t)$.
The full solution is
\bea
f_1(t) &= (d-2) t \partial_t C_{l_1}^{\left(\tfrac{d}{2}-1\right)} (t) + (t^2-1) \partial_t^2 C_{l_1}^{\left(\tfrac{d}{2}-1\right)} (t)\,,\\
f_2(t) &= -\partial_t^2  C_{l_1}^{\left(\tfrac{d}{2}-1\right)} (t)\,.
\eea{eq:f_10}
In the next section we describe the algorithm that was used to automatize the process of solving the differential equations in all other cases.

\subsubsection{Algorithm for solving the differential equations}
\label{sec:algorithm}

The algorithm for solving the systems of differential equations  we encounter is based on the assumption that
all  polynomials $f_s(t)$ can be written as a finite sum of derivatives of the Gegenbauer polynomial $C_{l_1}^{\left(\tfrac{d}{2}-1\right)} (t)$,
which will be denoted by
\beq
\calC_{l_1}^{(n)} (t) \equiv \partial_t^n C_{l_1}^{\left(\tfrac{d}{2}-1\right)} (t) = 2^n (\tfrac{d}{2}-1)_n C_{{l_1}-n}^{\left(\tfrac{d}{2}-1+n\right)}(t)\,,
\label{eq:harmonic_derivatives}
\eeq
with $l_1$-independent coefficients $w_{s,n}(d,t)$. That is, we write
\beq
f_s(t) = \sum\limits_n w_{s,n}(d,t) \calC_{l_1}^{(n)} (t) \,.
\label{eq:f_assumed_form}
\eeq
This assumption turns out to be true for all computed projectors, with the sum ranging from $n=l_2$ up to $n=2l_2$.
The overall normalization of all functions $f_s(t)$ in a projector will be chosen such that\footnote{
The functions $f_1(t)$ are special in that the corresponding $Q_1$ are chosen to contain only $H$ building blocks.}
\beq
w_{1,2 l_2}(d,t) = t^{2l_2} + \calO (t^{2l_2-2}) \,.
\label{eq:f_normalization}
\eeq
The independence on $l_1$ of the coefficients is relevant for the derivation
of recursion relations for conformal blocks.

We will use the following relation, which is a version of the Gegenbauer differential equation for the Gegenbauer polynomial appearing explicitly in
\eqref{eq:harmonic_derivatives},
\beq
(l_1 - n) (l_1 + n + d  -2 ) \calC_{l_1}^{(n)} (t) = (d + 2 n -1) t \calC_{l_1}^{(n+1)} (t) + (t^2-1) \calC_{l_1}^{(n+2)} (t)\,.
\label{eq:gegenbauer_relation}
\eeq
Using this relation repeatedly on a polynomial of the form \eqref{eq:f_assumed_form} of order $l_1$ or lower,
one can remove all but the two highest derivatives and write it in the following way
\beq
f_s(t) = \sum\limits_{i=0}^N u_{s,i}(d,l_1) t^i \calC_{l_1}^{(N)} (t) +  \sum\limits_{j=0}^{N-1} v_{s,j}(d,l_1) t^j \calC_{l_1}^{(N-1)} (t)\,,
\eeq
where $N$ is the highest $n$  appearing in the sum in \eqref{eq:f_assumed_form}.
Plugging this ansatz into the differential equations that arise from demanding tracelessness results in a system of linear equations. In all cases discussed below, these linear systems have a unique nontrivial solution for $u_{s,i}(d,l_1)$ and $v_{s,j}(d,l_1)$, as long as one chooses $N$ large enough. After finding the solution, \eqref{eq:gegenbauer_relation} can again be used to bring the solution into a form with $l_1$-independent coefficients.

\subsubsection{Projectors to the irreps $(l_1,2)$}

One can easily convince oneself that for Young diagrams of shape $\bts{young_hook2b}$ the possible combinations of the proposed building blocks are
\bea
Q_1 &= \ {\mathord{\vcenter{\hbox{\scalebox{0.8}{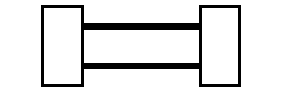}}}}} \ = H(z_2,\barz_2)^2 \,,\\
Q_2 &= \ {\mathord{\vcenter{\hbox{\scalebox{0.8}{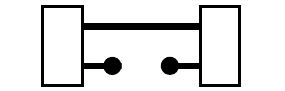}}}}} \ = H(z_2,\barz_2) V(z_2) \bar{V}(\barz_2)\,,\\
Q_3 &= \ {\mathord{\vcenter{\hbox{\scalebox{0.8}{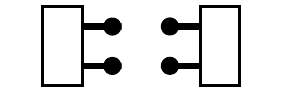}}}}} \ = V(z_2)^2 \bar{V}(\barz_2)^2\,,\\
Q_4 &= \ {\mathord{\vcenter{\hbox{\scalebox{0.8}{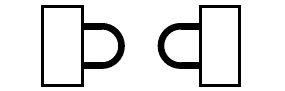}}}}} \ = T(z_2,z_2) \bar{T}(\barz_2, \barz_2)\,,\\
Q_5 &= \ {\mathord{\vcenter{\hbox{\scalebox{0.8}{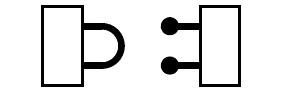}}}}} \ +  \ {\mathord{\vcenter{\hbox{\scalebox{0.8}{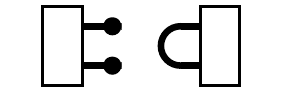}}}}} \ = T(z_2,z_2) \bar{V}(\barz_2)^2 + V(z_2)^2 \bar{T}(\barz_2, \barz_2)\,.
\eea{eq:structures_l2}
Using the algorithm for solving the tracelessness conditions one finds that the functions $f_i(t)$ can be expressed as \footnote{
Note that the overall denominator introduced in the first line is required for our normalization  \eqref{eq:f_normalization}.}
\beq
 f_i(t)= - \frac{\hat f_i(t)}{d-2}\,, \qquad  \forall i=1,\ldots,5\,,
\eeq
with
\begin{align}
 \hat f_1(t)&=(d-1) d \big(1-(d-2) t^2\big) \mathcal{C}_{l_1}^{(2)}(t)-2 (d-2) d t \big(t^2-1\big) \mathcal{C}_{l_1}^{(3)}(t)-(d-2) \big(t^2-1\big)^2
   \mathcal{C}_{l_1}^{(4)}(t)\,, \nonumber\\
 \hat f_2(t)&=2 (d-1) d \mathcal{C}_{l_1}^{(2)}(t)+2 (d-1) d t \mathcal{C}_{l_1}^{(3)}(t)+2 (d-2) \left(t^2-1\right) \mathcal{C}_{l_1}^{(4)}(t)\,, \nonumber\\
 \hat f_3(t)&=(d-1) d \mathcal{C}_{l_1}^{(2)}(t)+2 d t \mathcal{C}_{l_1}^{(3)}(t)+\left(2-d+t^2\right) \mathcal{C}_{l_1}^{(4)}(t)\,, \label{eq:f_20}\\
 \hat f_4(t)&=d \left(-2+(d-1) t^2\right) \mathcal{C}_{l_1}^{(2)}(t)+2 d t \left(t^2-1\right) \mathcal{C}_{l_1}^{(3)}(t)+\left(t^2-1\right)^2
   \mathcal{C}_{l_1}^{(4)}(t)\,, \nonumber\\
 \hat f_5(t)&=(d-1) d t \mathcal{C}_{l_1}^{(2)}(t)+2 d t^2 \mathcal{C}_{l_1}^{(3)}(t)+t \left(t^2-1\right) \mathcal{C}_{l_1}^{(4)}(t) \,.\nonumber
\end{align}
The projectors to the families of irreps $(l_1,3)$ and $(l_1,4)$ are given in Appendix \ref{sec:more_projectors}.

\subsection{Projectors for Young diagrams with three rows}

Next projectors to tensors corresponding to Young diagrams with three rows will be discussed.
Such tensors can be encoded with three auxiliary vectors $z_i$ $(i=1,2,3)$, hence  for the projector we will consider
\begin{align}
\label{eq:gen_poly_def_2}
& \pi_{\lambda}(\{z_i,\bar z_i\}) \equiv \pi_{\lambda}(z_1,z_2,z_3,\bar z_1, \bar z_2, \bar z_3) =\\
&z_1^{a_1}  \ldots  z_1^{a_{l_1}}
z_2^{a_{l_1 + 1}} \hspace{-1em} \ldots z_2^{a_{l_1+l_2}}
z_3^{a_{l_1 + l_2 + 1}} \hspace{-1em} \ldots z_3^{a_{|\lambda|}}
\pi_{\lambda\,a_1 \ldots a_{|\lambda|} , b_1 \ldots b_{|\lambda|}}
\barz_1^{b_1}  \ldots  \barz_1^{b_{l_1}}
\barz_2^{b_{l_1 + 1}} \hspace{-1em} \ldots \barz_2^{b_{l_1+l_2}}
\barz_3^{a_{l_1 + l_2 + 1}} \hspace{-1em} \ldots \barz_3^{a_{|\lambda|}} \,.
\nonumber
\end{align}
The mixed-symmetry property amounts to relations such as \eqref{eq:symmetrization_ex},
but now such relations hold between any two of the three sets of symmetrized indices.
With the same argument as in the case of two-row Young diagrams, this property can be imposed on
$\pi_{\lambda}$ by requiring
\bea
\left( z_1 \cdot \frac{\partial}{\partial z_2}\right)  \pi_{\lambda}(\{z_i,\bar z_i\}) =
\left( z_1 \cdot \frac{\partial}{\partial z_3}\right)  \pi_{\lambda}(\{z_i,\bar z_i\}) =
\left( z_2 \cdot \frac{\partial}{\partial z_3}\right)  \pi_{\lambda}(\{z_i,\bar z_i\}) = 0\,.
\eea{eq:mixed_sym_3rows}
The first two conditions can again be implemented by using the building blocks defined in \eqref{eq:building_blocks},
now also allowing $z_3, \barz_3$ in the place of $z_2, \barz_2$. The third condition can be implemented by Young symmetrizing in $z_2$ and $z_3$
(and $\barz_2, \barz_3$).
As before, this leads to the following ansatz separating structures depending on $z_2,z_3,\barz_2,\barz_3$ and functions of $t$
\beq
\pi_{\lambda}(\{z_i,\bar z_i\}) = c_\lambda \left( z_1^2 {\bar z}_1^2 \right)^{\frac{l_1}{2}} \sum\limits_s f_s(t) Q_s(z_1,z_2,z_3,\bar z_1, \bar z_2,\barz_3)\,. 
\label{eq:pi_ansatz_3rows}
\eeq
Tracelessness now amounts to the old equations \eqref{eq:laplac} and to the  three new ones 
\bea
\frac{\partial}{\partial z_1} \cdot \frac{\partial}{\partial z_3}  \pi_{\lambda}(\{z_i,\bar z_i\}) =
\frac{\partial}{\partial z_2} \cdot \frac{\partial}{\partial z_3}  \pi_{\lambda}(\{z_i,\bar z_i\}) =
\frac{\partial}{\partial z_3} \cdot \frac{\partial}{\partial z_3}  \pi_{\lambda}(\{z_i,\bar z_i\}) = 0 \;.
\eea{eq:laplac_3rows}

\subsubsection{Projectors to the irreps $(l_1,1,1)$}

For Young diagrams of shape $\bts{young_hook2a}$ there are only two possible structures, due to the antisymmetry between $z_2,z_3$ and $\barz_2,\barz_3$,
\begin{align}
Q_1 &= \ {\mathord{\vcenter{\hbox{\scalebox{0.8}{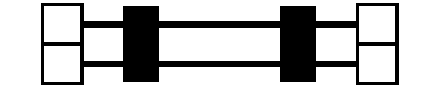}}}}} &&\!\!\!\!\!\!\!= \frac{1}{2} \Big( H(z_2,\barz_2)H(z_3,\barz_3) - H(z_2,\barz_3)H(z_3,\barz_2)\Big) \,,\nonumber\\ 
Q_2 &= \ {\mathord{\vcenter{\hbox{\scalebox{0.8}{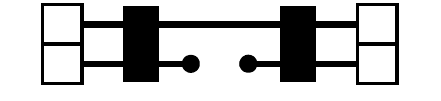}}}}} &&\!\!\!\!\!\!\!=\frac{1}{4} \Big(
H(z_2,\barz_2) V(z_3)\bar{V}(\barz_3) - H(z_2,\barz_3) V(z_3)\bar{V}(\barz_2) \label{eq:structures_l11}\\
&&&- H(z_3,\barz_2) V(z_2)\bar{V}(\barz_3)
+ H(z_3,\barz_3) V(z_2)\bar{V}(\barz_2)\Big)\,.\nonumber
\end{align}
The resulting functions $f_i(t)$ are similar as in the case of irreps $\bts{young_hook1}$ in \eqref{eq:f_10},
\bea
 f_1(t)&=(d-3) t \mathcal{C}_{l_1}^{(1)}(t)+\left(t^2-1\right) \mathcal{C}_{l_1}^{(2)}(t)\,, \\
 f_2(t)&=-2 \mathcal{C}_{l_1}^{(2)}(t) \,.
\eea{eq:f_11}

\subsubsection{Projectors to the irreps $(l_1,2,1)$}

Starting with this example, whose Young diagrams have shape $\bts{young_hook3a}$, it becomes helpful to consider some tensor products to make sure that one finds the correct number of structures. 
As explained above, the allowed structures are constructed from the building blocks in \eqref{eq:building_blocks}, including also the dependence on $z_3, \barz_3$ in the place of $z_2, \barz_2$. In order to satisfy the condition $\big( z_2 \cdot \frac{\partial}{\partial z_3} \big) Q_i = 0$ of \eqref{eq:mixed_sym_3rows} it is enough to
 Young  symmetrize in $z_2$ and $z_3$ according to the Young diagram $(l_2,l_3)=(2,1)$. In the case at hand
that means to consider
\beq
{\mathord{\vcenter{\hbox{\scalebox{0.8}{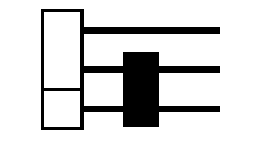}}}}}  
\ 
= z_2^{a_1} z_2^{b_1}z_3^{b_2}  \frac{1}{2}\left(    \delta^{b_1 a_2} \delta^{b_2 a_3} -\delta^{b_1 a_3} \delta^{b_2 a_2} \right) .
\label{eq:building_block_l21_1}
\eeq
Although this expression is a mixed-symmetry tensor, it is not traceless, hence it is in the reducible representation
\beq
\btm{young_11} \oplus \btm{young_1} \ .
\label{eq:reducible_rep_11}
\eeq
The only way to contract the building block $T_{a_1 a_2}$ to  expression \eqref{eq:building_block_l21_1} is
\beq
{\mathord{\vcenter{\hbox{\scalebox{0.8}{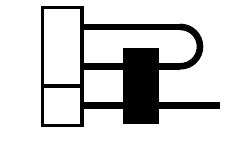}}}}} \ .
\label{eq:building_block_l21_2}
\eeq
This term contains a contraction with the primitive $SO(d)$ invariant $\delta_{a_1 a_2}$.
Thus, the number of possible contractions with $T_{a_1 a_2}$ is the same as with $\delta_{a_1 a_2}$. 
We conclude that 
the number of independent contractions of \eqref{eq:building_block_l21_1}
to the corresponding expression with $\barz_2$, $\barz_3$, 
with $H$ and $T$ building blocks in the middle,
 is given by the multiplicity of the scalar representation $\bullet$ in the 
$SO(d)$ tensor product\footnote{
All $SO(d)$ tensor products in this work are done assuming that $d$ is sufficiently large to make the tensor product $d$ independent, see \cite{Costa:2014rya}.}
\beq
\left(\, \btm{young_11} \oplus \btm{young_1}  \, \right)
\otimes
\left(\, \btm{young_11} \oplus \btm{young_1}  \, \right)\ .
\eeq
Since one can also use multiple copies of the building blocks $V$ and $\bar{V}$, which form a symmetric representation,
one can then take a further tensor product with a one-row Young diagram of any length. Thus,
the total number of birdtracks that one has to consider is the multiplicity of the scalar representation in the tensor product 
\beq
\label{eq:tensor_product_l21}
\left(\, \btm{young_11} \oplus \btm{young_1}  \, \right)
\otimes
\left(\, \btm{young_11} \oplus \btm{young_1}  \, \right)
\otimes
\sum\limits_{q=0}^\infty
\btm{young_hook_q} = 8 \bullet \oplus \ldots \,.
\eeq
Two of these eight birdtracks are combined into a single  structure ($Q_7$ below), when requiring the building blocks to respect
the left-right symmetry $z_i \leftrightarrow \barz_i$.
The resulting structures are
\begin{spreadlines}{20pt} 
\begin{gather}
Q_1 = \ {\mathord{\vcenter{\hbox{\scalebox{0.8}{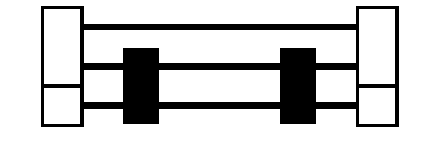}}}}} \ ,\ \ 
Q_2 = \ {\mathord{\vcenter{\hbox{\scalebox{0.8}{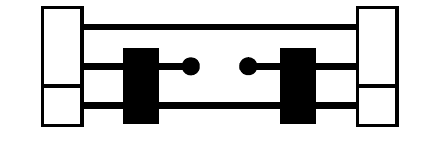}}}}} \ ,\ \ 
Q_3 = \ {\mathord{\vcenter{\hbox{\scalebox{0.8}{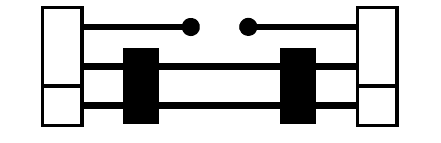}}}}} \ ,\nonumber\\
Q_4 = \ {\mathord{\vcenter{\hbox{\scalebox{0.8}{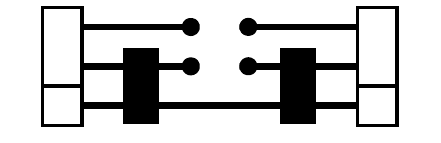}}}}} \ ,\ \ 
Q_5 = \ {\mathord{\vcenter{\hbox{\scalebox{0.8}{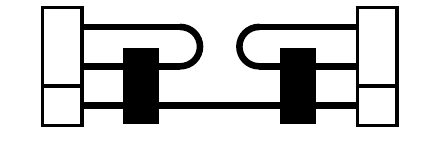}}}}} \ ,\ \ 
Q_6 = \ {\mathord{\vcenter{\hbox{\scalebox{0.8}{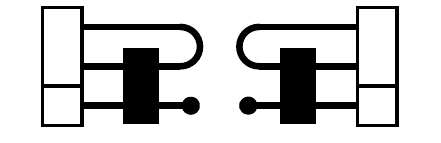}}}}} \ ,\ \nonumber\\
Q_7 = \ {\mathord{\vcenter{\hbox{\scalebox{0.8}{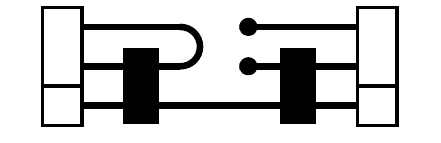}}}}} \ +  \ {\mathord{\vcenter{\hbox{\scalebox{0.8}{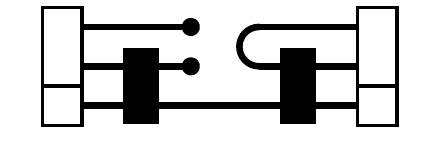}}}}} \quad.
\label{eq:structures_l21}
\end{gather}
\end{spreadlines}
It can be helpful to make a correspondence between the irreps in the tensor product
\eqref{eq:tensor_product_l21} and the expressions \eqref{eq:building_block_l21_1} and \eqref{eq:building_block_l21_2}
\beq
\btm{young_11} \ \rightarrow \ {\mathord{\vcenter{\hbox{\scalebox{0.8}{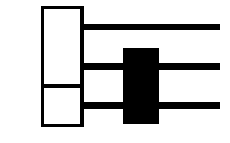}}}}} \ , \qquad
\btm{young_1} \ \rightarrow \ {\mathord{\vcenter{\hbox{\scalebox{0.8}{\input{figures/Q_l21_y1.pdf_t}}}}}} \ , \qquad
\btm{young_hook0} \ \rightarrow \ 
{\mathord{\vcenter{\hbox{\scalebox{0.8}{\begin{picture}(0,0)%
\includegraphics{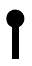}%
\end{picture}%
\setlength{\unitlength}{4144sp}%
\begingroup\makeatletter\ifx\SetFigFont\undefined%
\gdef\SetFigFont#1#2#3#4#5{%
  \reset@font\fontsize{#1}{#2pt}%
  \fontfamily{#3}\fontseries{#4}\fontshape{#5}%
  \selectfont}%
\fi\endgroup%
\begin{picture}(141,291)(-69,-2659)
\end{picture}%
}}}}} \, {\mathord{\vcenter{\hbox{\scalebox{0.8}{\begin{picture}(0,0)%
\includegraphics{Vs.pdf}%
\end{picture}%
\setlength{\unitlength}{4144sp}%
\begingroup\makeatletter\ifx\SetFigFont\undefined%
\gdef\SetFigFont#1#2#3#4#5{%
  \reset@font\fontsize{#1}{#2pt}%
  \fontfamily{#3}\fontseries{#4}\fontshape{#5}%
  \selectfont}%
\fi\endgroup%
\begin{picture}(141,291)(-69,-2659)
\end{picture}%
}}}}} \, {\mathord{\vcenter{\hbox{\scalebox{0.8}{\begin{picture}(0,0)%
\includegraphics{Vs.pdf}%
\end{picture}%
\setlength{\unitlength}{4144sp}%
\begingroup\makeatletter\ifx\SetFigFont\undefined%
\gdef\SetFigFont#1#2#3#4#5{%
  \reset@font\fontsize{#1}{#2pt}%
  \fontfamily{#3}\fontseries{#4}\fontshape{#5}%
  \selectfont}%
\fi\endgroup%
\begin{picture}(141,291)(-69,-2659)
\end{picture}%
}}}}} \cdots {\mathord{\vcenter{\hbox{\scalebox{0.8}{\begin{picture}(0,0)%
\includegraphics{Vs.pdf}%
\end{picture}%
\setlength{\unitlength}{4144sp}%
\begingroup\makeatletter\ifx\SetFigFont\undefined%
\gdef\SetFigFont#1#2#3#4#5{%
  \reset@font\fontsize{#1}{#2pt}%
  \fontfamily{#3}\fontseries{#4}\fontshape{#5}%
  \selectfont}%
\fi\endgroup%
\begin{picture}(141,291)(-69,-2659)
\end{picture}%
}}}}}\ .
\eeq
This is not a precise matching, for instance \eqref{eq:building_block_l21_1} is not traceless, however it can be helpful to find a set of independent structures
such as \eqref{eq:structures_l21}.
Using this correspondence one can map the structures to individual contributions in the tensor product \eqref{eq:tensor_product_l21} as follows
\bea
\btm{young_11} \otimes \btm{young_11} &= \bullet \oplus 2 \, \btm{young_2} \oplus \btm{young_4} \oplus \ldots &&\rightarrow \ \ Q_1,Q_2,Q_3,Q_4\,,\\
\btm{young_1} \otimes \btm{young_1} &= \bullet \oplus \btm{young_2} \oplus \ldots &&\rightarrow \ \ Q_5,Q_6\,,\\
\btm{young_11} \otimes \btm{young_1} &= \btm{young_2} \oplus \ldots &&\rightarrow \ \ Q_7\,.
\eea{eq:l21_tensor_product_contributions}
The dots here indicate Young diagrams with more than one row, which cannot contribute to the multiplicity of the scalar representation in \eqref{eq:tensor_product_l21}.
The solution of the tracelessness conditions can  then be written as
\beq
 f_i(t)= - \frac{ \hat f_i(t)}{d-3}\,, \qquad  \forall i=1,\ldots,7\,,
\eeq
with
\bea
 \hat f_1(t)={}&-(d-2) d \left(-1+(d-3) t^2\right) \mathcal{C}_{l_1}^{(2)}(t)-(d-3) (2d-1) t \left(t^2-1\right) \mathcal{C}_{l_1}^{(3)}(t)\\
   &-(d-3)
   \left(t^2-1\right)^2 \mathcal{C}_{l_1}^{(4)}(t)\,, \\
 \hat f_2(t)={}&2 (d-2) d \mathcal{C}_{l_1}^{(2)}(t)+2 \big(-5+(d-1) d\big) t \mathcal{C}_{l_1}^{(3)}(t)+2 (d-3) \left(t^2-1\right) \mathcal{C}_{l_1}^{(4)}(t)\,, \\
 \hat f_3(t)={}&(d-2) d \mathcal{C}_{l_1}^{(2)}(t)+\big(7+(d-4) d\big) t \mathcal{C}_{l_1}^{(3)}(t)+(d-3) \left(t^2-1\right) \mathcal{C}_{l_1}^{(4)}(t)\,, \\
 \hat f_4(t)={}&2 (d-2) d \mathcal{C}_{l_1}^{(2)}(t)+2 (2d-1) t \mathcal{C}_{l_1}^{(3)}(t)+2 \left(3-d+t^2\right) \mathcal{C}_{l_1}^{(4)}(t)\,, \\
 \hat f_5(t)={}&2 d \left(-2+(d-2) t^2\right) \mathcal{C}_{l_1}^{(2)}(t)+2 (2d-1) t \left(t^2-1\right) \mathcal{C}_{l_1}^{(3)}(t)+2 \left(t^2-1\right)^2
   \mathcal{C}_{l_1}^{(4)}(t) \,,\\
 \hat f_6(t)={}&-4 d \mathcal{C}_{l_1}^{(2)}(t)-2 (3+d) t \mathcal{C}_{l_1}^{(3)}(t)+\left(2-2 t^2\right) \mathcal{C}_{l_1}^{(4)}(t)\,, \\
 \hat f_7(t)={}&2 (d-2) d t \mathcal{C}_{l_1}^{(2)}(t)+2 (2d-1) t^2 \mathcal{C}_{l_1}^{(3)}(t)+2 t \left(t^2-1\right) \mathcal{C}_{l_1}^{(4)}(t)\,.
\eea{eq:f_21}

\subsubsection{Projectors to the irreps $(l_1,2,2)$}

The tensor structures for projectors to irreps $\bts{young_hook4a}$
are constructed with the reducible representation for the shape $\bts{young_02}$, given by
\beq
\btm{young_02} \oplus \btm{young_2} \oplus \bullet \,.
\eeq
Possible contractions with $T_{a_1 a_2}$ are
\beq
{\mathord{\vcenter{\hbox{\scalebox{0.8}{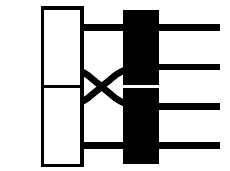}}}}}
\quad
,
\qquad
{\mathord{\vcenter{\hbox{\scalebox{0.8}{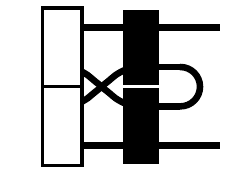}}}}}
\quad
,
\qquad
{\mathord{\vcenter{\hbox{\scalebox{0.8}{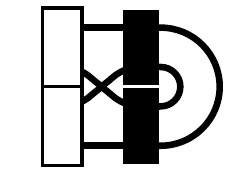}}}}}\quad
.
\eeq
Hence the number of birdtracks to consider is given by
\beq
\left(\, \btm{young_02} \oplus \btm{young_2} \oplus \bullet \, \right)
\otimes
\left(\, \btm{young_02} \oplus \btm{young_2} \oplus \bullet \, \right)
\otimes
\sum\limits_{q=0}^\infty
\btm{young_hook_q} = 11 \bullet \oplus \ldots\,.
\eeq
The individual contributions are
\bea
\btm{young_02} \otimes \btm{young_02} &= \bullet \oplus \btm{young_2} \oplus \btm{young_4} \oplus \ldots &&\rightarrow \ \ Q_1,Q_2,Q_3\,,\\
\btm{young_2} \otimes \btm{young_2} &= \bullet \oplus \btm{young_2} \oplus \btm{young_4} \oplus \ldots &&\rightarrow \ \ Q_4,Q_5,Q_6\,,\\
\bullet \otimes \bullet &= \bullet &&\rightarrow \ \ Q_7\,,\\
\btm{young_02} \otimes \btm{young_2} &= \btm{young_2} \oplus \ldots &&\rightarrow \ \ Q_8\,,\\
\btm{young_2} \otimes \bullet &= \btm{young_2} &&\rightarrow \ \ Q_9\,,
\eea{eq:l22_tensor_product_contributions}
with
\begin{spreadlines}{20pt} 
\begin{gather}
Q_1 = \ {\mathord{\vcenter{\hbox{\scalebox{0.8}{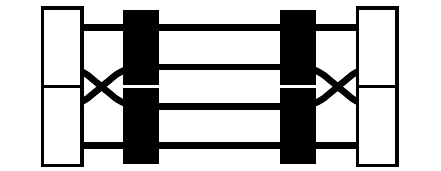}}}}} \ ,\ \
Q_2 = \ {\mathord{\vcenter{\hbox{\scalebox{0.8}{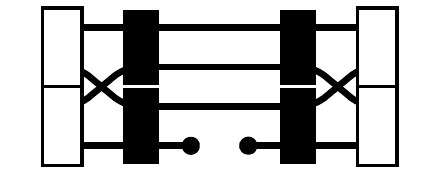}}}}} \ ,\ \
Q_3 = \ {\mathord{\vcenter{\hbox{\scalebox{0.8}{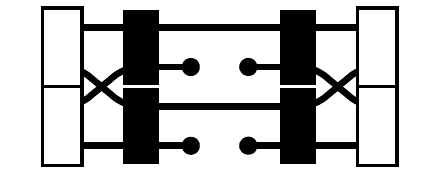}}}}} \ ,\nonumber\\
Q_4 = \ {\mathord{\vcenter{\hbox{\scalebox{0.8}{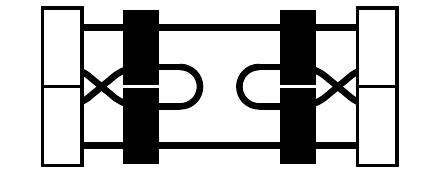}}}}} \ ,\ \
Q_5 = \ {\mathord{\vcenter{\hbox{\scalebox{0.8}{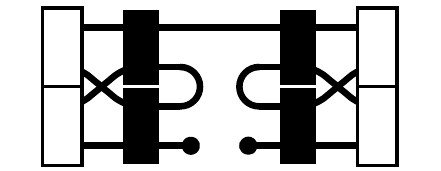}}}}} \ ,\ \
Q_6 = \ {\mathord{\vcenter{\hbox{\scalebox{0.8}{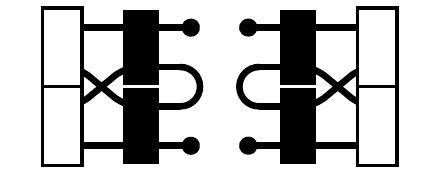}}}}} \ ,\nonumber\\
Q_7 = \ {\mathord{\vcenter{\hbox{\scalebox{0.8}{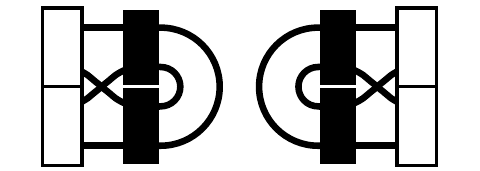}}}}} \ ,\quad \hspace{8pt} 
Q_8 = \ {\mathord{\vcenter{\hbox{\scalebox{0.8}{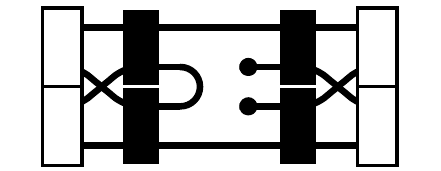}}}}} \ +  \ {\mathord{\vcenter{\hbox{\scalebox{0.8}{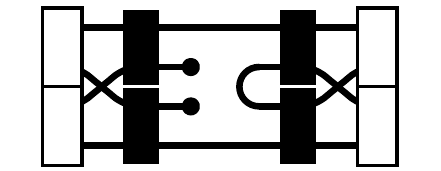}}}}} \ ,\nonumber\\
Q_9 = \ {\mathord{\vcenter{\hbox{\scalebox{0.8}{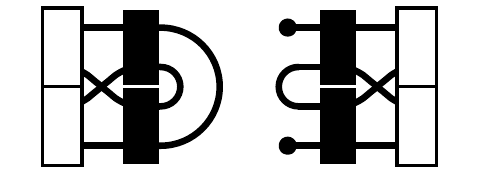}}}}} \ +  \ {\mathord{\vcenter{\hbox{\scalebox{0.8}{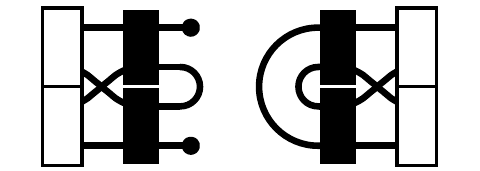}}}}} \ .
\label{eq:structures_l22}
\end{gather}
\end{spreadlines}
In this case the functions $f_i(t)$ can be written as
\beq
f_i(t)=-  \frac{\hat f_i(t)} {(d-4)(d-3)}\,, \qquad  \forall i=1,\ldots,9\,,
 \eeq
with
\begin{align}
 \hat f_1(t)={}&-(d-3)_3 \left(-1+(d-4) t^2\right) \mathcal{C}_{l_1}^{(2)}(t)-2 (d-4)_2 (d-1) t \left(t^2-1\right) \mathcal{C}_{l_1}^{(3)}(t)\nonumber\\
   &-(d-4)_2 \left(t^2-1\right)^2 \mathcal{C}_{l_1}^{(4)}(t)\,, \nonumber\\
 \hat f_2(t)={}&4 (d-3)_3 \mathcal{C}_{l_1}^{(2)}(t)+4 (d-3)^2 (d-1) t \mathcal{C}_{l_1}^{(3)}(t)+4 (d-4)_2 \left(t^2-1\right)
   \mathcal{C}_{l_1}^{(4)}(t)\,, \nonumber\\
 \hat f_3(t)={}&4 (d-3)_3 \mathcal{C}_{l_1}^{(2)}(t)+8 (d-3) (d-1) t \mathcal{C}_{l_1}^{(3)}(t)+4 (d-3) \left(4-d+t^2\right) \mathcal{C}_{l_1}^{(4)}(t)\,, \nonumber\\
 \hat f_4(t)={}&4 (d-2) (d-1) \left(-2+(d-3) t^2\right) \mathcal{C}_{l_1}^{(2)}(t)+8 (d-3) (d-1) t \left(t^2-1\right) \mathcal{C}_{l_1}^{(3)}(t)\nonumber\\
   &+4 (d-3) \left(t^2-1\right)^2 \mathcal{C}_{l_1}^{(4)}(t) \,,
\label{eq:f_22}\\
 \hat f_5(t)={}&-16 (d-2) (d-1) \mathcal{C}_{l_1}^{(2)}(t)-8 (d-1)^2 t \mathcal{C}_{l_1}^{(3)}(t)-8 (d-3) \left(t^2-1\right) \mathcal{C}_{l_1}^{(4)}(t)\,, \nonumber\\
 \hat f_6(t)={}&-8 (d-2) (d-1) \mathcal{C}_{l_1}^{(2)}(t)-16 (d-1) t \mathcal{C}_{l_1}^{(3)}(t)+4 \left(-3+d-2 t^2\right) \mathcal{C}_{l_1}^{(4)}(t)\,, \nonumber\\
 \hat f_7(t)={}&-2 (d-1) \left(-3+(d-2) t^2\right) \mathcal{C}_{l_1}^{(2)}(t)-4 (d-1) t \left(t^2-1\right) \mathcal{C}_{l_1}^{(3)}(t)-2 \left(t^2-1\right)^2
   \mathcal{C}_{l_1}^{(4)}(t)\,, \nonumber\\
 \hat f_8(t)={}&4 (d-3)_3 t \mathcal{C}_{l_1}^{(2)}(t)+8 (d-3) (d-1) t^2 \mathcal{C}_{l_1}^{(3)}(t)+4 (d-3) t \left(t^2-1\right)
   \mathcal{C}_{l_1}^{(4)}(t)\,, \nonumber\\
 \hat f_9(t)={}&-4 (d-2) (d-1) t \mathcal{C}_{l_1}^{(2)}(t)-8 (d-1) t^2 \mathcal{C}_{l_1}^{(3)}(t)-4 t \left(t^2-1\right) \mathcal{C}_{l_1}^{(4)}(t)\,.\nonumber
\end{align}
The projector to the family of irreps $(l_1,3,1)$ is given in Appendix \ref{sec:more_projectors}.

\subsection{Normalization of the projectors}
\label{sec:normalization}

The normalizations can be computed by using that the projectors have a term with a known normalization,
namely the term projecting to generic mixed-symmetry tensors, from which the traces are subtracted by further terms.
This is the only term in $\pi_\lambda(z_1,z_2,z_3,\barz_1,\barz_2,\barz_3)$ containing a factor $(z_1\cdot \barz_1)^{l_1}(z_2\cdot \barz_2)^{l_2}(z_3\cdot \barz_3)^{l_3}$.
For concreteness let us consider the Young diagram $\lambda=\bts{young_hook3a}$. In this section all lines in birdtracks denote just simple contractions, without any of the
definitions of \eqref{eq:birdtracks}
\beq
\pi_\lambda(z_1,z_2,z_3,\barz_1,\barz_2,\barz_3)=
n_\lambda
\quad
{\mathord{\vcenter{\hbox{\scalebox{0.8}{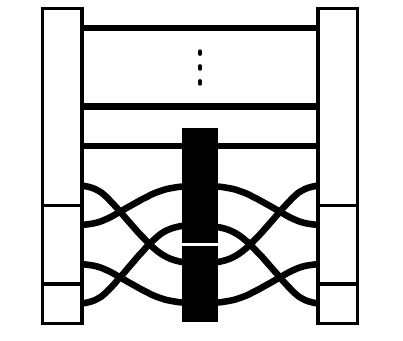}}}}}
\quad
+ \text{ trace subtractions}.
\label{eq:young_norm_bt}
\eeq
The antisymmetrizations appearing here are defined as
\bea
\btm{def_asym} &= \frac{1}{n!} \left\{ \btm{def_0twists} - \btm{def_1twists} + \btm{def_2twists} - \ldots \right\} ,
\eea{eq:def_bt_asym}
and the normalization of the first term is \cite{Cvitanovic:2008zz}
\beq
n_{\lambda} = \frac{\prod\limits_{i=1}^{l_1} h_i! \prod\limits_{j=1}^{h_1} l_j! }{H(\lambda)} \,, 
\quad 
H(\lambda) = \prod\limits_{i=1}^{l_1} \prod\limits_{j=1}^{h_i} (l_j - i + h_i - j + 1)\,,
\label{eq:n_lambda}
\eeq
where $h_i$ is the height of the $i^{\text{th}}$ column of the Young diagram $\lambda$.
Using the identity
\bea
\btm{def_asym} &= \frac{1}{n} \left\{ \btm{def_asym_rel1} - (n-1)\btm{def_asym_rel2}\right\} ,
\eea{eq:rel_bt_asym}
on both antisymmetrizations in \eqref{eq:young_norm_bt}, one can isolate the terms containing $(z_1 \cdot \barz_1)^{l_1}$,
\bea
{\mathord{\vcenter{\hbox{\scalebox{0.8}{\input{figures/norm1.pdf_t}}}}}}
\quad &= \frac{1}{3 \cdot 2} \quad
{\mathord{\vcenter{\hbox{\scalebox{0.8}{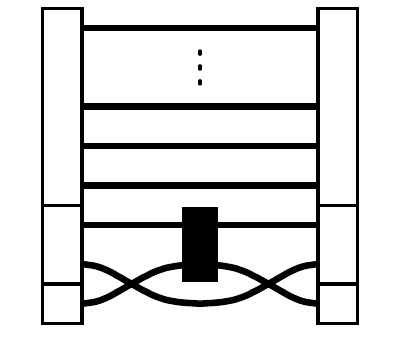}}}}} \quad + \calO\big( (z_1 \cdot \barz_1)^{l_1-1} \big)\\
&= \frac{1}{3 \cdot 2} (z_1 \cdot \barz_1)^{l_1} Q_1(z_2,z_3,\barz_2,\barz_3) + \calO\big( (z_1 \cdot \barz_1)^{l_1-1} \big)\,.
\eea{eq:norm_birdtrack_trick}
In general this factor $3\cdot 2$ will be $\prod\limits_{i=1}^{l_1} h_i$ and the part of the birdtrack with $z_2,z_3,\barz_2,\barz_3$ always matches the $z_1$ and $\barz_1$ independent part of $Q_1(z_2,z_3,\barz_2,\barz_3)$.
Comparing to \eqref{eq:pi_ansatz_3rows} this means that the normalization constant appearing in the projectors is
\beq
c_\lambda =  \frac{\prod\limits_{i=1}^{l_1} (h_i-1)! \prod\limits_{j=1}^{h_1} l_j! }
{b_\lambda H(\lambda)} \,,
\eeq
where $b_\lambda$ is the coefficient of $t^{l_1}$ in $f_1(t)$,
\beq
f_1(t) = b_\lambda t^{l_1} + \calO(t^{l_1-2})\,,
\eeq
and can be easily computed in each case using
\beq
C_{n}^{(\epsilon)} (t) = \frac{(\epsilon)_n}{n!} (2t)^n + \mathcal{O}(t^{n-2})\,.
\eeq
Doing this for each family of projectors one finds
\beq
b_\lambda = \frac{ 2^{l_1} (\frac{d}{2}-1)_{l_1} \prod\limits_{i=1}^{l_2}(l_1+d-h_i+i-2)}{(l_1-l_2)!}\,.
\eeq

\section{Recursion relations for seed conformal blocks}
\label{sec:recursion}

In this section recursion relations in $l_1$ will be read off from the integral representation of
conformal blocks in the shadow formalism. 
The recursion relations are based on the observation
that the projectors $\pi_\lambda$, as computed in the previous section, are linear combinations of just a few (at most $l_2 +1$) different Gegenbauer polynomials, 
\beq
\pi_{\lambda}(z_1,z_2,z_3,\bar z_1, \bar z_2,\barz_3) = c_\lambda \left( z_1^2 {\bar z}_1^2 \right)^{\frac{l_1}{2}} \sum\limits_{n=l_2}^{\text{min}(l_1,2l_2)} K^{\lambda}_n(z_1,z_2,z_3,\bar z_1, \bar z_2, \barz_3) \calC^{(n)}_{l_1} (t)\,, 
\label{eq:pi_Gegenbauer_sum}
\eeq
where $K^{\lambda}_n(z_1,z_2,z_3,\bar z_1, \bar z_2, \barz_3)$ does not depend on $l_1$,
and the sum stops at $\text{min}(l_1,2l_2)$ since $\calC^{(n)}_{l_1} (t) = 0$ for $n>l_1$.
We showed this for the projectors to the irreps that can appear in conformal blocks of four stress-tensors
by explicit computation.
This fact can be used to turn the recursion relation for the Gegenbauer polynomials
\beq
(l_1 - n) \calC_{l_1}^{(n)} (t) = (2l_1+d -4) t \calC_{l_1-1}^{(n)} (t) - (l_1 + n + d - 4) \calC_{l_1-2}^{(n)} (t)\,,
\label{eq:Gegenbauer_recursion}
\eeq
into recursion relations for the conformal blocks.

\subsection{Classification of seed conformal blocks}
\label{sec:classification_of_seed_conformal_blocks}
We will focus on conformal blocks that are unique  given the irreps of the external and exchanged operators.
These will be called \emph{seed conformal blocks}.
The origin of this terminology is the fact that conformal blocks involving three-point functions with multiple tensor structures
can be derived from these seed blocks by acting with differential operators, using the method of \cite{Costa:2011dw}.

Uniqueness of a conformal block exchanging the irrep $\lambda = (l_1,l_2,\ldots)$ 
in the channel $\lambda_1 \lambda_2 \to \lambda_3 \lambda_4$, means that the three-point functions $\bra \lambda_1 \lambda_2 \lambda \ket$ and
$\bra \lambda \lambda_3 \lambda_4 \ket$ both have a single OPE coefficient.
For such combinations of irreps the tensor product contains only one symmetric tensor (or as special cases  one scalar or  vector), with multiplicity one
\beq
\lambda_1 \otimes \lambda_2 \otimes \lambda  = \btm{young_hook0} \oplus \text{diagrams with }h_1 > 1 \, ,
\label{eq:single_structure_cond}
\eeq
and similar for $\lambda_3$, $\lambda_4$.
The most trivial case is when $\lambda_1$ and $\lambda_2$ are scalars, and $\lambda$ is a symmetric tensor.
We will consider the case when $\lambda_1$ and $\lambda_2$ are symmetric tensors, and $\lambda$ is a mixed-symmetry tensor.
In this case the lower rows of $\lambda$, which will be denoted by the Young diagram $\lambda^- = (l_2,l_3,\ldots)$,
must be removed by contraction to $\lambda_1$ and $\lambda_2$.
To understand this better let us consider three cases.
\begin{itemize}
\item[1.]
$|\lambda_1| + |\lambda_2| < | \lambda^-|$.
$\lambda_1$ and $\lambda_2$ do not have enough indices to remove all lower rows of $\lambda$. There is no solution to \eqref{eq:single_structure_cond}.
\item[2.]
$|\lambda_1| + |\lambda_2| = | \lambda^-|$.
If $\lambda^-$ appears in the tensor product of $\lambda_1$ and $\lambda_2$ with multiplicity one, then \eqref{eq:single_structure_cond}
is satisfied and the symmetric tensor on the right hand side is of rank $l_1$.
\item[3.]
$|\lambda_1| + |\lambda_2| > | \lambda^-|$.
If there are symmetric tensors in the tensor product \eqref{eq:single_structure_cond}, then there are more than one.
Consider for example $|\lambda_1| + |\lambda_2| = | \lambda^-|+1$. After removing the lower rows of $\lambda$, there
is still a tensor product of $\bts{young_1}$ and the remaining first row of $\lambda$, hence there will be symmetric tensors
of rank $l_1-1$ and $l_1+1$.
\end{itemize}
Hence the second case  is the only relevant one, leading to the necessary condition for seed conformal blocks
(for $\lambda_1$, $\lambda_2$, $\lambda_3$, $\lambda_4$ being symmetric tensors)
\beq
l_1 = |\lambda| - |\lambda_1| - |\lambda_2| = |\lambda| - |\lambda_3| - |\lambda_4| \,.
\label{eq:lrelation}
\eeq
To see this condition in action one can consider the OPE. For example,
the OPE of two vector operators 
has only a single term in the irreps $\lambda=(l_1,2)$ or $\lambda=(l_1,1,1)$,
both of the form
\beq
\calO_{\Delta_1,\bts{young_1}}^b (x_1) \calO_{\Delta_2,\bts{young_1}}^c (x_2)  
\underset{x_{12}\to 0}{\sim}
\frac{(x_{12})_{a_1} \ldots (x_{12})_{a_{l_1}} \delta_{a_{l_1+1}}^b \delta_{a_{l_1+2}}^c}{|x_{12}|^{\Delta_{1}+\Delta_2 - \Delta + l_1}}\,
\calO_{\Delta,\lambda}^{a_1 \ldots a_{|\lambda|}} (x_1) + \ldots \,.
\eeq

\subsection{Three-point functions}

We will use the embedding formalism and the methods to construct correlators of \cite{Costa:2011mg,Costa:2014rya}.
The embedding space is $d+2$ dimensional Minkowski space $\mathbb{R}^{d+1,1}$, where the conformal group $SO(d+1,1)$
acts linearly. Capital letters will denote vectors in this space $P_i, Z_{ij} \in \mathbb{R}^{d+1,1}$.
The second label on $Z_{ij}$ labels the different vectors required to encode mixed-symmetry tensors, e.g.\
$(z_1,z_2,z_3)$ from the previous section could be replaced by $(Z_{01},Z_{02},Z_{03})$ in embedding space.
Contractions of tensors can be written using derivatives acting on vectors
\beq
\delta_{B_1}^{(A_1}
\ldots
\delta_{B_{l_1}}^{A_{l_1})}
=\frac{1}{l_1!}
\partial_{Z_{i1}}^{A_1} \ldots
\partial_{Z_{i1}}^{A_{l_1}}
Z_{i1 \, B_{1}} \ldots
Z_{i1 \, B_{l_1}}
\,.
\label{eq:contraction}
\eeq
Furthermore, boldface letters indicate a set of vectors that are used to encode a mixed-symmetry tensor
and the corresponding sets of derivatives, normalized to include the factorial appearing in \eqref{eq:contraction},
\bea
\bZ_i &= (Z_{i1},Z_{i2},\ldots)\,,\\
\bdel_{\bZ_i} &= \left( (l_1!)^{-1/l_1} \partial_{Z_{i1}}, (l_2!)^{-1/l_2} \partial_{Z_{i2}}, \ldots \right) .
\eea{eq:boldface_notation}
It is convenient to consider polynomials that {\em encode} the correlators without fully implementing the
symmetry and tracelessness of the operators. The {\em full} correlators can then be obtained by projecting
\bea
\bra \calO_1(P_1,\bZ_1) \calO_2(P_2,\bZ_2) \calO(P_0,\bZ_{0}) \ket_\text{full} ={}& 
\pi_{\lambda_1} (\bZ_1, \bdel_{\bar \bZ_1})\,
\pi_{\lambda_2} (\bZ_2, \bdel_{\bar \bZ_2})\,
\pi_{\lambda} (\bZ_0, \bdel_{\bar \bZ_0})\\
&\bra \calO_1(P_1,\bar \bZ_1) \calO_2(P_2, \bar \bZ_2) \calO(P_0,\bar \bZ_0) \ket_{\text{enc}}\,.
\eea{eq:encoding_correlators}
Throughout the paper, the external operators will mostly be symmetric traceless tensors,
which we will encode by a single null vector instead of projecting to a traceless tensor.
Instead of \eqref{eq:encoding_correlators} we will thus use
\bea
\bra \calO_1(P_1,Z_1) \calO_2(P_2,Z_2) \calO(P_0,\bZ_{0}) \ket =
\pi_{\lambda} (\bZ_0, \bdel_{\bar \bZ_0})
\bra \calO_1(P_1,Z_1) \calO_2(P_2, Z_2) \calO(P_0,\bar \bZ_0) \ket_{\text{enc}} \,,
\eea{eq:encoding_correlators_2}
where $Z_1^2=Z_2^2=0$ will always be implied.
When considering conformal blocks for external operators that are either scalars, currents or stress-tensors, the three-point functions with a single tensor structure are
encoded by
\bea
\bra \calO_1 \big(P_1, Z_1\big) \calO_2 \big(P_2, Z_2\big) \calO \big(P_0, \bZ_0 \big) \ket_{\text{enc}}
= \frac{\left(V_{0,12}^{(Z_{01})} \right)^{l_1} N^{\lambda_1 \lambda_2 \lambda}_{120} (Z_{02},Z_{03}) }{
 (P_{12})^{\frac{\Delta_1+\Delta_2-\Delta}{2}}
(P_{20})^{\frac{\Delta_2+\Delta-\Delta_1}{2}} 
{(P_{01})^{\frac{\Delta+\Delta_1-\Delta_2}{2}}} }\,,
\eea{eq:3pt_functions_rec_12}
where $P_{ij}=-2 P_i \cdot P_j$ and
\bea
N^{\bullet \bullet \, \bts{young_hook0}}_{120}
&= 1\,,\\
N^{\bts{young_1} \, \bullet \, \bts{young_hook1}}_{120}(Z_{02})
&= H_{10}^{(Z_1,Z_{02})}\,,\\
N^{\bts{young_2} \, \bullet \, \bts{young_hook2b}}_{120}(Z_{02})
&= \left( H_{10}^{(Z_1,Z_{02})} \right)^2\,,\\
N^{\bts{young_1} \, \bts{young_1} \, \bts{young_hook2b}}_{120}(Z_{02})
&= H_{10}^{(Z_1,Z_{02})} H_{20}^{(Z_2,Z_{02})}\,,\\
N^{\bts{young_1} \, \bts{young_1} \, \bts{young_hook2a}}_{120}(Z_{02},Z_{03})
&= H_{10}^{(Z_1,Z_{02})} H_{20}^{(Z_2,Z_{03})}\,,\\
N^{\bts{young_2} \, \bts{young_1} \, \bts{young_hook3b}}_{120}(Z_{02})
&= \left( H_{10}^{(Z_1,Z_{02})} \right)^2 H_{20}^{(Z_2,Z_{02})}\,,\\
N^{\bts{young_2} \, \bts{young_1} \, \bts{young_hook3a}}_{120}(Z_{02},Z_{03})
&= \left( H_{10}^{(Z_1,Z_{02})} \right)^2 H_{20}^{(Z_2,Z_{03})}\,,\\
N^{\bts{young_2} \, \bts{young_2} \, \bts{young_hook4c}}_{120}(Z_{02})
&= \left( H_{10}^{(Z_1,Z_{02})} \right)^2 \left(H_{20}^{(Z_2,Z_{02})}\right)^2\,,\\
N^{\bts{young_2} \, \bts{young_2} \, \bts{young_hook4b}}_{120}(Z_{02},Z_{03})
&= \left( H_{10}^{(Z_1,Z_{02})} \right)^2 H_{20}^{(Z_2,Z_{02})} H_{20}^{(Z_2,Z_{03})}\,,\\
N^{\bts{young_2} \, \bts{young_2} \, \bts{young_hook4a}}_{120}(Z_{02},Z_{03})
&= \left( H_{10}^{(Z_1,Z_{02})} \right)^2 \left(H_{20}^{(Z_2,Z_{03})}\right)^2\,,
\eea{eq:N_examples}
with
\bea
H_{ij}^{(Z_i,Z_j)} 
&= \frac{\big(Z_i \cdot Z_j\big) \big(P_i \cdot P_j\big) -  \big(P_j \cdot Z_i\big) \big(P_i \cdot Z_j\big)}{P_i \cdot P_j} \,,\\
V_{i,jk}^{(Z)} 
&=\frac{\big(P_j \cdot Z\big)\big( P_{i} \cdot P_k \big)- \big(P_j \cdot P_{i}\big)\big( Z \cdot P_k\big)}
{\sqrt{ -2 \big(P_i \cdot P_j\big) \big(P_i \cdot P_k\big) \big(P_j \cdot P_k\big)}} \,.
\eea{eq:HV_building_blocks}

\subsection{Recursion relations from the shadow formalism}

Next we recall the formula for the conformal partial wave in the shadow formalism (an overview of the formalism in the case of mixed-symmetry tensors can be found in Section 5 of \cite{Costa:2014rya}). When using the three-point functions defined above,
the exchanged representation must be projected to an irreducible representation
\begin{align}
\label{eq:conf_partial_wave}
W_{\Delta,\lambda}^{\lambda_1 \lambda_2 \lambda_3 \lambda_4} ={}&  \frac{1}{\calS_{\tilde \Delta \Delta_{34}}^\lambda } \int D^d P_0 \,
 \bra \calO_1 \big(P_1, Z_1\big) \calO_2 \big(P_2, Z_2\big) \calO \big(P_0, \bdel_{\bZ_0}\big) \ket_{\text{enc}}\\
& \pi_\lambda (\bZ_0,\bdel_{\bar \bZ_0})
\bra \calO_3 \big(P_3; Z_3\big) \calO_4 \big(P_4; Z_4\big) \calO \big(P_0, \bar \bZ_0 \big) \ket_{\text{enc}}
\Big|_{ \Delta \to \tilde \Delta}\,.
\nonumber
\end{align}
Here $\tilde \Delta = d-\Delta$ is the conformal dimension of the shadow operator $\tilde \calO$
and $\calS_{\tilde \Delta \Delta_{34}}^\lambda = \calS_{\Delta \Delta_{34}}^\lambda |_{\Delta \to \tilde\Delta}$ is the constant that occurs when an operator in a three-point function is replaced by its shadow
\beq
\calS_{\Delta \Delta_{34}}^\lambda 
= \frac{\bra  \calO_3 \big(P_3; \bZ_3\big) \calO_4 \big(P_4; \bZ_4\big) \tilde  \calO \big(P_0; \bZ_0\big)
\ket }
{\bra \calO_3 \big(P_3; \bZ_3\big) \calO_4 \big(P_4; \bZ_4\big) \calO \big(P_0; \bZ_0\big) \ket \big|_{ \Delta \to  \tilde \Delta}} \,.
\label{eq:shadow_3pt_constant_def}
\eeq
It is computed in Appendix \ref{sec:shadow_constant} for any three-point function that can appear in a seed conformal block, with the result
\bea
\calS_{\tilde \Delta \Delta_{34}}^{\lambda} 
\!\!= \pi^{d/2} 
\prod\limits_{i=1}^{l_1} \!\left( \tilde \Delta - h_i + i - 1 \right)
\frac{\G \big(\tilde \Delta-\frac{d}{2}\big)\G\big(\frac{1}{2}(\Delta+\Delta_{34}+l_1)\big)\G\big(\frac{1}{2}(\Delta-\Delta_{34}+l_1)\big)}
{\G(\Delta+l_1)\G\big(\frac{1}{2}(\tilde \Delta+\Delta_{34}+l_1)\big)\G\big(\frac{1}{2}(\tilde\Delta -\Delta_{34}+l_1)\big)}\,.
\eea{eq:shadow_constant34}

We start by inserting the expressions for the three-point functions \eqref{eq:3pt_functions_rec_12}
into \eqref{eq:conf_partial_wave}, to find
\begin{align}
\label{eq:partial_wave_with_N}
&W_{\Delta,\lambda}^{\lambda_1 \lambda_2 \lambda_3 \lambda_4} =
\frac{1}{\calS_{\tilde \Delta \Delta_{34}}^{\lambda} (l_2!l_3!)^2}\\
& \int D^d P_0
\frac{
N^{\lambda_1 \lambda_2 \lambda}_{120}(\partial_{Z_{02}},\partial_{Z_{03}}) \,
\pi_\lambda \Big(\partial_{Z_{01}} V_{0,12}^{(Z_{01})},Z_{02},Z_{03},\partial_{{\bar Z_{01}}}V_{0,34}^{({\bar Z_{01}})} , \partial_{{\bar Z_{02}}}, \partial_{{\bar Z_{03}}}\Big)
N^{\lambda_3 \lambda_4 \lambda}_{340}({\bar Z_{02}},{\bar Z_{03}})
}{
{(P_{01})^{\frac{\Delta+\Delta_{12}}{2}}} (P_{20})^{\frac{\Delta-\Delta_{12}}{2}} 
 {(P_{03})^{\frac{\tilde \Delta+\Delta_{34}}{2}}}  (P_{40})^{\frac{\tilde \Delta-\Delta_{34}}{2}}
}\,.\nonumber
\end{align}
Next we wish to insert the expression \eqref{eq:pi_Gegenbauer_sum} for the projector $\pi_\lambda$.
The factor  $\left( z_1^2 {\bar z}_1^2 \right)^{\frac{l_1}{2}}$
in \eqref{eq:pi_Gegenbauer_sum} is equal to 1 due to
$
\left( \partial_{Z_{01}} V_{0,12}^{(Z_{01})} \right)^2 = 
\left( \partial_{{\bar Z_{01}}} V_{0,34}^{({\bar Z_{01}})} \right)^2 = 1
$.
Since the factors $\calS_{\tilde \Delta \Delta_{34}}^{\lambda}$ appearing in \eqref{eq:partial_wave_with_N} and $c_\lambda$ in \eqref{eq:pi_Gegenbauer_sum}
depend on $l_1$, it is convenient to use the following normalization for the conformal blocks
\footnote{As explained in \cite{SimmonsDuffin:2012uy}, to obtain the conformal block one needs to perform a monodromy projection of the integral (\ref{eq:conf_partial_wave}). This is assumed but not written explicitly because it is not important for the present discussion.}
\beq
\label{defGfromW}
G_{\Delta,\lambda}^{\lambda_1 \lambda_2 \lambda_3 \lambda_4}
= 
\frac{\calS_{\tilde \Delta \Delta_{34}}^\lambda}{c_\lambda}
(P_{12})^{\frac{\Delta}{2}} (P_{34})^{\frac{\tilde \Delta}{2}}
\left( \frac{P_{14}}{P_{24}} \right)^{\frac{\Delta_{12}}{2}} 
\left( \frac{P_{13}}{P_{14}} \right)^{\frac{\Delta_{34}}{2}} 
W_{\Delta,\lambda}^{\lambda_1 \lambda_2 \lambda_3 \lambda_4} .
\eeq
By inserting \eqref{eq:pi_Gegenbauer_sum} into \eqref{eq:partial_wave_with_N}
we can now write the conformal block as a sum of functions that only depend on a single Gegenbauer polynomial each,
\beq
G_{\Delta,\lambda}^{\lambda_1 \lambda_2 \lambda_3 \lambda_4}
= \sum\limits_{n=l_2}^{\text{min}(l_1,2l_2)} F_{\Delta,\lambda, n}^{\lambda_1 \lambda_2 \lambda_3 \lambda_4} (\Delta_{12}, \Delta_{34})\,,
\label{eq:cb_split}
\eeq
which are defined as
\begin{align}
\label{eq:f_integral}
&F_{\Delta,\lambda, n}^{\lambda_1 \lambda_2 \lambda_3 \lambda_4} (\Delta_{12}, \Delta_{34}) =
\frac{1}{ (l_2!l_3!)^2}
(P_{12})^{\frac{\Delta}{2}} (P_{34})^{\frac{\tilde \Delta}{2}}
\left( \frac{P_{14}}{P_{24}} \right)^{\frac{\Delta_{12}}{2}} 
\left( \frac{P_{13}}{P_{14}} \right)^{\frac{\Delta_{34}}{2}}
\int D^d P_0\\
&\frac{
N^{\lambda_1 \lambda_2 \lambda}_{120}(\partial_{Z_{02}},\partial_{Z_{03}})\,
K_n^\lambda \Big(\partial_{Z_{01}} V_{0,12}^{(Z_{01})},Z_{02},Z_{03},\partial_{{\bar Z_{01}}}V_{0,34}^{({\bar Z_{01}})} , \partial_{{\bar Z_{02}}}, \partial_{{\bar Z_{03}}}\Big)
N^{\lambda_3 \lambda_4 \lambda}_{340}({\bar Z_{02}},{\bar Z_{03}}) \calC^{(n)}_{l_1} (t)
}{
{(P_{01})^{\frac{\Delta+\Delta_{12}}{2}}} (P_{20})^{\frac{\Delta-\Delta_{12}}{2}} 
 {(P_{03})^{\frac{\tilde \Delta+\Delta_{34}}{2}}}  (P_{40})^{\frac{\tilde \Delta-\Delta_{34}}{2}}
}\,,\nonumber
\end{align}
where
\beq
t = \frac{1}{2 \sqrt{P_{12} P_{34}}}
\left(
 \sqrt{\frac{P_{02} P_{03}}{P_{01} P_{04}}}  P_{14}
-\sqrt{\frac{P_{02} P_{04}}{P_{01} P_{03}}}  P_{13}
-\sqrt{\frac{P_{01} P_{03}}{P_{02} P_{04}}}  P_{24}
+\sqrt{\frac{P_{01} P_{04}}{P_{02} P_{03}}}   P_{23}
\right) .
\eeq 
The Gegenbauer recursion relation \eqref{eq:Gegenbauer_recursion} implies the following recursion relations for each of these functions
\begin{align}
&(l_1-n) F_{\Delta,(l_1,l_2,l_3), n}^{\lambda_1 \lambda_2 \lambda_3 \lambda_4}(\Delta_{12}, \Delta_{34}) =
\nonumber\\
&\left(l_1+\tfrac{d}{2}-2\right) u^{-\frac{1}{2}} \Big( 
F_{\Delta,(l_1-1,l_2,l_3), n}^{\lambda_1 \lambda_2 \lambda_3 \lambda_4}(\Delta_{12}+1, \Delta_{34}-1) 
-F_{\Delta,(l_1-1,l_2,l_3), n}^{\lambda_1 \lambda_2 \lambda_3 \lambda_4}(\Delta_{12}+1, \Delta_{34}+1) \nonumber\\
&{\hskip 75pt}
-F_{\Delta,(l_1-1,l_2,l_3), n}^{\lambda_1 \lambda_2 \lambda_3 \lambda_4}(\Delta_{12}-1, \Delta_{34}-1) 
+v F_{\Delta,(l_1-1,l_2,l_3), n}^{\lambda_1 \lambda_2 \lambda_3 \lambda_4}(\Delta_{12}-1, \Delta_{34}+1) 
\Big)\nonumber\\
&-(l_1+n+d-4)F_{\Delta,(l_1-2,l_2,l_3), n}^{\lambda_1 \lambda_2 \lambda_3 \lambda_4}(\Delta_{12}, \Delta_{34}) \,,
\label{eq:recursion_f}
\end{align}
where $u = \frac{P_{12} P_{34}}{P_{13} P_{24}}, v = \frac{P_{14} P_{23}}{P_{13} P_{24}}$ are the conformal cross-ratios.
To use the recursion relations, it is only necessary to know the conformal blocks for $l_1=l_2$ up to $l_1 = 2 l_2$
as seeds. Once these are known, one can forget about the conformal integrals that were used for the derivation of the recursion relations
and even the precise form of the functions $K_n^\lambda$. Any family of conformal blocks can be mapped to the functions $F_{\Delta,\lambda, i}^{\lambda_1 \lambda_2 \lambda_3 \lambda_4}$
using that
\bea
G_{\Delta,(l_2,l_2,l_3)}^{\lambda_1 \lambda_2 \lambda_3 \lambda_4} 
&= F_{\Delta,(l_2,l_2,l_3), l_2}^{\lambda_1 \lambda_2 \lambda_3 \lambda_4} ,\\
G_{\Delta,(l_2+1,l_2,l_3)}^{\lambda_1 \lambda_2 \lambda_3 \lambda_4} 
&= F_{\Delta,(l_2+1,l_2,l_3), l_2}^{\lambda_1 \lambda_2 \lambda_3 \lambda_4} + F_{\Delta,(l_2+1,l_2,l_3), l_2+1}^{\lambda_1 \lambda_2 \lambda_3 \lambda_4},\\
G_{\Delta,(l_2+2,l_2,l_3)}^{\lambda_1 \lambda_2 \lambda_3 \lambda_4} 
&= F_{\Delta,(l_2+2,l_2,l_3), l_2}^{\lambda_1 \lambda_2 \lambda_3 \lambda_4} + F_{\Delta,(l_2+2,l_2,l_3), l_2+1}^{\lambda_1 \lambda_2 \lambda_3 \lambda_4}
+ F_{\Delta,(l_2+2,l_2,l_3), l_2+2}^{\lambda_1 \lambda_2 \lambda_3 \lambda_4},\\
& \vdotswithin{=}
\eea{eq:conformal_blocks_to_f}
Hence the conformal blocks for $l_1=l_2,\ldots,2l_2$ can be used in conjunction with the recursion relation to compute the
seeds
\beq
F_{\Delta,(n,l_2,l_3), n}\,, \qquad n = l_2, \ldots,  2l_2\,.
\label{eq:seeds_F}
\eeq
For example, the conformal blocks for exchange of $\bts{young_01}$ and $\bts{young_11}$ act as seeds for the family $\bts{young_hook1}$ in the following way
\bea
G^{\, \bts{young_1}\, \bullet\, \bts{young_1}\, \bullet}_{\Delta,(1,1)} ={}& F^{\, \bts{young_1}\, \bullet\, \bts{young_1}\, \bullet}_{\Delta,(1,1), 1}\\
& \quad \downarrow\\
G^{\, \bts{young_1}\, \bullet\, \bts{young_1}\, \bullet}_{\Delta,(2,1)} ={}& F^{\, \bts{young_1}\, \bullet\, \bts{young_1}\, \bullet}_{\Delta,(2,1), 1} + F^{\, \bts{young_1}\, \bullet\, \bts{young_1}\, \bullet}_{\Delta,(2,1), 2}\\
& \quad \downarrow \qquad \qquad \ \downarrow\\
G^{\, \bts{young_1}\, \bullet\, \bts{young_1}\, \bullet}_{\Delta,(3,1)} ={}
& F^{\, \bts{young_1}\, \bullet\, \bts{young_1}\, \bullet}_{\Delta,(3,1), 1} + F^{\, \bts{young_1}\, \bullet\, \bts{young_1}\, \bullet}_{\Delta,(3,1), 2}\\
& \quad \vdotswithin{\downarrow}  \qquad \qquad \, \vdots
\eea{eq:recursion_diagram}
The recursion relation (\ref{eq:recursion_f}) allows us to move down along the arrows.
Note that in order for the conformal blocks to satisfy the stated recursion relations, it is crucial that their normalization depends on $l_1$, $\Delta_{12}$ and $\Delta_{34}$
as defined in (\ref{defGfromW}). A good method to compare normalizations for conformal blocks obtained via different methods is to consider the OPE limit, which is done for the
partial wave \eqref{eq:conf_partial_wave} in Appendix \ref{sec:ope_limit}.

\subsection{Solution of the recursion relation in terms of scalar conformal blocks}
\label{sec:solution_with_scalar}
For $n=0$ the recursion relation \eqref{eq:recursion_f} is equivalent to the one for scalar conformal blocks that was solved for $d=2,4$ in \cite{Dolan:2000ut}.
The new parameter $n$ can be removed from the prefactors by using the variables $l_1'=l_1-n$ and $d'=d+2n$,
\begin{align}
& l_1' F_{\Delta,(l_1,l_2,l_3), n}^{\lambda_1 \lambda_2 \lambda_3 \lambda_4}(\Delta_{12}, \Delta_{34})= \nonumber\\
&\left(l_1'+\tfrac{d'}{2}-2\right) u^{-\frac{1}{2}} \Big( 
F_{\Delta,(l_1-1,l_2,l_3), n}^{\lambda_1 \lambda_2 \lambda_3 \lambda_4}(\Delta_{12}+1, \Delta_{34}-1) 
-F_{\Delta,(l_1-1,l_2,l_3), n}^{\lambda_1 \lambda_2 \lambda_3 \lambda_4}(\Delta_{12}+1, \Delta_{34}+1) \nonumber\\
&{\hskip 80pt}-F_{\Delta,(l_1-1,l_2,l_3), n}^{\lambda_1 \lambda_2 \lambda_3 \lambda_4}(\Delta_{12}-1, \Delta_{34}-1) 
+v F_{\Delta,(l_1-1,l_2,l_3), n}^{\lambda_1 \lambda_2 \lambda_3 \lambda_4}(\Delta_{12}-1, \Delta_{34}+1) 
\Big)\nonumber\\
&-\big(l_1'+d'-4\big)F_{\Delta,(l_1-2,l_2,l_3), n}^{\lambda_1 \lambda_2 \lambda_3 \lambda_4}(\Delta_{12}, \Delta_{34}) \,.
\label{eq:recursion_f_prime}
\end{align}
This implies that the scalar conformal block in $d'$ dimensions
\beq
G^{\Delta_{12},\Delta_{34}}_{\Delta, l_1',d'} (u,v) \equiv
G^{\bullet \bullet \bullet \bullet}_{\Delta, (l_1')} \text{ in } d' \text{ dimensions}  \,,
\eeq
is a solution of the recurrence relation for seed conformal blocks.
Maybe this can be used to solve the recursion relations in terms of known functions.
While we do not know if this is actually possible, assume for a moment
that the seeds for the recursion relation can be written in terms of scalar conformal blocks as
\beq
F_{\Delta,(n,l_2,l_3), n} = \sum_{i,j,k} f_{i,j,k} G^{\Delta_{12}+i,\Delta_{34}+j}_{\Delta+k,0,d+2n}(u,v) \,, \qquad n = l_2, \ldots,  2l_2\,.
\label{eq:scalar_block_speculation}
\eeq
We allowed for arbitrary shifts $i,j,k$ in the three parameters $\Delta_{12}$, $\Delta_{34}$ and $\Delta$
and  functions $f_{i,j,k}$ that do not influence the recursion relation,
i.e. $f_{i,j,k}$ can depend on $\Delta$, $u$, $v$ and the polarizations of external operators,
but not on $l_1$, $d$, $\Delta_{12}$ or $\Delta_{34}$.
Then the result for arbitrary $l_1$ would be the same linear combination of scalar conformal blocks with appropriately increased spin $l_1$,
\beq
F_{\Delta,(l_1,l_2,l_3), n} = \sum_{i,j,k} f_{i,j,k} G^{\Delta_{12}+i,\Delta_{34}+j}_{\Delta+k,l_1-n,d+2n}(u,v)\,.
\eeq
In \cite{SimmonsDuffin:2012uy,Costa:2014rya} the two initial blocks of the family $(l_1,1)$
were computed explicitly in terms of conformal blocks with $l_1=0$ in higher dimensions,
however the block $G^{\, \bts{young_1}\, \bullet\, \bts{young_1}\, \bullet}_{\Delta,(1,1)}=F_{\Delta,(1,1),1}$
was given in terms of scalar blocks in 8 dimensions instead of the $d'=6$ required by \eqref{eq:scalar_block_speculation}.
While the simple relation to scalar conformal blocks assumed in \eqref{eq:scalar_block_speculation}
might not be true, it is likely that the recursion relations should be solvable in terms of functions
similar to scalar conformal blocks in higher dimensions.  
This correspondence was also observed in the
recent paper \cite{Echeverri:2016dun}, where the mixed-symmetry seed conformal blocks in four dimensions were computed using a twistor formalism.

\subsection{Recursion relation for radial coordinates}

Here we will discuss how to use these recursion relations for expansions of conformal blocks in the radial coordinates $(r,\eta)$ of \cite{Hogervorst:2013sma,Penedones:2015aga,Costa:2016other}.
These coordinates are related to the cross-ratios $(u,v)$ by
\beq
u= \frac{16 r^2}{\left(r^2+2 \eta  r+1\right)^2} \ , 
\qquad 
v= \frac{\left(r^2-2 \eta  r+1\right)^2}{\left(r^2+2 \eta  r+1\right)^2} \ ,
\eeq
and one typically considers expansions of the conformal blocks up to some order $m$ in $r$
\bea
G_{\Delta,\lambda}^{\lambda_1 \lambda_2 \lambda_3 \lambda_4} (r) ={}&
\sum\limits_{m'=0}^m r^{\Delta+m'} H_{m'} + \calO(r^{\Delta+m+1})\\
\equiv{}&
G_{\Delta,\lambda}^{\lambda_1 \lambda_2 \lambda_3 \lambda_4} (r, m) + \calO(r^{\Delta+m+1})\,.
\eea{eq:r_expansion}
For such expansions the recursion relations can be used by splitting them into contributions which satisfy the different recursion relations, as in \eqref{eq:cb_split}
\beq
G_{\Delta,\lambda}^{\lambda_1 \lambda_2 \lambda_3 \lambda_4} (r, m)
= \sum\limits_{n=l_2}^{2 l_2} F_{\Delta,\lambda, n}^{\lambda_1 \lambda_2 \lambda_3 \lambda_4} (\Delta_{12}, \Delta_{34}; r, m)\,.
\label{eq:cb_split_r}
\eeq
The identification of the different parts of the conformal blocks can be performed as illustrated in the diagram \eqref{eq:recursion_diagram},
by starting at the lowest allowed $l_1$ and using the recursion to find the term that has to be subtracted from the next conformal block to get the seed for the next recursion relation.
To derive the recursion relations in $(r,\eta)$ one has to expand $u^{-\frac{1}{2}}$ and $v$ in $r$
\beq
u^{-\frac{1}{2}} = \frac{1}{4r} + \frac{\eta}{2} + \frac{r}{4}\,,\qquad
v = 1 + \calO(r)\,.
\eeq
Unfortunately the appearance of the term of order $r^{-1}$ means that the recursion relations decrease the order of the expansion in $r$. They read
\begin{align}
& (l_1-n) F_{\Delta,(l_1,l_2,l_3), n}^{\lambda_1 \lambda_2 \lambda_3 \lambda_4}(\Delta_{12}, \Delta_{34}; r, m) =
\nonumber\\
&\left(l_1+\tfrac{d}{2}-2\right) \left(  \frac{1}{4r} + \frac{\eta}{2} + \frac{r}{4} \right) \Big( 
F_{\Delta,(l_1-1,l_2,l_3), n}^{\lambda_1 \lambda_2 \lambda_3 \lambda_4}(\Delta_{12}+1, \Delta_{34}-1; r, m+1)
\label{eq:recursion_f_r}-\\ 
&
F_{\Delta,(l_1-1,l_2,l_3), n}^{\lambda_1 \lambda_2 \lambda_3 \lambda_4}(\Delta_{12}+1, \Delta_{34}+1;r, m+1) 
-F_{\Delta,(l_1-1,l_2,l_3), n}^{\lambda_1 \lambda_2 \lambda_3 \lambda_4} (\Delta_{12}-1, \Delta_{34}-1; r, m+1) +
\nonumber\\
&
 v F_{\Delta,(l_1-1,l_2,l_3), n}^{\lambda_1 \lambda_2 \lambda_3 \lambda_4} (\Delta_{12}-1, \Delta_{34}+1; r, m+1) 
\!\Big)
\!-\!
(l_1+n+d-4)F_{\Delta,(l_1-2,l_2,l_3), n}^{\lambda_1 \lambda_2 \lambda_3 \lambda_4}(\Delta_{12}, \Delta_{34}; r, m) \,.
\nonumber
\end{align}
These relations were checked with the $r$ expansions from \cite{Costa:2016other} for the case of two external vectors and exchange of the irrep $\bts{young_hook1}$,
which is depicted in the diagram \eqref{eq:recursion_diagram}.
To this end the normalizations were matched in the OPE limit.

\section{Concluding remarks}
\label{sec:remarks}

In this paper the projectors to traceless mixed-symmetry tensors that appear in the correlator of four stress tensors
were derived in terms of Gegenbauer polynomials. Knowledge of the explicit form of the projectors led us to a single universal
recursion relation in $l_1$ for seed conformal blocks, given by \eqref{eq:recursion_f}.
Interestingly, the existence of the recursion relation does not rely on the
complete expressions for the projectors, but only requires that their dependence on $l_1$ is according to \eqref{eq:pi_Gegenbauer_sum}.
Of course this implies that, to show that the recursion relation is truly universal (i.e.\ holds for any seed conformal block of bosonic operators),
it is enough to prove that all projectors to traceless mixed-symmetry tensors can be written as in \eqref{eq:pi_Gegenbauer_sum}.

In order for the conformal blocks to obey the recursion relation, it is required that they are normalized in a particular way.
In particular, the normalization constant of the projector as well as the normalization of the shadow operator need to be canceled from the
integral expression of the conformal partial wave. To this end the normalization of the shadow was computed.
Furthermore, the OPE limit of the shadow integral was analyzed to allow for comparisons to other results.

Another remark concerns the solution of the recursion relations. 
The new recursion relations \eqref{eq:recursion_f} are generalizations of the recursion relation for scalar conformal blocks of \cite{Dolan:2011dv}
with a new parameter $n$.
This new parameter can be absorbed into $l_1$ and $d$ by using shifted parameters
\beq
l_1' = l_1 - n \,, \qquad d' = d + 2n \,.
\eeq
As a result the recursion relations for general seed conformal blocks
are solved by scalar conformal blocks in higher dimensions,
suggesting that seed conformal blocks can generally be expressed in terms
of such scalar blocks.
A similar correspondence was also observed in the
recent paper \cite{Echeverri:2016dun} for the case $d=4$.

Beside this application, we want to stress that the projectors actually play a very important role in CFTs. The most basic example in which they appear is the two point functions of mixed-symmetry operators,  
\begin{equation}
\bra \calO_{\Delta,\lambda}^{a_1\dots a_{|\lambda|}}(x) \calO_{\Delta,\lambda}^{b_1 \dots b_{|\lambda|}}(0)\ket=|x|^{-2\Delta} 
 \,\pi_{\lambda}^{c_1 \dots c_{|\lambda|},b_1 \dots b_{|\lambda|}} \prod_{i=1}^{|\lambda|} \left(\delta^{a_i}_{c_i}-2 \frac{ x^{a_i}x_{c_i}}{x^2}\right) .
\end{equation}
Moreover, the projector $\pi_{\lambda}$ is a necessary ingredient to compute a conformal block for the exchange of an irrep $\lambda$. 
For example, the OPE limit of any seed conformal block with external operators $\calO_i$ of spins $\ell_i$ is always written
(here without being precise on the placement of the indices on the $\pi_{\ell_i}$) in terms of the two-point function of the exchanged operator as
\beq
 \frac{
x_{12 \,a_1} .. x_{12 \,a_{l_1}}
\left( \pi_{\ell_1} \pi_{\ell_2} \right)^{c_1 \ldots c_{\ell_1} e_1 \ldots e_{\ell_2}}_{a_{l_1+1} \ldots a_{|\lambda|}}
x_{34 \,b_1} .. x_{34 \,b_{l_1}}
\left( \pi_{\ell_3} \pi_{\ell_4} \right)^{f_1 \ldots f_{\ell_3} g_1 \ldots g_{\ell_4}}_{b_{l_1+1} \ldots b_{|\lambda|}}
\bra \calO_{\Delta,\lambda}^{a_1\dots a_{|\lambda|}}(x_2) \calO_{\Delta,\lambda}^{b_1 \dots b_{|\lambda|}}(x_4)\ket\,}{|x_{12}|^{\D_{1}+\D_{2}-\D+l_1}|x_{34}|^{\D_{3}+\D_{4}-\D+l_1}} \nonumber
\eeq
therefore it involves the projector $\pi_{\lambda}$  (see appendix \ref{sec:ope_limit}).
Of course, also the leading OPE of any other conformal block is written in terms of the projectors since it can be generated by acting with some derivatives on the seed block.
From this remark it is clear that the knowledge of $\pi_{\lambda}$ is needed to compute conformal blocks from their radial coordinate expansion \cite{Hogervorst:2013sma,Penedones:2015aga,Costa:2016other}, since the leading term of the expansion is the leading OPE.

To compute conformal blocks with the shadow formalism it is crucial to know the form of $\pi_{\lambda}$ as well, since it appears explicitly in the shadow integral. In this paper we expressed $\pi_{\lambda}$  in terms of derivatives of the Gegenbauer polynomial, which encodes the projector to traceless symmetric irreps. Therefore, it should be possible to compute any family of seed conformal blocks (for generic $l_1$) in terms of scalar conformal blocks by a direct computation.
To achieve this, it is enough to rewrite the integrand of the shadow integral of the seed conformal block \eqref{eq:partial_wave_with_N}
in terms of derivatives of integrands of scalar conformal blocks. The general result for seed conformal blocks of the family $\lambda=(l_1,1)$ was obtained in this way in \cite{Rejon-Barrera:2015bpa}.

\section*{Acknowledgements}
This research received funding from the [European Union] 7th Framework Programme (Marie Curie Actions) under grant agreements No 269217 and 317089 (GATIS), and from the research grant CERN/FIS-NUC/0045/2015.
T.H.\ was supported by the German Science Foundation (DFG) within the Collaborative Research Center 676 ``Particles, Strings and the Early Universe''.
The work of E.T.\ has been supported by the Portuguese Fundac\~ao para a Ci\^encia e a Tecnologia (FCT) through the fellowship SFRH/BD/51984/2012. His research was partially supported by Perimeter Institute for Theoretical Physics. Research at Perimeter Institute is supported by the Government of Canada through Industry Canada and by the Province of Ontario though the Ministry of Economic Development \& innovation.

\appendix

\section{A mixed-symmetry differential operator}\label{sec:todorov}
In this section we find an alternative way to generate the projectors into traceless mixed-symmetry tensors for Young diagrams with two rows. The main idea is to generalize the result of  \cite{Dobrev:1975ru}, where the differential operator
\beq
D^a_z = \left( \frac{d}{2} - 1 + z \cdot \frac{\partial}{\partial z} \right) \frac{\partial}{\partial z_a} - \frac{1}{2} z^a \frac{\partial^2}{\partial z \cdot \partial z}\,,
\eeq
was defined in order to generate projectors to traceless symmetric tensor representations
\beq
\pi_{(l)}^{a_1 \ldots a_l, b_1 \ldots b_l} = \frac{1}{l!(\frac{d}{2}-1)_l} D^{a_1}_z \ldots D^{a_l}_z z^{b_1} \ldots z^{b_l}\,.
\label{eq:todorov_proj}
\eeq
One can construct this operator by looking for an operator of weight $-1$ in $z$ that preserves the space defined by $z^2 = 0$, i.e.
\beq
D^{a}_{z} \big(z^2 f(z) \big) =\calO\big(z^2\big) \,.
\eeq
This ensures tracelessness of the expression \eqref{eq:todorov_proj}, since contracting with $\delta_{b_1 b_2}$ and acting
with all the operators yields something that is $\calO(z^2)$ and of degree zero in $z$, hence $0$.

A generalized differential operator for Young diagrams with two rows should generate the projectors into traceless mixed-symmetry tensors in a similar way
by acting on a combination of two different vectors
\begin{align}
&\pi_{\lambda}^{[a_1 a_2] \ldots [a_{2 l_2 - 1} a_{2 l_2}] a_{2 l_2+1} \ldots a_{l_1+l_2},
[b_1 b_2] \ldots [b_{2 l_2 - 1} b_{2 l_2}] b_{2 l_2+1} \ldots b_{l_1+l_2}}
\label{eq:gen_todorov_proj}
\\
={}&\Ncal_{l_1,l_2}
D^{[a_1}_1 D^{a_2]}_2 \ldots D^{[a_{2 l_2 - 1}}_1 D^{a_{2l_2}]}_2 D^{a_{2 l_2 + 1}}_1  \ldots D^{a_{l_1 + l_2}}_1
 z_1^{[b_1} z_2^{b_2]} \ldots z_1^{[b_{2l_2-1}} z_2^{b_{2l_2}]} z_1^{b_{2l_2+1}} \ldots z_1^{b_{l_1+l_2}} \,.
\nonumber
\end{align}
Clearly this expression is Young symmetrized (in the antisymmetric representation) and traceless, provided the differential operators $D_1$ and $D_2$
preserve the space defined by $z_1^2=0$, $z_2^2=0$ and $z_1\cdot z_2=0$, namely
\begin{align}
\begin{split}
D^{a}_{1,2}\big(z_1^2 f(z_1,z_2)\big)&=\calO\big(z_1^2,z_1\cdot z_2,z_2^2\big) \,,\\
D^{a}_{1,2}\big( z^2_2  f(z_1,z_2)\big)&=\calO\big(z_1^2,z_1\cdot z_2,z_2^2\big)\,, \\
D^{a}_{1,2}\big(z_1\cdot z_2 f(z_1,z_2)\big)&=\calO\big(z_1^2,z_1\cdot z_2,z_2^2\big)\,.
\end{split}
\end{align}
Furthermore, $D_1$ must have weight $-1$ in $z_1$ and $0$ in $z_2$, and $D_2$ must have weight $0$  in $z_1$ and $-1$ in $z_2$.
These requirements fix the operators completely as \footnote{There exist higher order differential operators that satisfy the same requirements. The ones that we found are the lowest order ones.} 
\bea
D^a_1 &\equiv D^{a}_{z_1,z_2}\equiv d_{00} \partial^{a}_{z_1}+ d_{-1 1} \partial^{a}_{z_2}+z^{a}_{1} d_{-2 0} +z^{a}_{2} d_{-1 -1}\,,\\
D^a_2 &\equiv D^{a}_{z_2,z_1}\,.
\eea{eq:gen_todorov}
where $d_{m n}$ are differential operators with weight $m$ in the variable $z_1$ and $n$ in the variable $z_2$, defined by
\begin{align}
 d_{00}& \equiv  \left(1-\tfrac{d}{2}\right) \big[ (d-3)+3  (z_1\cdot \partial_{z_1})+  (z_2\cdot \partial_{z_2})\big]-(z_1\cdot \partial_{z_1})(z_2\cdot \partial_{z_2})-z_1^{a}(z_1\cdot \partial_{z_1}) \partial_{z_1\, a} \,,
 \nonumber\\
 d_{-2 0}& \equiv   \frac{1}{2} \big[ 2 \left(\tfrac{d}{2}-1\right)+(z_2\cdot \partial_{z_2}) +(z_1\cdot \partial_{z_1})\big](\partial_{z_1}\cdot \partial_{z_1}) \,,
 \nonumber\\
d_{-1 1} &\equiv- (d-2) (z_2\cdot \partial_{z_1})-(z_2\cdot \partial_{z_1})(z_2\cdot \partial_{z_2})-(z_1\cdot \partial_{z_1})(z_2\cdot \partial_{z_1})\,,
\\
d_{-1 -1}& \equiv \big[ \left(\tfrac{d}{2}-1\right) +(z_1\cdot \partial_{z_1}) \big] (\partial_{z_1}\cdot \partial_{z_2})+
\frac{1}{2}\big[ (z_2\cdot \partial_{z_1})(\partial_{z_2}\cdot \partial_{z_2})- (z_1\cdot \partial_{z_2}) (\partial_{z_1}\cdot \partial_{z_1})\big]\,.
\nonumber
\end{align}
Moreover, these operators automatically satisfy
\beq
\delta_{a b}D^{a}_i D^{b}_{i}=0\,, \qquad
[D^{a}_{i},D^{b}_{i}]=0\,,  \qquad
[D^{a}_{1},D^{b}_{2}]=0  \,,
\eeq
for $ i \in \{1,2\}$.
 The general normalization factor appearing in \eqref{eq:gen_todorov_proj} is fixed asking for the idempotence of the projector,
\beq
\Ncal_{l_1, l_2}=
\frac{(-1)^{l_2-l_1} 2^{l_2} (l_1-l_2+1) }{
\Gamma (l_1+2) \Gamma (l_2+1)
 \left(\frac{d}{2}-1\right)_{l_1} \left(\frac{d}{2}-2\right)_{l_2} (d-3)_{l_1+l_2}} \,.
\eeq
As an example we construct the projector into the representations $\bts{young_hook1}$,
\beq
\pi_{(l_1,1)}^{[a_1 a_2] a_3 \ldots a_{l_1+1},[b_1 b_2] b_3 \ldots b_{l_1+1}}= \Ncal_{l_1,1} \;\; D^{[a_1}_{1} D^{a_2]}_{2} D^{a_3}_{1} \cdots D^{a_{l_1+1}}_{1} \; \;
 z_1^{[b_1} z_2^{b_2]} z_1^{b_3} \cdots z_1^{b_{l_1+1}} \ .
\eeq

The projector that is generated in this way is in the antisymmetric representation. It is related to the other expressions derived in the main text by the contraction
\bea
\pi_\lambda (z_1, z_2, \barz_1, \barz_2) ={}&
n_\lambda z_1^{a_1} z_2^{a_2} \ldots z_1^{a_{2l_2-1}} z_2^{a_{2l_2}} z_1^{a_{2l_2+1}} \ldots z_1^{a_{l_1+l_2}}\\
&\pi_{\lambda}^{[a_1 a_2] \ldots [a_{2 l_2 - 1} a_{2 l_2}] a_{2 l_2+1} \ldots a_{l_1+l_2},
[b_1 b_2] \ldots [b_{2 l_2 - 1} b_{2 l_2}] b_{2 l_2+1} \ldots b_{l_1+l_2}}\\
& \barz_1^{b_1} \barz_2^{b_2} \ldots \barz_1^{b_{2l_2-1}} \barz_2^{b_{2l_2}} \barz_1^{b_{2l_2+1}} \ldots \barz_1^{b_{l_1+l_2}},
\eea{eq:antisym_proj_contraction}
where the factor $n_\lambda$ appears due to the change from the antisymmetric to the symmetric representation and is defined in \eqref{eq:n_lambda}.

It is in principle possible to generalize this method in order to study Young diagrams with more than two rows. However this requires to find a new set of differential operators, which we expect to be lengthy and therefore not very efficient. 

\section{Relating different projectors}\label{ref:relating_similar_projectors}
This section introduces a useful relation between similar projectors. 
The main observation is that a contraction between corresponding indices
on the left and right of a projector leads to an object that is still traceless
and has mixed-symmetry. Hence it should be proportional to another projector, e.g.\
\beq
\delta_{a_1 b_1} \pi_{(l_1,l_2,l_3)}^{(a_1 \ldots a_{l_1}) \ldots ( \ldots a_{l_1+l_2+l_3}),
(b_1 \ldots b_{l_1}) \ldots(\ldots b_{l_1+l_2+l_3})}
\propto
\pi_{(l_1-1,l_2,l_3)}^{(a_2 \ldots a_{l_1}) \ldots ( \ldots a_{l_1+l_2+l_3}),
(b_2 \ldots b_{l_1}) \ldots(\ldots b_{l_1+l_2+l_3})}\,.
\eeq
Using auxiliary vectors $\{z_i,\bar z_i\}$  $(i=1,2,3)$,  
this equation can be written as
\beq
\frac{1}{(l_1)^2}
\left(
\frac{\partial}{\partial z_1} \cdot \frac{\partial}{\partial \barz_1}
\right)
\pi_{(l_1,l_2,l_3)} (\{z_i,\bar z_i\})
\propto
\pi_{(l_1-1,l_2,l_3)} (\{z_i,\bar z_i\})\,.
\label{eq:projectors_proportional}
\eeq
The proportionality factor can be found by using that the full trace of a projector is given by the dimension $d_\lambda$ of the $SO(d)$
irrep,
\beq
\frac{1}{(l_1! l_2! l_3!)^2}
\left(
\frac{\partial}{\partial z_1} \cdot \frac{\partial}{\partial \barz_1}
\right)^{l_1}
\hspace{-3pt}
\left(
\frac{\partial}{\partial z_2} \cdot \frac{\partial}{\partial \barz_2}
\right)^{l_2}
\hspace{-3pt}
\left(
\frac{\partial}{\partial z_3} \cdot \frac{\partial}{\partial \barz_3}
\right)^{l_3}
\hspace{-3pt}
\pi_{(l_1,l_2,l_3)} (\{z_i,\bar z_i\})
=
d_{(l_1,l_2,l_3)} \,,
\eeq
which is given by
\cite{MR552445}
\beq
d_\lambda = \frac{1}{H(\lambda)}
\prod\limits_{i \geq j}^{h_1} (d + l_i +l_j-i-j)
\prod\limits_{i < j}^{l_1} (d - h_i -h_j+i+j-2)\,,
\eeq
where $H(\lambda)$ was defined in \eqref{eq:n_lambda}.
Since both sides in \eqref{eq:projectors_proportional} can be reduced to the corresponding
dimensions by doing full contractions, the missing constant is given by the ratios of dimensions
\beq
\frac{1}{(l_1)^2}
\left(
\frac{\partial}{\partial z_1} \cdot \frac{\partial}{\partial \barz_1}
\right)
\pi_{(l_1,l_2,l_3)} (\{z_i,\bar z_i\})
=
\frac{d_{(l_1,l_2,l_3)}}{d_{(l_1-1,l_2,l_3)}}
\,\pi_{(l_1-1,l_2,l_3)} (\{z_i,\bar z_i\})\,.
\eeq
Of course the same relation holds also for the other rows in the Young diagram, e.g.\
\beq
\frac{1}{(l_2)^2}
\left(
\frac{\partial}{\partial z_2} \cdot \frac{\partial}{\partial \barz_2}
\right)
\pi_{(l_1,l_2,l_3)} (\{z_i,\bar z_i\})
=
\frac{d_{(l_1,l_2,l_3)}}{d_{(l_1,l_2-1,l_3)}}
\,\pi_{(l_1,l_2-1,l_3)} (\{z_i,\bar z_i\})\,.
\eeq

\section{More projectors}
\label{sec:more_projectors}

In this appendix we state the remaining projectors to irreducible representations of $SO(d)$
that can appear in a three-point function with two stress-tensors.

\subsection{Projectors to the irreps $(l_1,3)$}
For Young diagrams of shape $\bts{young_hook3b}$  the possible combinations of the proposed building blocks are
\bea
Q_1 &= H(z_2,\barz_2)^3\,,\\
Q_2 &= H(z_2,\barz_2)^2 V(z_2) \bar{V}(\barz_2)\,,\\
Q_3 &= H(z_2,\barz_2) V(z_2)^2 \bar{V}(\barz_2)^2\,,\\
Q_4 &= V(z_2)^3 \bar{V}(\barz_2)^3\,,\\
Q_5 &= H(z_2,\barz_2) T(z_2,z_2) \bar{T}(\barz_2, \barz_2)\,,\\
Q_6 &= V(z_2) \bar{V}(\barz_2) T(z_2,z_2) \bar{T}(\barz_2, \barz_2)\,,\\
Q_7 &= H(z_2,\barz_2) \left( T(z_2,z_2) \bar{V}(\barz_2)^2 + V(z_2)^2 \bar{T}(\barz_2, \barz_2) \right)\,,\\
Q_8 &= V(z_2) \bar{V}(\barz_2) \left( T(z_2,z_2) \bar{V}(\barz_2)^2 + V(z_2)^2 \bar{T}(\barz_2, \barz_2) \right)\,.
\eea{eq:structures_l3}
Imposing the tracelessness conditions (\ref{eq:laplac}) one finds that the functions $f_i(t)$ that appear in the projector  (\ref{eq:pi_ansatz_2rows}) can then be written as 
\beq
f_i(t)= - \frac{\hat f_i(t)}{d} \,, \qquad  \forall i=1,\ldots,8\,,
\eeq
with
\begin{align}
 \hat f_1(t)={}&-d (1+d) (2+d) t \left(-3+d t^2\right) \mathcal{C}_{l_1}^{(3)}(t)-3 (1+d) (2+d) \left(t^2-1\right) \left(d t^2-1\right) \mathcal{C}_{l_1}^{(4)}(t)\nonumber \\
   &-3
   d (2+d) t \left(t^2-1\right)^2 \mathcal{C}_{l_1}^{(5)}(t)-d \left(t^2-1\right)^3 \mathcal{C}_{l_1}^{(6)}(t)\,, \nonumber \\
 \hat f_2(t)={}&6 d (1+d) (2+d) t \mathcal{C}_{l_1}^{(3)}(t)+3 (1+d) (2+d) \left(-3+(4+d) t^2\right) \mathcal{C}_{l_1}^{(4)}(t)\nonumber \\
   &+6 (1+d) (2+d) t \left(t^2-1\right)
   \mathcal{C}_{l_1}^{(5)}(t)+3 d \left(t^2-1\right)^2 \mathcal{C}_{l_1}^{(6)}(t)\,, \nonumber \\
 \hat f_3(t)={}&3 d (1+d) (2+d) t \mathcal{C}_{l_1}^{(3)}(t)+9 (1+d) (2+d) \left(t^2-1\right) \mathcal{C}_{l_1}^{(4)}(t)\nonumber \\
   &-3 (2+d) t \left(4+d-3 t^2\right)
   \mathcal{C}_{l_1}^{(5)}(t)-3 \left(d-t^2\right) \left(t^2-1\right) \mathcal{C}_{l_1}^{(6)}(t)\,, \nonumber \\
 \hat f_4(t)={}&-3 (1+d) (2+d) \mathcal{C}_{l_1}^{(4)}(t)
   -6 (2+d) t \mathcal{C}_{l_1}^{(5)}(t)+\left(d-3 t^2\right) \mathcal{C}_{l_1}^{(6)}(t)\,, \label{eq:f_30} \\
 \hat f_5(t)={}&3 d (2+d) t \left(-4+(1+d) t^2\right) \mathcal{C}_{l_1}^{(3)}(t)+3 (2+d) \left(t^2-1\right) \left(-4+3 (1+d) t^2\right) \mathcal{C}_{l_1}^{(4)}(t)\nonumber \\
   &+9
   (2+d) t \left(t^2-1\right)^2 \mathcal{C}_{l_1}^{(5)}(t)+3 \left(t^2-1\right)^3 \mathcal{C}_{l_1}^{(6)}(t)\,, \nonumber \\
 \hat f_6(t)={}&-3 (2+d) \left(-4+(1+d) t^2\right) \mathcal{C}_{l_1}^{(4)}(t)
   -6 (2+d) t \left(t^2-1\right) \mathcal{C}_{l_1}^{(5)}(t)-3 \left(t^2-1\right)^2
   \mathcal{C}_{l_1}^{(6)}(t)\,, \nonumber \\
 \hat f_7(t)={}&3 d (1+d) (2+d) t^2 \mathcal{C}_{l_1}^{(3)}(t)+3 (1+d) (2+d) t \left(-1+3 t^2\right) \mathcal{C}_{l_1}^{(4)}(t)\nonumber \\
   &+9 (2+d) t^2 \left(t^2-1\right)
   \mathcal{C}_{l_1}^{(5)}(t)+3 t \left(t^2-1\right)^2 \mathcal{C}_{l_1}^{(6)}(t)\,, \nonumber \\
 \hat f_8(t)={}&-3 (1+d) (2+d) t \mathcal{C}_{l_1}^{(4)}(t)
   -6 (2+d) t^2 \mathcal{C}_{l_1}^{(5)}(t)-3 t \left(t^2-1\right) \mathcal{C}_{l_1}^{(6)}(t)\,.
\nonumber
\end{align}

\subsection{Projectors to the irreps $(l_1,4)$}
For Young diagrams of shape $\bts{young_hook4c}$  the required building blocks are
\bea
Q_1 &= H(z_2,\barz_2)^4\,,\\
Q_2 &= H(z_2,\barz_2)^3 V(z_2) \bar{V}(\barz_2)\,,\\
Q_3 &= H(z_2,\barz_2)^2 V(z_2)^2 \bar{V}(\barz_2)^2\,,\\
Q_4 &= H(z_2,\barz_2) V(z_2)^3 \bar{V}(\barz_2)^3\,,\\
Q_5 &= V(z_2)^4 \bar{V}(\barz_2)^4\,,\\
Q_6 &= H(z_2,\barz_2)^2 T(z_2,z_2) \bar{T}(\barz_2, \barz_2)\,,\\
Q_7 &= H(z_2,\barz_2) V(z_2) \bar{V}(\barz_2) T(z_2,z_2) \bar{T}(\barz_2, \barz_2)\,,\\
Q_8 &= V(z_2)^2 \bar{V}(\barz_2)^2 T(z_2,z_2) \bar{T}(\barz_2, \barz_2)\,,\\
Q_9 &= T(z_2,z_2)^2 \bar{T}(\barz_2, \barz_2)^2\,,\\
Q_{10} &= H(z_2,\barz_2)^2 \left( T(z_2,z_2) \bar{V}(\barz_2)^2 + V(z_2)^2 \bar{T}(\barz_2, \barz_2) \right)\,,\\
Q_{11} &= H(z_2,\barz_2) V(z_2) \bar{V}(\barz_2) \left( T(z_2,z_2) \bar{V}(\barz_2)^2 + V(z_2)^2 \bar{T}(\barz_2, \barz_2) \right)\,,\\
Q_{12} &= V(z_2)^2 \bar{V}(\barz_2)^2 \left( T(z_2,z_2) \bar{V}(\barz_2)^2 + V(z_2)^2 \bar{T}(\barz_2, \barz_2) \right)\,,\\
Q_{13} &= T(z_2,z_2)^2 \bar{V}(\barz_2)^4 + V(z_2)^4 \bar{T}(\barz_2, \barz_2)^2 \,,\\
Q_{14} &= T(z_2,z_2) \bar{T}(\barz_2, \barz_2) \left( T(z_2,z_2) \bar{V}(\barz_2)^2 + V(z_2)^2 \bar{T}(\barz_2, \barz_2) \right)\,.
\eea{eq:structures_l4}
In this case the functions $f_i(t)$ that appear in the projector  (\ref{eq:pi_ansatz_2rows}) can then be written as
\beq
f_i(t)=-  \frac{\hat f_i(t)}{d(d+2)}\,, \qquad  \forall i=1,\ldots,14\,,\\
\eeq
with
\begin{align}
 \hat f_1(t)={}&(d+1)_4 \left(6 d t^2-3-d (2+d) t^4\right) \mathcal{C}_{l_1}^{(4)}(t)-4 d (d+2)_3 t \left(t^2-1\right) \left((2+d)
   t^2 - 3\right) \mathcal{C}_{l_1}^{(5)}(t)\nonumber \\
   &-6 d (d+3)_2 \left(t^2-1\right)^2 \left(-1+(2+d) t^2\right) \mathcal{C}_{l_1}^{(6)}(t)-4 d (2+d) (4+d) t
   \left(t^2-1\right)^3 \mathcal{C}_{l_1}^{(7)}(t)\nonumber \\
   &-d (2+d) \left(t^2-1\right)^4 \mathcal{C}_{l_1}^{(8)}(t) \nonumber \,, \\
 \hat f_2(t)={}&12 (d+1)_4 \left(d t^2-1\right) \mathcal{C}_{l_1}^{(4)}(t)+4 (d+2)_3 t \left(-3-9 d+d (11+d) t^2\right)
   \mathcal{C}_{l_1}^{(5)}(t)\nonumber \\
   &+12 d (d+3)_2 \left(t^2-1\right) \left(-2+(5+d) t^2\right) \mathcal{C}_{l_1}^{(6)}(t)+12 d (d+3)_2 t
   \left(t^2-1\right)^2 \mathcal{C}_{l_1}^{(7)}(t)\nonumber \\
   &+4 d (2+d) \left(t^2-1\right)^3 \mathcal{C}_{l_1}^{(8)}(t) \nonumber \,, \\
 \hat f_3(t)={}&6 (d+1)_4 \left(-3+d t^2\right) \mathcal{C}_{l_1}^{(4)}(t)+12 (d+2)_3 t \left(-3+d \left(-3+2 t^2\right)\right)
   \mathcal{C}_{l_1}^{(5)}(t)\nonumber \\
   &-6 (d+3)_2 \left(-6 d+(3+d (16+d)) t^2-6 d t^4\right) \mathcal{C}_{l_1}^{(6)}(t)\nonumber \\
   &-12 d (4+d) t \left(5+d-2 t^2\right)
   \left(t^2-1\right) \mathcal{C}_{l_1}^{(7)}(t)-6 d \left(2+d-t^2\right) \left(t^2-1\right)^2 \mathcal{C}_{l_1}^{(8)}(t) \nonumber \,,
   \nonumber\\
 \hat f_4(t)={}&-12 (d+1)_4 \mathcal{C}_{l_1}^{(4)}(t)-12 (2+d) (3+d)^2 (4+d) t \mathcal{C}_{l_1}^{(5)}(t)\nonumber \\
   &-12 (d+3)_2 \left(-2 d+3 (1+d)
   t^2\right) \mathcal{C}_{l_1}^{(6)}(t)+4 (4+d) t \left(d (11+d)-3 (1+3 d) t^2\right) \mathcal{C}_{l_1}^{(7)}(t)\nonumber \\
   &+4 d \left(2+d-3 t^2\right)
   \left(t^2-1\right) \mathcal{C}_{l_1}^{(8)}(t) \nonumber \,,
   \\   
 \hat f_5(t)={}&-3 (d+1)_4 \mathcal{C}_{l_1}^{(4)}(t)-12 (d+2)_3 t \mathcal{C}_{l_1}^{(5)}(t)+6 (d+3)_2 \left(d-3 t^2\right)
   \mathcal{C}_{l_1}^{(6)}(t)\nonumber \\
   &+12 (4+d) t \left(d-t^2\right) \mathcal{C}_{l_1}^{(7)}(t)+\left(-d (2+d)+6 d t^2-3 t^4\right) \mathcal{C}_{l_1}^{(8)}(t) \nonumber \,, \\
 \hat f_6(t)={}&6 (1+d) (2+d) (4+d) \left(4-(1+7 d) t^2+d (3+d) t^4\right) \mathcal{C}_{l_1}^{(4)}(t)\nonumber \\
   &+12 (2+d) (4+d) t \left(t^2-1\right) \left(-1-7 d+2 d (3+d)
   t^2\right) \mathcal{C}_{l_1}^{(5)}(t)\nonumber \\
   &+6 (4+d) \left(t^2-1\right)^2 \left(-1-7 d+6 d (3+d) t^2\right) \mathcal{C}_{l_1}^{(6)}(t)+24 d (4+d) t
   \left(t^2-1\right)^3 \mathcal{C}_{l_1}^{(7)}(t)\nonumber \\
   &+6 d \left(t^2-1\right)^4 \mathcal{C}_{l_1}^{(8)}(t) \nonumber \,, \\
 \hat f_7(t)={}&-12 (1+d) (2+d) (4+d) \left(-4+(3+d) t^2\right) \mathcal{C}_{l_1}^{(4)}(t)\nonumber 
\\
   &-12 (2+d) (4+d) t \left(-8 (1+d)+(3+d)^2 t^2\right)
   \mathcal{C}_{l_1}^{(5)}(t)\nonumber \\
   &-12 (4+d) \left(t^2-1\right) \left(-1-7 d+3 (1+d) (3+d) t^2\right) \mathcal{C}_{l_1}^{(6)}(t)\nonumber \\
   &-12 (4+d) (1+3 d) t
   \left(t^2-1\right)^2 \mathcal{C}_{l_1}^{(7)}(t)-12 d \left(t^2-1\right)^3 \mathcal{C}_{l_1}^{(8)}(t)\,,
\label{eq:f_40_1}
\\
 \hat f_8(t)={}&-12 (1+d) (2+d) (4+d) \left(-2+(3+d) t^2\right) \mathcal{C}_{l_1}^{(4)}(t)\nonumber \\
   &-12 (2+d) (4+d) t \left(-7-d+4 (3+d) t^2\right)
   \mathcal{C}_{l_1}^{(5)}(t)\nonumber \\
   &+6 (4+d) \left(-1-7 d+(4+d) (7+d) t^2-12 (3+d) t^4\right) \mathcal{C}_{l_1}^{(6)}(t)\nonumber \\
   &+12 (4+d) t \left(1+d-4 t^2\right)
   \left(t^2-1\right) \mathcal{C}_{l_1}^{(7)}(t)+6 \left(d-2 t^2\right) \left(t^2-1\right)^2 \mathcal{C}_{l_1}^{(8)}(t) \nonumber \,, 
   \end{align}
\begin{align}    
 \hat f_9(t)={}&-3 (2+d) (4+d) \left(8+(1+d) t^2 \left(-8+(3+d) t^2\right)\right) \mathcal{C}_{l_1}^{(4)}(t)\nonumber \\
   &-12 (2+d) (4+d) t \left(t^2-1\right) \left(-4+(3+d)
   t^2\right) \mathcal{C}_{l_1}^{(5)}(t)\nonumber \\
   &-6 (4+d) \left(t^2-1\right)^2 \left(-4+3 (3+d) t^2\right) \mathcal{C}_{l_1}^{(6)}(t)-12 (4+d) t \left(t^2-1\right)^3
   \mathcal{C}_{l_1}^{(7)}(t)\nonumber \\
   &-3 \left(t^2-1\right)^4 \mathcal{C}_{l_1}^{(8)}(t) \nonumber \,, \\
 \hat f_{10}(t)={}&6 (d+1)_4 t \left(d t^2-1\right) \mathcal{C}_{l_1}^{(4)}(t)+12 (d+2)_3 t^2 \left(-1+d \left(-1+2 t^2\right)\right)
   \mathcal{C}_{l_1}^{(5)}(t)\nonumber \\
   &+6 (d+3)_2 t \left(t^2-1\right) \left(-1+d \left(-1+6 t^2\right)\right) \mathcal{C}_{l_1}^{(6)}(t)+24 d (4+d) t^2
   \left(t^2-1\right)^2 \mathcal{C}_{l_1}^{(7)}(t)\nonumber \\
   &+6 d t \left(t^2-1\right)^3 \mathcal{C}_{l_1}^{(8)}(t) \nonumber \,, \\
 \hat f_{11}(t)={}&-12 (d+1)_4 t \mathcal{C}_{l_1}^{(4)}(t)-12 (2+d) (3+d)^2 (4+d) t^2 \mathcal{C}_{l_1}^{(5)}(t)\nonumber \\
   &-12 (1+d) (d+3)_2 t \left(-1+3
   t^2\right) \mathcal{C}_{l_1}^{(6)}(t)-12 (4+d) (1+3 d) t^2 \left(t^2-1\right) \mathcal{C}_{l_1}^{(7)}(t)\nonumber \\
   &-12 d t \left(t^2-1\right)^2
   \mathcal{C}_{l_1}^{(8)}(t) \nonumber \,, \\
 \hat f_{12}(t)={}&-6 (d+1)_4 t \mathcal{C}_{l_1}^{(4)}(t)-24 (d+2)_3 t^2 \mathcal{C}_{l_1}^{(5)}(t)+6 (d+3)_2 t \left(1+d-6
   t^2\right) \mathcal{C}_{l_1}^{(6)}(t)\nonumber \\
   &+12 (4+d) t^2 \left(1+d-2 t^2\right) \mathcal{C}_{l_1}^{(7)}(t)+6 t \left(d-t^2\right) \left(t^2-1\right)
   \mathcal{C}_{l_1}^{(8)}(t) \nonumber \,, \\
 \hat f_{13}(t)={}&-3 (d+1)_4 t^2 \mathcal{C}_{l_1}^{(4)}(t)-12 (d+2)_3 t^3 \mathcal{C}_{l_1}^{(5)}(t)-6 (d+3)_2 t^2 \left(-1+3
   t^2\right) \mathcal{C}_{l_1}^{(6)}(t)\nonumber \\
   &-12 (4+d) t^3 \left(t^2-1\right) \mathcal{C}_{l_1}^{(7)}(t)-3 t^2 \left(t^2-1\right)^2 \mathcal{C}_{l_1}^{(8)}(t)
   \nonumber \,, \\
 \hat f_{14}(t)={}&-6 (1+d) (2+d) (4+d) t \left(-4+(3+d) t^2\right) \mathcal{C}_{l_1}^{(4)}(t)\nonumber \\
   &-12 (2+d) (4+d) t^2 \left(-7-d+2 (3+d) t^2\right)
   \mathcal{C}_{l_1}^{(5)}(t)\nonumber \\
   &-6 (4+d) t \left(t^2-1\right) \left(-7-d+6 (3+d) t^2\right) \mathcal{C}_{l_1}^{(6)}(t)\nonumber \\
   &-24 (4+d) t^2 \left(t^2-1\right)^2
   \mathcal{C}_{l_1}^{(7)}(t)-6 t \left(t^2-1\right)^3 \mathcal{C}_{l_1}^{(8)}(t)\,.
\nonumber
\end{align}

\subsection{Projectors to the irreps $(l_1,3,1)$}
To construct the  tensor structures for projectors to irreps $\bts{young_hook4b}$ we need to consider the 
reducible representation for the shape $\bts{young_21}$, given by
\beq
\btm{young_21} \oplus \btm{young_2} \oplus \btm{young_01}\ ,
\eeq
with the building blocks
\beq
{\mathord{\vcenter{\hbox{\scalebox{0.8}{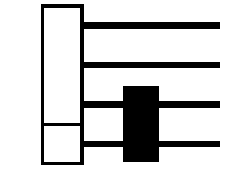}}}}}
\quad
,
\quad
{\mathord{\vcenter{\hbox{\scalebox{0.8}{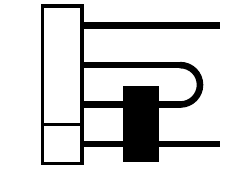}}}}}
\quad
,
\quad
{\mathord{\vcenter{\hbox{\scalebox{0.8}{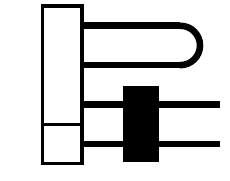}}}}}\quad
.
\eeq
Note that the second expression here is not symmetric in its indices. However, the third expression is antisymmetric and the second one has a symmetric part, so this will suffice as a basis. From the tensor product
\beq
\left(\, \btm{young_21} \oplus \btm{young_2} \oplus \btm{young_01} \, \right)
\otimes
\left(\, \btm{young_21} \oplus \btm{young_2} \oplus \btm{young_01} \, \right)
\otimes
\sum\limits_{q=0}^6
\btm{young_hook_q} = 21 \bullet \oplus \ldots\,,
\eeq
we conclude that in this case there will be 21 allowed structures, which can be identified with different terms in the tensor product
\bea
\btm{young_21} \otimes \btm{young_21} &= \bullet \oplus 2 \, \btm{young_2} \oplus 2 \, \btm{young_4} \oplus \btm{young_6} \oplus \ldots &&\rightarrow \ Q_1,\ldots,Q_6\,,\\
\btm{young_2} \otimes \btm{young_2} &= \bullet \oplus \btm{young_2} \oplus \btm{young_4} \oplus \ldots &&\rightarrow \ Q_7,Q_8,Q_9\,,\\
\btm{young_01} \otimes \btm{young_01} &= \bullet \oplus \btm{young_2} \oplus \ldots &&\rightarrow \ Q_{10},Q_{11}\,,\\
\btm{young_21} \otimes \btm{young_2} &= \btm{young_2} \oplus \btm{young_4} \oplus \ldots &&\rightarrow \ Q_{12},Q_{13}\,,\\
\btm{young_21} \otimes \btm{young_01} &= \btm{young_2} \oplus \btm{young_4} \oplus \ldots &&\rightarrow \ Q_{14},Q_{15}\,,\\
\btm{young_2} \otimes \btm{young_01} &= \btm{young_2} \oplus \ldots &&\rightarrow \ Q_{16}\,.
\eea{eq:l31_tensor_product_contributions}
Using the  birdtrack  notation (\ref{eq:birdtracks}), each  $Q_i$ has the form
\begin{spreadlines}{20pt} 
\begin{gather}
Q_1 = \ {\mathord{\vcenter{\hbox{\scalebox{0.8}{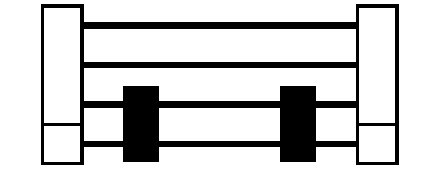}}}}} \ ,\ 
Q_2 = \ {\mathord{\vcenter{\hbox{\scalebox{0.8}{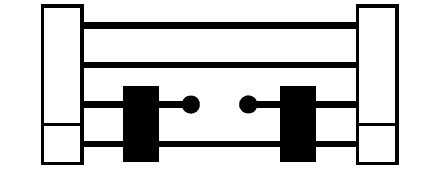}}}}} \ ,\ 
Q_3 = \ {\mathord{\vcenter{\hbox{\scalebox{0.8}{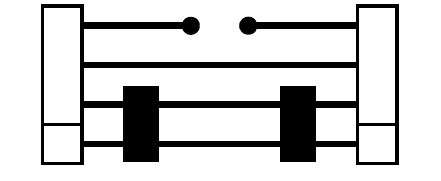}}}}} \ ,\nonumber\\
Q_4 = \ {\mathord{\vcenter{\hbox{\scalebox{0.8}{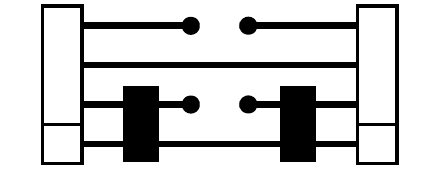}}}}} \ ,\ 
Q_5 = \ {\mathord{\vcenter{\hbox{\scalebox{0.8}{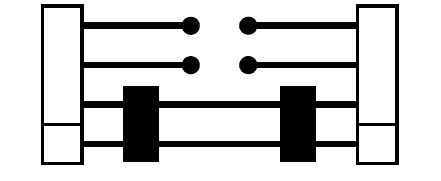}}}}} \ ,\ 
Q_6 = \ {\mathord{\vcenter{\hbox{\scalebox{0.8}{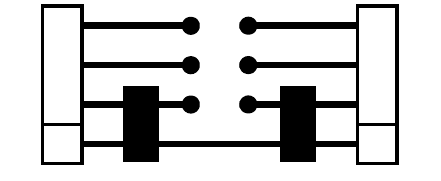}}}}} \ ,\ \nonumber\\
Q_{7} = \ {\mathord{\vcenter{\hbox{\scalebox{0.8}{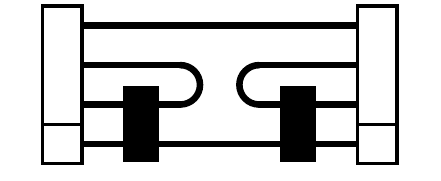}}}}} \ ,\ 
Q_{8} = \ {\mathord{\vcenter{\hbox{\scalebox{0.8}{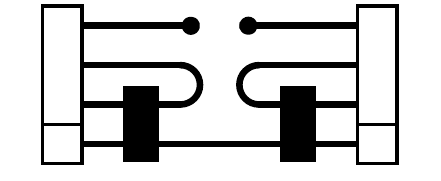}}}}} \ ,\ 
Q_{9} = \ {\mathord{\vcenter{\hbox{\scalebox{0.8}{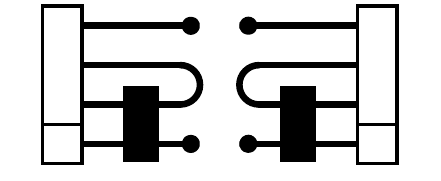}}}}} \ ,\nonumber\\
Q_{10} = \ {\mathord{\vcenter{\hbox{\scalebox{0.8}{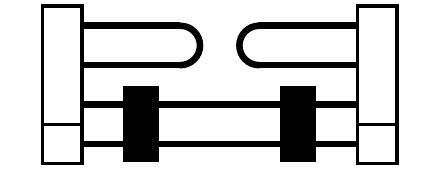}}}}} \ ,\ 
Q_{11} = \ {\mathord{\vcenter{\hbox{\scalebox{0.8}{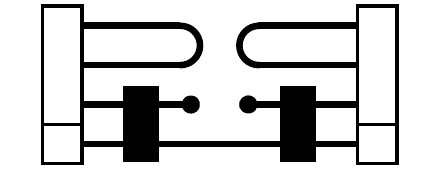}}}}} \ ,\nonumber\\
Q_{12} = \ {\mathord{\vcenter{\hbox{\scalebox{0.8}{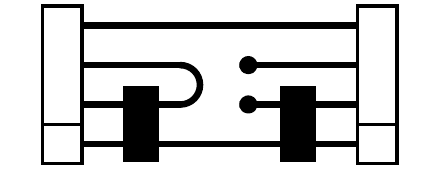}}}}} \ +  \ {\mathord{\vcenter{\hbox{\scalebox{0.8}{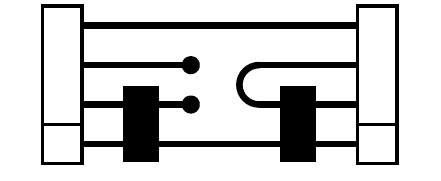}}}}} \quad,
\label{eq:structures_l31}\\
Q_{13} = \ {\mathord{\vcenter{\hbox{\scalebox{0.8}{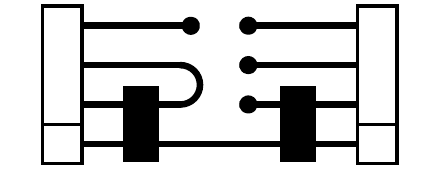}}}}} \ +  \ {\mathord{\vcenter{\hbox{\scalebox{0.8}{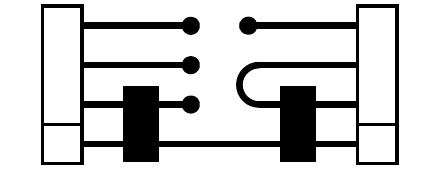}}}}} \quad,\nonumber\\
Q_{14} = \ {\mathord{\vcenter{\hbox{\scalebox{0.8}{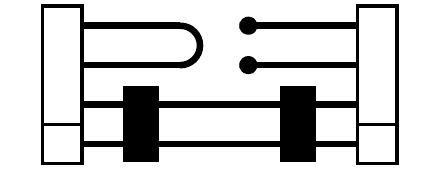}}}}} \ +  \ {\mathord{\vcenter{\hbox{\scalebox{0.8}{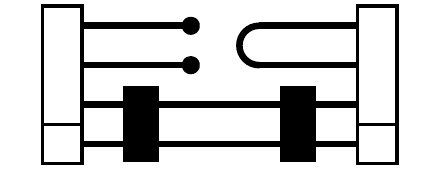}}}}} \quad,\nonumber
\end{gather}
\begin{gather}
Q_{15} = \ {\mathord{\vcenter{\hbox{\scalebox{0.8}{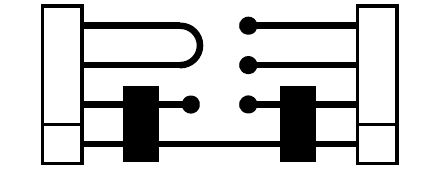}}}}} \ +  \ {\mathord{\vcenter{\hbox{\scalebox{0.8}{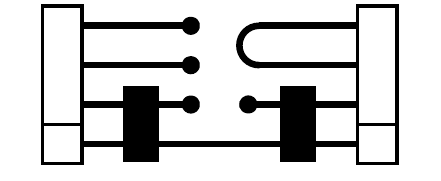}}}}} \quad,\nonumber\\
Q_{16} = \ {\mathord{\vcenter{\hbox{\scalebox{0.8}{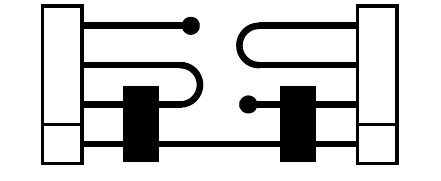}}}}} \ +  \ {\mathord{\vcenter{\hbox{\scalebox{0.8}{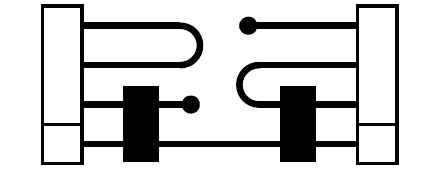}}}}} \quad.
\nonumber
\end{gather}
\end{spreadlines}
The functions $f_i(t)$ can now be written as 
\beq
f_i(t)= - \frac{\hat f_i(t)}{(d-2)d}\,, \qquad  \forall i=1,\ldots,16\,,
\eeq
with
\begin{align}
 \hat f_1(t)={}&-(d-1) (1+d)_2 t \left(4-3 d+(d-2) d t^2\right) \mathcal{C}_{l_1}^{(3)}(t)\nonumber \\
   &-(2+d) \left(t^2-1\right) \left(4-3 (d-1) d+(d-2) d (1+3 d) t^2\right)
   \mathcal{C}_{l_1}^{(4)}(t)\nonumber \\
   &-(d-2) d (5+3 d) t \left(t^2-1\right)^2 \mathcal{C}_{l_1}^{(5)}(t)-(d-2) d \left(t^2-1\right)^3 \mathcal{C}_{l_1}^{(6)}(t) \nonumber \,, \\
 \hat f_2(t)={}&4 (d-1)_4  t \mathcal{C}_{l_1}^{(3)}(t)+2 (2+d) \left(4-3 (d-1) d+d (-10+d (5+d)) t^2\right) \mathcal{C}_{l_1}^{(4)}(t)\nonumber \\
   &+4 d (-7+d (2+d)) t
   \left(t^2-1\right) \mathcal{C}_{l_1}^{(5)}(t)+2 (d-2) d \left(t^2-1\right)^2 \mathcal{C}_{l_1}^{(6)}(t) \nonumber \,, \\
 \hat f_3(t)={}&4 \left(4-5 d^2+d^4\right) t \mathcal{C}_{l_1}^{(3)}(t)+2 (2+d) \left(4-4 t^2+(d-1) d \left(-3+(2+d) t^2\right)\right) \mathcal{C}_{l_1}^{(4)}(t)\nonumber \\
   &+4
   \left(-4-d+d^3\right) t \left(t^2-1\right) \mathcal{C}_{l_1}^{(5)}(t)+2 (d-2) d \left(t^2-1\right)^2 \mathcal{C}_{l_1}^{(6)}(t) \nonumber \,, \\
 \hat f_4(t)={}&4 (d-1)_4 t \mathcal{C}_{l_1}^{(3)}(t)+4 (2+d) \left(4+d \left(3+t^2+3 d \left(t^2-1\right)\right)\right) \mathcal{C}_{l_1}^{(4)}(t)\nonumber \\
   &+4 t
   \left(8+d \left(6+5 t^2-d \left(5+d-3 t^2\right)\right)\right) \mathcal{C}_{l_1}^{(5)}(t)+4 d \left(-2+d+t^2-d t^2+t^4\right) \mathcal{C}_{l_1}^{(6)}(t) \nonumber \,, 
\\
 \hat f_5(t)={}&(d-4) (d-1) (1+d)_2 t \mathcal{C}_{l_1}^{(3)}(t)+(2+d) \left(4+3 d-3 d^2+(d-4) (1+3 d) t^2\right) \mathcal{C}_{l_1}^{(4)}(t)\nonumber \\
   &+t \left(16-d
   \left(6+d+d^2\right)+(d-4) (5+3 d) t^2\right) \mathcal{C}_{l_1}^{(5)}(t)\nonumber \\
   &+\left(t^2-1\right) \left(-(d-2) d+(d-4) t^2\right) \mathcal{C}_{l_1}^{(6)}(t) \nonumber \,, \\
 \hat f_6(t)={}&2 (2+d) (4+3 d-3 d^2) \mathcal{C}_{l_1}^{(4)}(t)
   -4 \left(3 d^2 +d-8\right) t \mathcal{C}_{l_1}^{(5)}(t)\nonumber \\
   &+\left(2 (d-2) d+2 (4-3 d) t^2\right)
   \mathcal{C}_{l_1}^{(6)}(t) \nonumber \,, \\
 \hat f_7(t)={}&4 (d-1) (1+d)_2 t \left(-4+d t^2\right) \mathcal{C}_{l_1}^{(3)}(t)+4 d (2+d) \left(t^2-1\right) \left(-4+(1+3 d) t^2\right)
   \mathcal{C}_{l_1}^{(4)}(t)\nonumber \\
   &+4 d (5+3 d) t \left(t^2-1\right)^2 \mathcal{C}_{l_1}^{(5)}(t)+4 d \left(t^2-1\right)^3 \mathcal{C}_{l_1}^{(6)}(t) \nonumber \,, \\
 \hat f_8(t)={}&-16 (d-1) (1+d)_2 t \mathcal{C}_{l_1}^{(3)}(t)+8 (2+d) \left(4 d-(-2+d (5+d)) t^2\right) \mathcal{C}_{l_1}^{(4)}(t)
\nonumber
  \\
   &-16 d (3+d) t
   \left(t^2-1\right) \mathcal{C}_{l_1}^{(5)}(t)-8 d \left(t^2-1\right)^2 \mathcal{C}_{l_1}^{(6)}(t) \,,
    \label{eq:f_31} \\
 \hat f_9(t)={}&16 d (2+d) \mathcal{C}_{l_1}^{(4)}(t)
   +4 d (7+d) t \mathcal{C}_{l_1}^{(5)}(t)+4 d \left(t^2-1\right) \mathcal{C}_{l_1}^{(6)}(t) \nonumber \,, \\
 \hat f_{10}(t)={}&(d-4) \left(d^2+d-2\right) t \left((1+d) t^2-4\right) \mathcal{C}_{l_1}^{(3)}(t)
   +(d-4) (2+d)  
   \left((1+3 d) t^2-4\right)\nonumber \\
   &\times \left(t^2-1\right) \mathcal{C}_{l_1}^{(4)}(t)
   +(d-4) (5+3 d) t \left(t^2-1\right)^2 \mathcal{C}_{l_1}^{(5)}(t)+(d-4) \left(t^2-1\right)^3 \mathcal{C}_{l_1}^{(6)}(t) \nonumber \,, 
   \end{align}
\begin{align}
 \hat f_{11}(t)={}&-16 (d-1) (1+d)_2 t \mathcal{C}_{l_1}^{(3)}(t)-2 (2+d) \left(16-12 d+(-28+3 d (7+d)) t^2\right) \mathcal{C}_{l_1}^{(4)}(t)\nonumber \\
   &-4 (-16+3 d (3+d)) t
   \left(t^2-1\right) \mathcal{C}_{l_1}^{(5)}(t)-2 (3d-4) \left(t^2-1\right)^2 \mathcal{C}_{l_1}^{(6)}(t) \nonumber \,, \\
 \hat f_{12}(t)={}&4 (d-1)_4 t^2 \mathcal{C}_{l_1}^{(3)}(t)+4 (2+d) t \left(-4+d-d^2+d (1+3 d) t^2\right) \mathcal{C}_{l_1}^{(4)}(t)\nonumber \\
   &+4 d (5+3 d) t^2
   \left(t^2-1\right) \mathcal{C}_{l_1}^{(5)}(t)+4 d t \left(t^2-1\right)^2 \mathcal{C}_{l_1}^{(6)}(t) \nonumber \,, \\
 \hat f_{13}(t)={}&-4 \left(8+d \left(2+d+d^2\right)\right) t \mathcal{C}_{l_1}^{(4)}(t)
   -8 \left(2+d+d^2\right) t^2 \mathcal{C}_{l_1}^{(5)}(t)-4 d t
   \left(t^2-1\right) \mathcal{C}_{l_1}^{(6)}(t) \nonumber \,, \\
 \hat f_{14}(t)={}&(d-4) (d-1) (1+d)_2 t^2 \mathcal{C}_{l_1}^{(3)}(t)+(d-4) (2+d) t \left(-3+t^2+d \left(-1+3 t^2\right)\right) \mathcal{C}_{l_1}^{(4)}(t)\nonumber \\
   &+(d-4)
   (5+3 d) t^2 \left(t^2-1\right) \mathcal{C}_{l_1}^{(5)}(t)+(d-4) t \left(t^2-1\right)^2 \mathcal{C}_{l_1}^{(6)}(t) \nonumber \,, \\
 \hat f_{15}(t)={}&2 (4-d) (2+d) (3+d) t \mathcal{C}_{l_1}^{(4)}(t)
   -4 (d-4) (3+d) t^2 \mathcal{C}_{l_1}^{(5)}(t)-2 (d-4) t \left(t^2-1\right)
   \mathcal{C}_{l_1}^{(6)}(t) \nonumber \,, \\
 \hat f_{16}(t)={}&16 (d-1) (1+d)_2 t \mathcal{C}_{l_1}^{(3)}(t)+4 (2+d) \left(-4 d+(-8+d (11+d)) t^2\right) \mathcal{C}_{l_1}^{(4)}(t)\nonumber \\
   &+8 (-2+d (5+d)) t
   \left(t^2-1\right) \mathcal{C}_{l_1}^{(5)}(t)+4 d \left(t^2-1\right)^2 \mathcal{C}_{l_1}^{(6)}(t)\,.
\nonumber
\end{align}

\section{Computation of the constants $\calS_{\Delta \Delta_{12}}^{\lambda}$ from the shadow formalism}
\label{sec:shadow_constant}

In this section the constant 
\beq
\calS_{\Delta \Delta_{12}}^\lambda 
= \frac{ \bra  \calO_1 \big(P_1; \bZ_1\big) \calO_2 \big(P_2; \bZ_2\big) \tilde  \calO \big(P_3; \bZ_3\big)
\ket }
{\bra \calO_1 \big(P_1; \bZ_1\big) \calO_2 \big(P_2; \bZ_2\big) \calO \big(P_3; \bZ_3\big) \ket \big|_{\Delta \to \tilde \Delta}} \,,
\label{eq:shadow_3pt_constant_def_2}
\eeq
 is computed.
Previous known results are the cases $\bra \bullet \bullet \, \bts{young_hook0} \, \ket$ \cite{Dolan:2011dv}
and $\bra \bullet \, \bts{young_1} \  \bts{young_hook1} \, \ket$ \cite{Rejon-Barrera:2015bpa}.
In this appendix, the constant is computed for any three-point function that has a single tensor structure and obeys $|\lambda_1|+|\lambda_2| = |\lambda|-l_1$
(as discussed in section \ref{sec:classification_of_seed_conformal_blocks}).
We start in subsection \ref{sec:spin_transfer_operator} by showing that $\calS_{\Delta \Delta_{12}}^\lambda$ does not depend on the $SO(d)$ irreps
$\lambda_1$ and $\lambda_2$, by constructing a differential operator that transfers spin between the operators 
$\calO_1$ and $\calO_2$ without violating the condition 
$|\lambda_1|+|\lambda_2| = |\lambda|-l_1$.
In the following subsections $\calS_{\Delta \Delta_{12}}^\lambda$ is  computed for the case
of an arbitrary irrep $\lambda = (l_1,l_2,l_3,\ldots)$ and $\lambda_1=\bullet$, $\lambda_2=(l_2,l_3,\ldots)$.

\subsection{Spin transfer operator}
\label{sec:spin_transfer_operator}

The shadow operator is
\beq
\tilde \calO (P_3; \bZ_3) 
=
 \int D^d P_0 
\left.
\calO \big(P_0; \bdel_{\bZ_0}\big)
\pi_\lambda(\bZ_0; \bdel_{\bar \bZ_0})
 \bra \calO \big(P_0; \bar \bZ_0\big) \calO \big(P_3; \bZ_3\big)\ket \right|_{\Delta \to \tilde \Delta}
 \,,
\label{eq:shadow}
\eeq
hence the three-point function of the shadow is
\bea
\bra \calO_1 \big(P_1; \bZ_1\big) \calO_2 \big(P_2; \bZ_2\big) \tilde \calO (P_3; \bZ_3) \ket
={}&
 \int D^d P_0 
\bra \calO_1 \big(P_1; \bZ_1\big) \calO_2 \big(P_2; \bZ_2\big) \calO \big(P_0; \bdel_{\bZ_0}\big)\ket\\
&\quad
\pi_\lambda(\bZ_0; \bdel_{\bar \bZ_0})
 \bra \calO \big(P_0; \bar \bZ_0\big) \calO \big(P_3; \bZ_3\big)\ket \big|_{\Delta \to \tilde \Delta}
 \,.
\eea{eq:shadow_3pt}
Our strategy will be to find a differential operator which transforms a three-point function with
operators $\calO$, $\calO_1$ and $\calO_2$
in the representations $\lambda = (l_1,l_2,\ldots,l_{h_1})$ and $\lambda_1=\bullet$, $\lambda_2=(l_2,\ldots,l_{h_1})$
\beq
\bra \calO_1(P_1) \calO_2(P_2,\bZ_2) \calO(P_3,\bZ_{3}) \ket_{\text{enc}} = \frac{
\left(V_{(3,12)}^{(Z_{31})}\right)^{l_1}
\left(H_{23}^{(Z_{21}, Z_{32})}\right)^{l_2}
\ldots
\left(H_{23}^{(Z_{2 (h_1-1)}, Z_{3 h_1})}\right)^{l_{h_1}}
}{
( P_{13})^{\frac{\Delta_1+\Delta - \Delta_2}{2}}
( P_{23})^{\frac{ \Delta_2+\Delta - \Delta_1}{2}}
( P_{12})^{\frac{ \Delta_1+\Delta_2 - \Delta}{2}}}\,,
\label{eq:3point_es}
\eeq
into another three-point function with a single tensor structure
by transferring spin from the operator $\calO_2$ to $\calO_1$. This operator must decrease the homogeneity of a polynomial
in $Z_{2i}$ and increase it in $Z_{1j}$ by one each, while not changing the homogeneity in $P_1$ and $P_2$.
Furthermore, it must preserve the transverseness of the functions, that is
\beq
F(P_i, Z_{ij}+c P_i) = F(P_i, Z_{ij}).
\eeq
This last requirement means that the operator should transfer terms of order
\beq
\calO(Z_{ij}^2, Z_{ij} \cdot P_i, P_i^2) \,, \qquad i \in \{1,2 \} \,,
\eeq
into terms of the same kind.
Such an operator is given in terms of the differential operator derived in Appendix \ref{sec:todorov} in $d+2$ dimensions (i.e. each $d$ in its definition must be replaced by $d+2$),
\beq
{\mathcal D}_{Z_{2i},Z_{1j},P_2,P_1} = \frac{(Z_{1j} \cdot P_2) P_{1A} - (P_1 \cdot P_2) Z_{1jA}}{P_1 \cdot P_2}  \,D_{Z_{2i},P_2}^A\,.
\eeq
The term that is contracted into the operator $D_{Z_{2i},P_2}^A$ ensures transverseness in $Z_{1j}$.
By construction, the action on \eqref{eq:3point_es} must yield a transverse function
\begin{align}
&{\mathcal D}_{Z_{2i},Z_{1j},P_2,P_1}
\frac{
\left(V_{(3,12)}^{(Z_{31})}\right)^{l_1}
\left(H_{23}^{(Z_{21}, Z_{32})}\right)^{l_2}
\ldots
\left(H_{23}^{(Z_{2 (h_1-1)}, Z_{3 h_1})}\right)^{l_{h_1}}
}{
( P_{13})^{\frac{1}{2}( \Delta_1+\Delta - \Delta_2)}
( P_{23})^{\frac{1}{2}( \Delta_2+\Delta - \Delta_1)}
( P_{12})^{\frac{1}{2}( \Delta_1+\Delta_2 - \Delta)}}
\label{eq:spin_trans_ex}
\\
\propto{}&
\left(
\frac{H_{13}^{(Z_{1j}, Z_{3(i+1)})}
+ c
V_{(1,23)}^{(Z_{1j})}
V_{(3,12)}^{(Z_{3(i+1)})}
}{H_{23}^{(Z_{2i}, Z_{3(i+1)})}}
\right)
\frac{
\left(V_{(3,12)}^{(Z_{31})}\right)^{l_1}
\left(H_{23}^{(Z_{21}, Z_{32})}\right)^{l_2}
\ldots
\left(H_{23}^{(Z_{2 (h_1-1)}, Z_{3 h_1})}\right)^{l_{h_1}}
}{
( P_{13})^{\frac{1}{2}( \Delta_1+\Delta - \Delta_2)}
( P_{23})^{\frac{1}{2}( \Delta_2+\Delta - \Delta_1)}
( P_{12})^{\frac{1}{2}( \Delta_1+\Delta_2 - \Delta)}}
\nonumber\\
&+\calO(P_3^2, P_3 \cdot Z_{3k}, Z_{3k}\cdot Z_{3l})\,,
\nonumber
\end{align}
where $c$ is an unspecified constant. The term containing $V_{(3,12)}^{(Z_{3(i+1)})}$ vanishes upon Young symmetrization with $\left(V_{(3,12)}^{(Z_{31})}\right)^{l_1}$. By repeated use of such operators,
any three-point function with a single tensor structure can be generated
\begin{align}
\bra \calO_1(P_1,\bZ_2) \calO_2(P_2,\bZ_2)& \calO(P_3,\bZ_{3}) \ket_\text{full}
\propto
\pi_{\lambda_1} (\bZ_1, \bdel_{\bar \bZ_1}) 
\,\pi_{\lambda_2} (\bZ_2, \bdel_{\bar \bZ_2})
\,\pi_{\lambda} (\bZ_3, \bdel_{\bar \bZ_3})
\label{eq:3pt_function_from_op}
\\
&{\mathcal D}_{\bar Z_{2i}, \bar Z_{1j},P_2,P_1}
\ldots
{\mathcal D}_{\bar Z_{2k}, \bar Z_{1l},P_2,P_1}
\bra \hat \calO_1(P_1) \hat \calO_2(P_2,\bar \bZ_2) \calO(P_3,\bar \bZ_{3}) \ket_{\text{enc}}\,,
\nonumber
\end{align}
where $\hat \calO_1(P_1)$ is in the scalar representation $\hat \lambda_1 = \bullet$ and 
$\hat \calO_2(P_2, \bar \bZ_2)$ in an irrep $\hat \lambda_2$ satisfying $|\hat \lambda_2| = |\lambda_1| + |\lambda_2|$. 
In order to compute $\calS_{\Delta \Delta_{12}}^\lambda$ one can insert \eqref{eq:3pt_function_from_op}
into \eqref{eq:shadow_3pt_constant_def_2}. The action of the operator on  both three-point functions is obviously 
the same, and the resulting $\calS_{\Delta \Delta_{12}}^\lambda$ is the same that one gets when using the three-point function
given in \eqref{eq:3point_es}. In the next two subsections, $\calS_{\Delta \Delta_{12}}^\lambda$
is computed using this three-point function.

\subsection{Young diagrams with two rows}

The computations here are done using the conventions of \cite{Rejon-Barrera:2015bpa} to allow reusing some of their computations.
We start with the case of $\lambda=(l_1,l_2)$. To compute the constant we will consider a three-point function of an operator $\calO$ in this representation with a scalar $\calO_1$
and a symmetric tensor $\calO_2$ of spin $l_2$. This correlator can be written as
\beq
\bra \calO_1(x_1) \calO_2(x_2,z_2) \calO(x_3,\bz_{3}) \ket = \frac{
\pi_\lambda (\bz_{3}; \bdel_{\bar \bz_{3}})
\left(k^{(312)}(\bar z_{31})\right)^{l_1} \left(m^{(23)}(z_2, \bar z_{32})\right)^{l_2}}{
\big( x_{13}^2\big)^{\frac{ \Delta_1-\Delta_2+\Delta_3}{2}}
\big( x_{23}^2\big)^{\frac{-\Delta_1+\Delta_2+\Delta_3}{2}}
\big( x_{12}^2\big)^{\frac{\Delta_1+\Delta_2-\Delta_3}{2}}}\,,
\label{eq:3point}
\eeq
where the building blocks $m^{(ij)}$ and $k^{(ijk)}$ are physical space variants of the building blocks $H_{ij}$ and $V_{i,jk}$ defined in \eqref{eq:HV_building_blocks}, given by
\beq
m^{(ij)}_{ab} = \delta_{ab} - \frac{2}{x_{ij}^2} (x_{ij})_a  (x_{ij})_b \,, \qquad
k^{(ijk)}_{a} = \frac{x_{ij}^2 (x_{ik})_a - x_{ik}^2 (x_{ij})_a}{\big(x_{ij}^2 x_{ik}^2 x_{jk}^2\big)^{1/2}}\,.
\eeq
We will also use the notation
\beq
m^{(ij)} (z_k,z_l) = z_k^a m^{(ij)}_{ab} z_l^b \,, \qquad k^{(ijk)} (z) = k^{(ijk)}_a z^a\,.
\eeq
To compute the ratio in \eqref{eq:shadow_3pt_constant_def_2} it is not necessary to keep track of terms
containing $z_{31}^2$, $z_{32}^2$ or $z_{31}\cdot z_{32}$. Those will be collectively  denoted by $\calO(z_{3i}\cdot z_{3j})$.
Furthermore, it is enough to consider the term of order $l_2$ in $z_2 \cdot z_{32}$, so terms of order $\calO\big((z_{2}\cdot z_{32})^{l_2-1}\big)$ can be dropped.

We need to compute the three-point function of the shadow operator. This is given by
\begin{align}
&\bra \calO_1(x_1) \calO_2(x_2,z_2) \tilde \calO(x_3,\bz_{3}) \ket
\nonumber\\
={}& 
\frac{1}{l_1! l_2!}
\,\pi_\lambda (\bz_{3}; \bdel_{\bar \bz_{3}})
\int 
\frac{d^d x_0}{\big(x_{03}^2\big)^{d-\Delta}
\big( x_{01}^2\big)^{\frac{ \Delta_1-\Delta_2+\Delta}{2}}\big( x_{02}^2\big)^{\frac{-\Delta_1+\Delta_2+\Delta}{2}}\big( x_{12}^2\big)^{\frac{\Delta_1+\Delta_2-\Delta}{2}}}
\label{eq:shadow_3point}
\\
&
\left( m^{(30)}(\bar z_{31}, \partial_{z_{01}}) \right)^{l_1}
\left( m^{(30)}(\bar z_{32}, \partial_{z_{02}}) \right)^{l_2}
\pi_\lambda (\bz_{0}; \bdel_{\bar \bz_{0}})
\left( k^{(012)}(\bar z_{01}) \right)^{l_1}
\left( m^{(02)}(\bar z_{02},z_2) \right)^{l_2}\nonumber\\
={}&
\pi_\lambda (\bz_{3}; \bdel_{\bar \bz_{3}})
\int \frac{d^d x_0}{\big(x_{03}^2\big)^{d-\Delta}}
\frac{\big( y(\bar z_{31}) \big)^{l_1}
\big( y_2(\bar z_{32},z_2) \big)^{l_2}}{
\big( x_{01}^2\big)^{\frac{\Delta_1-\Delta_2+\Delta}{2}}\big( x_{02}^2\big)^{\frac{-\Delta_1+\Delta_2+\Delta}{2}}\big( x_{12}^2\big)^{\frac{ \Delta_1+\Delta_2-\Delta}{2}}}\,,
\nonumber
\end{align}
where we defined
\begin{align}
&y(z_{31})= z_{31}^a m^{(30)}_{ab}k^{(012)\,b}=
\left(\sqrt{\frac{x_{01}^2x_{23}^2}{x_{03}^2x_{12}^2}}-\frac{x_{02}^2x_{13}^2}{\sqrt{x_{01}^2x_{03}^2x_{12}^2x_{23}^2}}\right) k^{(302)}(z_{31})+\sqrt{\frac{x_{02}^2x_{13}^2}{x_{01}^2x_{23}^2}}k^{(312)} (z_{31})\,,
\nonumber\\
&y_2(z_{32},z_2)= z_{32}^a m^{(30)}_{ab} m^{(02)}_{bc} z_2^c = m^{(32)}(z_{32},z_2) - 2 \, k^{(302)} (z_{32}) k^{(203)} (z_2) \,.
\label{eq:y_def}
\end{align}
The reason why the trace subtracting terms of the projector $\pi_\lambda (\bz_{0}; \bdel_{\bar \bz_{0}})$ in \eqref{eq:shadow_3point}
do not contribute is that $m^{(30)}_{ac}m^{(30)}_{bc} = \delta_{ab}$, so these terms are annihilated by the other projector.
Not keeping track of terms containing
$z_{31}^2$, $z_{32}^2$ or $z_{31}\cdot z_{32}$ means that the trace subtracting terms of the other projector can be ignored as well and
we can use instead the Young symmetrizer
\beq
Y_\lambda (z_{1},z_{2}; \bar z_{1}, \bar z_{2})
= \pi_\lambda (z_{1},z_{2}; \bar z_{1}, \bar z_{2}) \big|_{z_i^2 = z_1 \cdot z_2 = \bar z_i^2 = \bar z_1 \cdot \bar z_2 = 0} \,,
\eeq
leading to
\begin{align}
\label{eq:shadow_3point_with_Y}
&\bra \calO_1(x_1) \calO_2(x_2,z_2) \tilde \calO(x_3,\bz_{3}) \ket\\
&= 
Y_\lambda (\bz_{3}; \bdel_{\bar \bz_{3}})
\int \frac{d^d x_0}{\big(x_{03}^2\big)^{d-\Delta}}
\frac{\big( y(\bar z_{31}) \big)^{l_1}
\big( y_2(\bar z_{32},z_2) \big)^{l_2}}{
\big( x_{01}^2\big)^{\frac{\Delta_1-\Delta_2+\Delta}{2}}\big( x_{02}^2\big)^{\frac{-\Delta_1+\Delta_2+\Delta}{2}}\big( x_{12}^2\big)^{\frac{ \Delta_1+\Delta_2-\Delta}{2}}}
+ \calO(z_{3i}\cdot z_{3j})\,.
\nonumber
\end{align}
One needs to compute the conformal integral
\bea
&I_{\alpha, \beta, \gamma}^{n,m}(x_1,x_2,x_3,z_{31},z_{32},z_2)
= \int \frac{d^d x_0 \ \left(k^{(302)} (z_{31})\right)^n \left(k^{(302)} (z_{32}) k^{(203)} (z_2)\right)^m}{\big(x_{01}^2\big)^{\alpha} \big(x_{02}^2\big)^{\beta} \big(x_{03}^2\big)^{\gamma}} \,,
\eea{eq:vector_integral2}
which is done in subsection \ref{sec:conformal_integrals} below. 
Next we do a trinomial expansion of $y(z_{31})$ and a binomial expansion of $y_2(z_{32},z_2)$ in \eqref{eq:shadow_3point_with_Y}.
Note that we consider only the term proportional to $(z_{2}\cdot z_{32})^{l_2}$, for which the action of the Young symmetrizer results only in
the factor $S_{l_1,k+b,i}$, which will be explained below
\bea
&\bra \calO_1(x_1) \calO_2(x_2,z_2) \tilde \calO(x_3,\bz_{3}) \ket \\
={}&\sum_{k=0}^{l_1}\sum_{b=0}^{l_1-k}\frac{l_1!}{k!b!(l_1-k-b)!}( -1)^b
\sum_{i=0}^{l_2} \frac{l_2!}{i!(l_2-i)!}(-2)^i S_{l_1,k+b,i}\\
& \big( x_{12}^2\big)^{\frac{ -\Delta_1-\Delta_2+\Delta-k-b}{2}}\big( x_{13}^2\big)^{\frac{l_1-k+b}{2}}\big( x_{23}^2\big)^{k-\frac{l_1}{2}} 
\left(k^{(312)} (z_{31}) \right)^{l_1-k-b} \left(z_2 \cdot z_{32}\right)^{l_2-i} \\
& I^{k+b,i}_{\frac{1}{2}(\Delta_{12}+\Delta+l_1-2k),\frac{1}{2}( -\Delta_{12}+\Delta-l_1+k-b),d-\Delta+\frac{k+b}{2}}(x_1,x_2,x_3,z_{31},z_{32},z_2)\\
& + \calO(z_{3i}\cdot z_{3j}) + \calO\big((z_{2}\cdot z_{32})^{l_2-1}\big)\\
={}&\pi^{d/2} \left(k^{(312)} (z_{31}) \right)^{l_1} \left(z_2 \cdot z_{32}\right)^{l_2}\big( x_{12}^2\big)^{\frac{\tilde \Delta -\Delta_1-\Delta_2}{2}}\big( x_{13}^2\big)^{\frac{\Delta_2 -\Delta_1-\tilde \Delta}{2}}\big( x_{23}^2\big)^{\frac{\Delta_1-\Delta_2-\tilde \Delta}{2}}\\
& \sum_{i=0}^{l_2}\sum_{k=0}^{l_1-i}\sum_{b=0}^{l_1-k-i} \frac{(l_1-i)! l_2!}{k!b!(l_1-k-b-i)!(l_2-i)!}( -1)^{b+i}\\
& \frac{\G\big(\frac{1}{2}( d-\Delta_{12}-\Delta-l_1)+k+i\big) \G\big(\frac{1}{2}( d+\Delta_{12}-\Delta+l_1)+b\big)\G\big(\Delta-\frac{d}{2}\big)}{\G\big(\frac{1}{2}(\Delta_{12}+\Delta+l_1)-k\big)\G\big(\frac{1}{2}( -\Delta_{12}+\Delta-l_1)+k+i\big)\G(d-\Delta+k+b+i)}\\
& + \calO(z_{3i}\cdot z_{3j}) + \calO\big((z_{2}\cdot z_{32})^{l_2-1}\big)\\
={}&\pi^{d/2} \left(k^{(312)} (z_{31}) \right)^{l_1} \left(z_2 \cdot z_{32}\right)^{l_2}\big( x_{12}^2\big)^{\frac{\tilde \Delta -\Delta_1-\Delta_2}{2}}\big( x_{13}^2\big)^{\frac{\Delta_2 -\Delta_1-\tilde \Delta}{2}}\big( x_{23}^2\big)^{\frac{\Delta_1-\Delta_2-\tilde \Delta}{2}}\\
& \frac{(\Delta-2)_{l_1+1}}{(\Delta-2+l_2)} \frac{\G\big(\Delta-\frac{d}{2}\big)\G\big(\frac{1}{2}( d+\Delta_{12}-\Delta+l_1)\big)\G\big(\frac{1}{2}( d-\Delta_{12}-\Delta+l_1)\big)}{\G(d-\Delta+l_1)\G\big(\frac{1}{2}(\Delta_{12}+\Delta+l_1)\big)\G\big(\frac{1}{2}( -\Delta_{12}+\Delta+l_1)\big)}\\
& + \calO(z_{3i}\cdot z_{3j}) + \calO\big((z_{2}\cdot z_{32})^{l_2-1}\big)\,.
\eea{eq:shadow_computation1}
The sums were evaluated by using first that
\beq
\sum\limits_{b=0}^{N} \frac{N!}{b!(N-b)!}(-1)^b \frac{\G(x+b)}{\G(y+b)}
=
\frac{\G(x)\G(y-x+N)}{\G(y+N) \G(y-x)}\,,
\eeq
then 
\beq
\sum\limits_{k=0}^{N} \frac{N!}{k!(N-k)!} \frac{1}{\G(x+k)\G(y-k)}
=
\frac{\G(x+y+N-1)}{\G(x+N)\G(y)\G(x+y-1)}\,,
\eeq
and finally
\beq
\sum\limits_{i=1}^{l_2} \frac{l_2! (-1)^i}{(l_2-i)!} \frac{\G(\Delta - 1 + l_1)}{\G(\Delta - 1 + i)}
= \frac{(\Delta-2)_{l_1+1}}{(\Delta-2+l_2)}\,.
\eeq
The factor $S_{l_1,k+b,i}$ appearing in \eqref{eq:shadow_computation1} is given by
\beq
S_{l_1,k+b,i} = \frac{(l_1-k-b)_{(i)}}{(l_1)_{(i)}}\,,
\eeq
where we used the following notation for the falling factorial
\beq 
(x)_{(n)} \equiv (x-n+1)_n = (x)(x-1)\ldots(x-n+1)\,.
\eeq
This factor arises from the Young symmetrization of the tensor encoded by $(z_{31},z_{32})$.
This can be understood by drawing birdtracks, where the $(z_{31},z_{32})$ are antisymmetrized
according to the Young symmetrizer.
The following birdtracks represent the term
\beq
\left( k^{(312)} (z_{31}) \right)^{l_1-k-b}
\left( k^{(302)} (z_{31}) \right)^{k+b}
\left( k^{(302)} (z_{32}) \right)^{i},
\eeq
which appears in \eqref{eq:shadow_computation1} before integration. The contraction with multiple copies of $k^{(302)}_a$
results in a symmetrization which cancels all terms in which two antisymmetrized indices get symmetrized.
Consider for example the case with $i=1$, where the cancellations lead to a factor of $S_{l_1,k+b,1} =\frac{l_1-k-b}{l_1}$,
\beq
 {\mathord{\vcenter{\hbox{\scalebox{0.8}{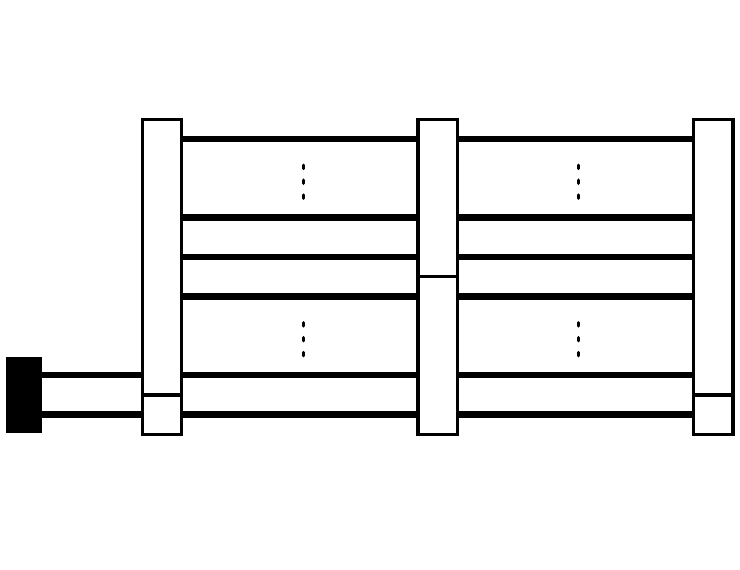}}}}}
\ = \ 
\frac{l_1-k-b}{l_1}
 {\mathord{\vcenter{\hbox{\scalebox{0.8}{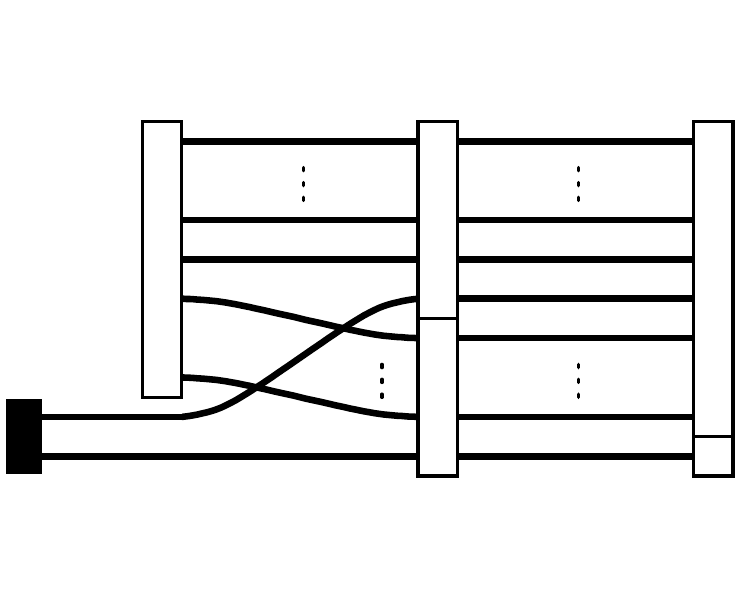}}}}}\ \ .
\nonumber
\eeq
Consider also the next case $i=2$,
\beq
 {\mathord{\vcenter{\hbox{\scalebox{0.8}{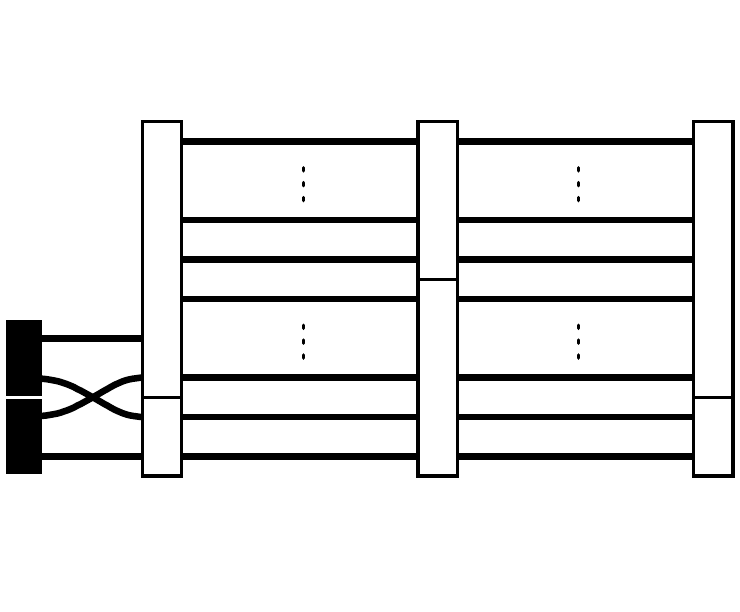}}}}}
\ =\ 
\frac{(l_1-k-b)_{(2)}}{(l_1)_{(2)}}
 {\mathord{\vcenter{\hbox{\scalebox{0.8}{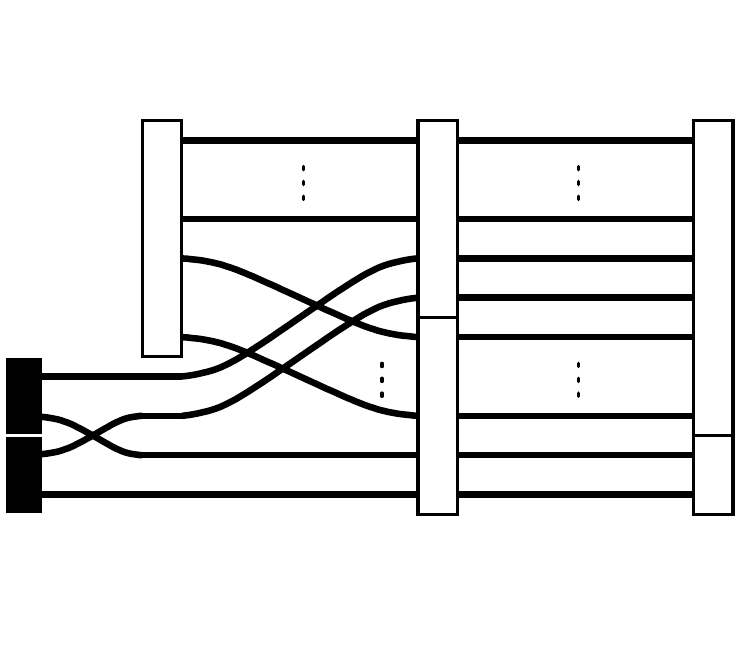}}}}}\ \ .
\nonumber
\eeq
The constant $\calS$ is now found by comparing \eqref{eq:shadow_computation1}
with the corresponding term in \eqref{eq:3point}
\beq
\calS_{\Delta \Delta_{12}}^{(l_1,l_2)} 
= \pi^{d/2} 
\frac{(\Delta-2)_{l_1+1}}{(\Delta-2+l_2)} 
\frac{\G\big(\Delta-\frac{d}{2}\big)\G\big(\frac{1}{2}( \tilde \Delta+\Delta_{12}+l_1)\big)\G\big(\frac{1}{2}( \tilde\Delta-\Delta_{12}+l_1)\big)}
{\G(\tilde\Delta+l_1)\G\big(\frac{1}{2}(\Delta+\Delta_{12}+l_1)\big)\G\big(\frac{1}{2}(\Delta -\Delta_{12}+l_1)\big)}\,.
\eeq

\subsection{Arbitrary representations}

As a next example we will compute the constant for the representations $\lambda=(l_1,l_2,l_3)$, $\lambda_1 = \bullet$, $\lambda_2 = (l_2, l_3)$.
From the derivation it will also be clear what the result is for arbitrary $\lambda$.
The three-point function is in this case
\bea
&\bra \calO_1(x_1) \calO_2(x_2,\bz_{2}) \calO(x_3,\bz_{3}) \ket_\text{full} \\
={}& \pi_\lambda (\bz_{2}; \bdel_{\bar \bz_{2}})
\pi_\lambda (\bz_{3}; \bdel_{\bar \bz_{3}})
\frac{
\left(k^{(312)}(\bar z_{31})\right)^{l_1} 
\left(m^{(23)}(\bar z_{21}, \bar z_{32})\right)^{l_2} 
\left(m^{(23)}(\bar z_{22}, \bar z_{33})\right)^{l_3}}{
\big( x_{13}^2\big)^{\frac{\Delta_1-\Delta_2+\Delta_3}{2}}
\big( x_{23}^2\big)^{\frac{-\Delta_1+\Delta_2+\Delta_3}{2}}
\big( x_{12}^2\big)^{\frac{ \Delta_1+\Delta_2-\Delta_3}{2}}}\\
={}& 
\pi_\lambda (\bz_{3}; \bdel_{\bar \bz_{3}})
\frac{
\left(k^{(312)}(\bar z_{31})\right)^{l_1} 
\left(m^{(23)}(z_{21}, \bar z_{32})\right)^{l_2} 
\left(m^{(23)}(z_{22}, \bar z_{33})\right)^{l_3}}{
\big( x_{13}^2\big)^{\frac{\Delta_1-\Delta_2+\Delta_3}{2}}
\big( x_{23}^2\big)^{\frac{-\Delta_1+\Delta_2+\Delta_3}{2}}
\big( x_{12}^2\big)^{\frac{\Delta_1+\Delta_2-\Delta_3}{2}}}\\
&+ \calO(z_{2i}\cdot z_{2j}) + \calO\big((z_{21}\cdot z_{32})^{l_2-1}\big) + \calO\big((z_{22}\cdot z_{33})^{l_3-1}\big)\,.
\eea{eq:3point2}
The three-point function of the shadow operator is then
\bea
&\bra \calO_1(x_1) \calO_2(x_2,\bz_{2}) \tilde \calO(x_3,\bz_{3}) \ket_\text{full}\\
={}& 
\int \frac{d^d x_0}{(x_{03}^2)^{d-\Delta}}
\frac{Y_\lambda (\bz_{3}; \bdel_{\bar \bz_{31}}) 
\left( y(\bar z_{31}) \right)^{l_1}
\left( y_2(\bar z_{32},z_{21}) \right)^{l_2}
\left( y_2(\bar z_{33},z_{22}) \right)^{l_3}}{
\big( x_{01}^2\big)^{\frac{\Delta_1-\Delta_2+\Delta}{2}}\big( x_{02}^2\big)^{\frac{-\Delta_1+\Delta_2+\Delta}{2}}\big( x_{12}^2\big)^{\frac{\Delta_1+\Delta_2-\Delta}{2}}}\\
&+ \calO(z_{2i}\cdot z_{2j})+ \calO(z_{3i}\cdot z_{3j}) + \calO\big((z_{21}\cdot z_{32})^{l_2-1}\big) + \calO\big((z_{22}\cdot z_{33})^{l_3-1}\big)\,.
\eea{eq:shadow_3point2}
To compute this expression one needs the following integral, which can be immediately read off from the result \eqref{eq:Inm2}
\bea
&I_{\alpha, \beta, \gamma}^{n,m,o}(x_1,x_2,x_3,z_{31},z_{32},z_{21},z_{33},z_{22}) \\
\equiv{}& \int \frac{d^d x_0 \ \left(k^{(302)} (z_{31})\right)^n \left(k^{(302)} (z_{32}) k^{(203)} (z_{21})\right)^m
\left(k^{(302)} (z_{33}) k^{(203)} (z_{22})\right)^o}
{\big(x_{01}^2\big)^{\alpha} \big(x_{02}^2\big)^{\beta} \big(x_{03}^2\big)^{\gamma}}\\
={}&\pi^{d/2}  \frac{m!o!}{2^{m+o}}
\frac{\Gamma\big(\frac{d}{2}-\alpha+m+o\big) \Gamma\big(\frac{d}{2}-\beta+\frac{n}{2}\big) \Gamma\big(\frac{d}{2}-\gamma+\frac{n}{2}\big)}{\Gamma(\alpha) \Gamma\big(\beta+\frac{n}{2}+m+o\big) \Gamma\big(\gamma+\frac{n}{2}+m+o\big)}\\
&\big(x_{23}^2\big)^{\alpha-\frac{d}{2}} \big(x_{13}^2\big)^{\beta-\frac{d}{2}} \big(x_{12}^2\big)^{\gamma-\frac{d}{2}} 
\left(k^{(312)} (z_{31})\right)^{n} \left(z_{21}\cdot z_{32}\right)^m \left(z_{22}\cdot z_{33}\right)^o\\
&+ \calO(z_{3i}\cdot z_{3j}) + \calO\big((z_{21}\cdot z_{32})^{m-1}\big) + \calO\big((z_{22}\cdot z_{33})^{o-1}\big).
\eea{eq:Inmo}
Next, the computation of \eqref{eq:shadow_computation1} is repeated
\bea
&\bra \calO_1(x_1) \calO_2(x_2,\bz_{2}) \tilde \calO(x_3,\bz_{3}) \ket_\text{full} \\
={}&\sum_{k=0}^{l_1}\sum_{b=0}^{l_1-k}\frac{l_1!}{k!b!(l_1-k-b)!}( -1)^b
\sum_{i=0}^{l_2} \frac{l_2!}{i!(l_2-i)!}(-2)^i
\sum_{j=0}^{l_3} \frac{l_3!}{j!(l_3-j)!}(-2)^j S_{l_1,k+b,i+j} S_{l_2,i,j}\\
&\big( x_{12}^2\big)^{\frac{ -\Delta_1-\Delta_2+\Delta-k-b}{2}}\big( x_{13}^2\big)^{\frac{l_1-k+b}{2}}\big( x_{23}^2\big)^{k-\frac{l_1}{2}} 
\left(k^{(312)} (z_{31}) \right)^{l_1-k-b}
 \left(z_{21} \cdot z_{32}\right)^{l_2-i} \left(z_{22} \cdot z_{33}\right)^{l_3-j}\\
& I^{k+b,i,j}_{\frac{1}{2}(\Delta_{12}+\Delta+l_1-2k),\frac{1}{2}( -\Delta_{12}+\Delta-l_1+k-b),d-\Delta+\frac{k+b}{2}}(x_1,x_2,x_3,z_{31},z_{32},z_{21},z_{33},z_{22})\\
&+ \calO(z_{2i}\cdot z_{2j})+ \calO(z_{3i}\cdot z_{3j}) + \calO\big((z_{21}\cdot z_{32})^{l_2-1}\big) + \calO\big((z_{22}\cdot z_{33})^{l_3-1}\big)\\
={}&\left(k^{(312)} (z_{31}) \right)^{l_1}  \left(z_{21} \cdot z_{32}\right)^{l_2} \left(z_{22} \cdot z_{33}\right)^{l_3} \big( x_{12}^2\big)^{\frac{\tilde \Delta -\Delta_1-\Delta_2}{2}}\big( x_{13}^2\big)^{\frac{\Delta_2 -\Delta_1-\tilde \Delta}{2}} \big( x_{23}^2\big)^{\frac{\Delta_1-\Delta_2-\tilde \Delta}{2}}\\
&\pi^{d/2} \sum_{j=0}^{l_3}\sum_{i=0}^{l_2-j} \sum_{k=0}^{l_1-i-j}\sum_{b=0}^{l_1-k-i-j}   \frac{(l_1-i-j)! (l_2-j)!l_3!}{k!b!(l_1-k-b-i-j)!(l_2-i-j)!(l_3-j)!} ( -1)^{b+i+j}\\
&\frac{\G \big(\frac{1}{2}( d-\Delta_{12}-\Delta-l_1)+k+i+j\big)\G\big(\frac{1}{2}( d+\Delta_{12}-\Delta+l_1)+b\big)\G\big(\Delta-\frac{d}{2}\big)}{\G\big(\frac{1}{2}(\Delta_{12}+\Delta+l_1)-k\big)\G\big(\frac{1}{2}( -\Delta_{12}+\Delta-l_1)+k+i+j\big)\G(d-\Delta+k+b+i+j)}\\
&+ \calO(z_{2i}\cdot z_{2j})+ \calO(z_{3i}\cdot z_{3j}) + \calO\big((z_{21}\cdot z_{32})^{l_2-1}\big) + \calO\big((z_{22}\cdot z_{33})^{l_3-1}\big)\\
={}&\left(k^{(312)} (z_{31}) \right)^{l_1}  \left(z_{21} \cdot z_{32}\right)^{l_2} \left(z_{22} \cdot z_{33}\right)^{l_3}\big( x_{12}^2\big)^{\frac{\tilde \Delta -\Delta_1-\Delta_2}{2}}\big( x_{13}^2\big)^{\frac{\Delta_2 -\Delta_1-\tilde \Delta}{2}} \big( x_{23}^2\big)^{\frac{\Delta_1-\Delta_2-\tilde \Delta}{2}}\\
& \pi^{d/2} \frac{(\Delta-3)_{l_1+2}}{(\Delta-2+l_2)(\Delta-3+l_3)} \frac{\G \big(\Delta-\frac{d}{2}\big)\G\big(\frac{1}{2}( d+\Delta_{12}-\Delta+l_1)\big)\G\big(\frac{1}{2}( d-\Delta_{12}-\Delta+l_1)\big)}{\G(d-\Delta+l_1)\G\big(\frac{1}{2}(\Delta_{12}+\Delta+l_1)\big)\G\big(\frac{1}{2}( -\Delta_{12}+\Delta+l_1)\big)}\\
&+ \calO(z_{2i}\cdot z_{2j})+ \calO(z_{3i}\cdot z_{3j}) + \calO\big((z_{21}\cdot z_{32})^{l_2-1}\big) + \calO\big((z_{22}\cdot z_{33})^{l_3-1}\big)\,.
\eea{eq:shadow_computation2}
The only non-trivial new ingredient here is the factor $S_{l_2,i,j}$. Since the vectors $z_{31}$ are treated as before (with the factor $S_{l_1,k+b,i+j}$), consider only the vectors $z_{32}$ and $z_{33}$. They appear in the form of
\beq
\left(z_{21} \cdot z_{32}\right)^{l_2-i} \left(z_{22} \cdot z_{33}\right)^{l_3-j} 
\left( k^{(302)} (z_{32}) \right)^i \left( k^{(302)} (z_{33}) \right)^j .
\eeq
One can now draw a birdtrack containing only the antisymmetrizations of the $j$ copies of $z_{33}$ that are contracted to $k^{(302)}$.
All of the indices that are antisymmetrized with those cannot be contracted to $k^{(302)}$ as well. Consider for example the case $j=2$,
\beq
 {\mathord{\vcenter{\hbox{\scalebox{0.8}{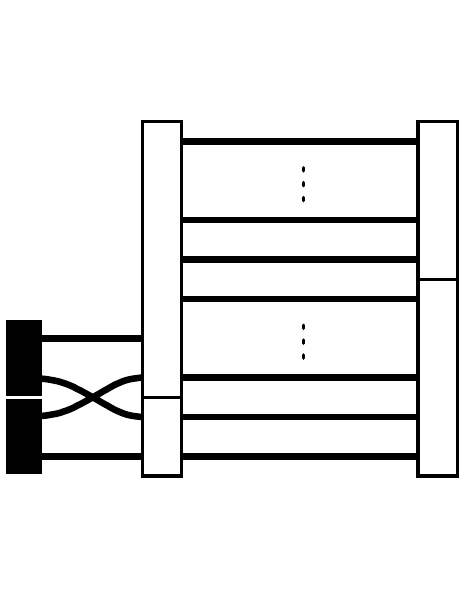}}}}}
\ =\ 
\frac{(l_2-i)_{(j)}}{(l_2)_{(j)}}
 {\mathord{\vcenter{\hbox{\scalebox{0.8}{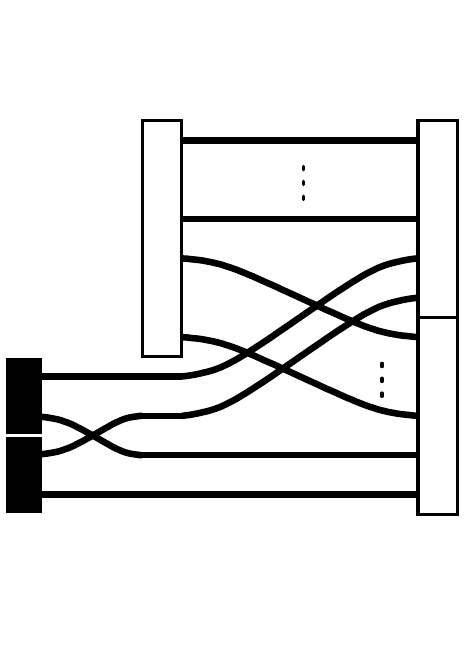}}}}} \quad.
\eeq
Hence the factor $S_{l_2,i,j}=\frac{(l_2-i)_{(j)}}{(l_2)_{(j)}}$ has to be included.

By now it is clear that the computation for Young diagrams with more rows will work out analogously, and that the constant relating a three-point  functions of an operator and the one of its shadow for a Young diagram with row lengths $(l_1,l_2,\ldots)$ and column heights $(h_1,h_2,\ldots)$ is
\begin{align}
\calS_{\Delta \Delta_{12}}^{\lambda} ={}&\pi^{d/2} 
\frac{(\Delta-h_1)_{l_1+h_1-1}}{\prod\limits_{i=2}^{h_1}(\Delta-i+l_i)} 
\frac{\G\big(\Delta-\frac{d}{2}\big)\G\big(\frac{1}{2}( \tilde \Delta+\Delta_{12}+l_1)\big)\G\big(\frac{1}{2}( \tilde\Delta-\Delta_{12}+l_1)\big)}
{\G(\tilde\Delta+l_1)\G\big(\frac{1}{2}(\Delta+\Delta_{12}+l_1)\big)\G\big(\frac{1}{2}(\Delta -\Delta_{12}+l_1)\big)}
\label{eq:shadow_constant}
\\
={}& \pi^{d/2} 
\prod\limits_{i=1}^{l_1} \left( \Delta - h_i + i - 1 \right)
\frac{\G\big(\Delta-\frac{d}{2}\big)\G\big(\frac{1}{2}( \tilde \Delta+\Delta_{12}+l_1)\big)\G\big(\frac{1}{2}( \tilde\Delta-\Delta_{12}+l_1)\big)}
{\G(\tilde\Delta+l_1)\G\big(\frac{1}{2}(\Delta+\Delta_{12}+l_1)\big)\G\big(\frac{1}{2}(\Delta -\Delta_{12}+l_1)\big)}\,.
\nonumber
\end{align}

\subsection{Conformal integrals}
\label{sec:conformal_integrals}
We will use the following conformal integral (for $\alpha + \beta + \gamma = d$)
\bea
I_{\alpha, \beta, \gamma}(x_1,x_2,x_3)
={}&\int \frac{d^d x_0}{(x_{01}^2)^{\alpha} (x_{02}^2)^{\beta} (x_{03}^2)^{\gamma}}\\
={}& \pi^{d/2} \frac{\Gamma(\tfrac{d}{2}-\alpha) \Gamma(\tfrac{d}{2}-\beta) \Gamma(\tfrac{d}{2}-\gamma)}{\Gamma(\alpha) \Gamma(\beta) \Gamma(\gamma)}
(x_{23}^2)^{\alpha-\tfrac{d}{2}} (x_{13}^2)^{\beta-\tfrac{d}{2}} (x_{12}^2)^{\gamma-\tfrac{d}{2}}\,,
\eea{eq:scalar_integral}
to compute the following integral (for $\alpha + \beta + \gamma = d$)
\begin{align}
&I_{\alpha, \beta, \gamma}^{n,m}(x_1,x_2,x_3,z_{31},z_{32},z_2) 
=\sum\limits_{k=0}^n \frac{n!}{k!(n-k)!} (-1)^k \big(x_{23}^2\big)^{\frac{n}{2}-k+m} (x_{23}\cdot z_{31})^k
\label{eq:Inm}
\\
&\qquad\qquad\qquad\qquad\quad
\int \frac{d^d x_0 \ \left(x_{03}\cdot z_{31}\right)^{n-k} \left(x_{03} \cdot z_{32}\right)^m \left(x_{02} \cdot z_2\right)^m}{\big(x_{01}^2\big)^{\alpha} \big(x_{02}^2\big)^{\beta + \frac{n}{2}+m} \big(x_{03}^2\big)^{\gamma + \frac{n}{2}-k+m}}+ \calO\big((z_{2}\cdot z_{32})^{m-1}\big)
\nonumber\\
={}&\sum\limits_{k=0}^n \frac{n!}{k!(n-k)!} (-1)^k \big(x_{23}^2\big)^{\frac{n}{2}-k+m} (x_{23}\cdot z_{31})^k
\frac{\Gamma\big(\gamma-\frac{n}{2}\big)}{2^{n-k+m}\Gamma\big(\gamma+\frac{n}{2}-k+m\big)}
\frac{\Gamma\big(\beta+\frac{n}{2}\big)}{2^m\Gamma\big(\beta+\frac{n}{2}+m\big)}
\nonumber\\
&\left( z_2 \cdot \partial_{x_2} \right)^m
\left( z_{32} \cdot \partial_{x_3} \right)^m
\left( z_{31} \cdot \partial_{x_3} \right)^{n-k}
I_{\alpha, \beta+\frac{n}{2}, \gamma-\frac{n}{2}}(x_1,x_2,x_3)
+ \calO\big((z_{2}\cdot z_{32})^{m-1}\big)\,.
\nonumber
\end{align}
The computation of the derivative 
also simplifies significantly when considering only terms of order $\calO((z_2 \cdot z_{32})^m)$.
It is enough to consider the contribution of $\left( z_2 \cdot \partial_{x_2} \right)^m
\left( z_{32} \cdot \partial_{x_3} \right)^m$ acting on $(x_{23}^2)^{\alpha-h-(n-k-b)}$,
\begin{align}
&\left( z_2 \cdot \partial_{x_2} \right)^m
\left( z_{32} \cdot \partial_{x_3} \right)^m
\left( z_{31} \cdot \partial_{x_3} \right)^{n-k}
I_{\alpha, \beta+\frac{n}{2}, \gamma-\frac{n}{2}}(x_1,x_2,x_3) 
\label{eq:deriv_of_int}
\\
={}& \pi^{d/2} \sum\limits_{b=0}^{n-k} \frac{(n-k)!m!}{b!(n-k-b)!} 2^{n-k+m}
\frac{\Gamma\big(\frac{d}{2}-\alpha+n-k-b+m\big) \Gamma\big(\frac{d}{2}-\beta-\frac{n}{2}+b\big) \Gamma\big(\frac{d}{2}-\gamma+\frac{n}{2}\big)}{\Gamma(\alpha) \Gamma\big(\beta+\frac{n}{2}\big) \Gamma\big(\gamma-\frac{n}{2}\big)}
\nonumber\\
&\big(x_{23}^2\big)^{\alpha-\frac{d}{2}-(n-k-b)-m} \big(x_{13}^2\big)^{\beta-\frac{d}{2}+\frac{n}{2}-b} \big(x_{12}^2\big)^{\gamma-\frac{d}{2}-\frac{n}{2}} 
\left(x_{13}\cdot z_{31}\right)^b \left(x_{23}\cdot z_{31}\right)^{n-k-b} \left(z_2\cdot z_{32}\right)^m
\nonumber\\
&+ \calO(z_{3i}\cdot z_{3j}) + \calO\big((z_{2}\cdot z_{32})^{m-1}\big) \,.
\nonumber
\end{align}
By inserting this in the last expression for $I_{\alpha, \beta, \gamma}^{n,m}$ in (\ref{eq:Inm}) , one finds
\begin{align}
&I_{\alpha, \beta, \gamma}^{n,m}(x_1,x_2,x_3,z_{31},z_{32},z_2)
\nonumber \\
={}&\pi^{d/2} \sum\limits_{k=0}^n \sum\limits_{b=0}^{n-k} \frac{n!m!(-1)^k}{k!b!(n-k-b)!2^m} 
\frac{\Gamma\big(\frac{d}{2}-\alpha+n-k-b+m\big) \Gamma\big(\frac{d}{2}-\beta-\frac{n}{2}+b\big) \Gamma\big(\frac{d}{2}-\gamma+\frac{n}{2}\big)}{\Gamma(\alpha) \Gamma(\beta+\frac{n}{2}+m) \Gamma(\gamma+\frac{n}{2}-k+m)}
\nonumber\\
&\big(x_{23}^2\big)^{\alpha-\frac{d}{2}-\frac{n}{2}+b} \big(x_{13}^2\big)^{\beta-\frac{d}{2}+\frac{n}{2}-b} \big(x_{12}^2\big)^{\gamma-\frac{d}{2}-\frac{n}{2}} 
\left(x_{13}\cdot z_{31}\right)^b \left(x_{23}\cdot z_{31}\right)^{n-b} \left(z_2\cdot z_{32}\right)^m
\nonumber\\
&+ \calO(z_{3i}\cdot z_{3j}) + \calO\big((z_{2}\cdot z_{32})^{m-1}\big)
\nonumber\\
={}&\pi^{d/2} \sum\limits_{b=0}^{n} \frac{n!m!}{b!(n-b)!} \frac{(-1)^{n-b}}{2^m}
\frac{\Gamma\big(\frac{d}{2}-\alpha+m\big) \Gamma\big(\frac{d}{2}-\beta+\frac{n}{2}\big) \Gamma\big(\frac{d}{2}-\gamma+\frac{n}{2}\big)}{\Gamma(\alpha) \Gamma\big(\beta+\frac{n}{2}+m\big) \Gamma\big(\gamma+\frac{n}{2}+m\big)}
\label{eq:Inm2}
\\
&\big(x_{23}^2\big)^{\alpha-\frac{d}{2}-\frac{n}{2}+b} \big(x_{13}^2\big)^{\beta-\frac{d}{2}+\frac{n}{2}-b} \big(x_{12}^2\big)^{\gamma-\frac{d}{2}-\frac{n}{2}} 
\left(x_{13}\cdot z_{31}\right)^b \left(x_{23}\cdot z_{31}\right)^{n-b} \left(z_2\cdot z_{32}\right)^m
\nonumber\\
&
+ \calO(z_{3i}\cdot z_{3j}) + \calO\big((z_{2}\cdot z_{32})^{m-1}\big)
\nonumber\\
={}&\pi^{d/2}  \frac{m!}{2^m}
\frac{\Gamma\big(\frac{d}{2}-\alpha+m\big) \Gamma\big(\frac{d}{2}-\beta+\frac{n}{2}\big) \Gamma\big(\frac{d}{2}-\gamma+\frac{n}{2}\big)}{\Gamma(\alpha) \Gamma\big(\beta+\frac{n}{2}+m\big) \Gamma\big(\gamma+\frac{n}{2}+m\big)}
\nonumber\\
&\big(x_{23}^2\big)^{\alpha-\frac{d}{2}} \big(x_{13}^2\big)^{\beta-\frac{d}{2}} \big(x_{12}^2\big)^{\gamma-\frac{d}{2}} 
\left(k^{(312)} (z_{31})\right)^{n} \left(z_2\cdot z_{32}\right)^m 
+ \calO(z_{3i}\cdot z_{3j}) + \calO\big((z_{2}\cdot z_{32})^{m-1}\big)\,.
\nonumber
\end{align}
The sum over $k$ was evaluated using
\beq
\sum\limits_{k=0}^{N} \frac{N!}{k!(N-k)!}(-1)^k \frac{\G(x-k)}{\G(y-k)}
=
(-1)^N
\frac{\G(x-N)\G(y-x+N)}{\G(y) \G(y-x)}\,.
\eeq

\section{OPE limit of conformal blocks in the shadow formalism}
\label{sec:ope_limit}
In order for the conformal blocks to satisfy the recursion relation derived above, it is crucial that they have the correct normalization.
To compare to other results it is a good idea to consider the OPE limit $x_{12}^a \rightarrow 0, x_{34}^a \rightarrow 0$.
This can be done in physical space by generalizing a trick from \cite{Dolan:2011dv}.
To this end let us work again in physical space, as in the previous appendix.
The shadow operator of an operator $\calO$ in the irrep $(\Delta,\lambda)$ is given by
\beq
\tilde \calO (x_0, \bz_0) =
\frac{\pi_{\lambda}(\bz_0, \bdel_{\bz_4})}{l_1!l_2! \ldots l_{h_1}!}
\int\frac{d^d x_4}{(x_{04}^2)^{d-\Delta}}
\left(
m^{(04)}(z_{41},\partial_{\bar z_{41}})
\right)^{l_1}
\ldots
\left(
m^{(04)}(z_{4 h_1},\partial_{\bar z_{4 h_1}})
\right)^{l_{h_1}}
\calO (x_4, \bar \bz_4) \,.
\label{eq:shadow_phys}
\eeq
For a lighter notation we will choose for the remainder of this appendix to consider Young diagrams with at most three rows $\lambda = (l_1,l_2,l_3)$.
Furthermore, we will compute the OPE limit for conformal blocks with $\lambda_1 = \lambda_3 = \bullet$ and $\lambda_2 = \lambda_4 = (l_2,l_3)$.
Other configurations can be treated analogously by replacing some of the vectors $z_{2i}$ by  $z_{1j}$
or $z_{4i}$ by  $z_{3j}$.

We want to compute the OPE limit of the conformal partial wave
\bea
W_{\Delta,\lambda}^{\lambda_1 \lambda_2 \lambda_3 \lambda_4} =
\frac{1}{\calS_{\Delta \Delta_{34}}^{\lambda}
\calS_{\tilde \Delta \Delta_{34}}^{\lambda}}
 \int \!d^d x_0
\bra \calO_1(x_1) \calO_2(x_2,\bz_{2}) \calO(x_0,\bdel_{\bz_{0}}) \ket
\bra \calO_3(x_3) \calO_4(x_4,\bz_{4}) \tilde \calO(x_0,\bz_{0}) \ket\,,
\eea{eq:partial_wave_phys}
where the three-point functions are given by\footnote{We are omitting the projector $\pi_{\lambda_2}$
which has no impact on this computation.}
\bea
\bra \calO_1(x_1) \calO_2(x_2,\bz_{2}) \calO(x_0,\bz_{0}) \ket
\!=\! \pi_\lambda (\bz_{0},\bdel_{\bar\bz_{0}}) 
\frac{
\left(k^{(012)}(\barz_{01})\right)^{l_1} \hspace{-1pt}
\left(m^{(20)}(z_{21}, \barz_{02})\right)^{l_2}   \hspace{-1pt}
\left(m^{(20)}(z_{22},\barz_{03})\right)^{l_3}}{
\big( x_{12}^2\big)^{\frac{\Delta_1+\Delta_2-\Delta}{2}}
\big( x_{01}^2\big)^{\frac{\Delta+\Delta_1-\Delta_2}{2}}
\big( x_{02}^2\big)^{\frac{\Delta+\Delta_2-\Delta_1}{2}}}\,.
\eea{eq:3point2.2}
To perform the OPE limit one can use
\beq
k^{(012)}(z_{01}) \underset{x_{12}\to 0}{\sim}
\frac{1}{\sqrt{x_{12}^2}} \,m^{(20)} (x_{12},z_{01})\,,
\eeq
and hence
\bea
&\bra \calO_1(x_1) \calO_2(x_2,\bz_{2}) \calO(x_0,\bz_{0}) \ket
\underset{x_{12}\to 0}{\sim}\\
&\qquad \qquad\pi_\lambda (\bz_{0},\bdel_{\bar\bz_{0}}) 
\frac{
\left(m^{(20)} (x_{12},\barz_{01})\right)^{l_1} 
\left(m^{(20)}(z_{21}, \barz_{02})\right)^{l_2} 
\left(m^{(20)}(z_{22}, \barz_{03})\right)^{l_3}}{
\big( x_{12}^2\big)^{\frac{\Delta_1+\Delta_2-\Delta+l_1}{2}}
\big( x_{02}^2\big)^{\Delta}}\,.
\eea{eq:3point2.2_limit}
Inserting this into \eqref{eq:partial_wave_phys}, and using the definition of the shadow operator \eqref{eq:shadow_phys},
one finds
\begin{align}
W_{\Delta,\lambda}^{\lambda_1 \lambda_2 \lambda_3 \lambda_4}
&\underset{x_{12}\to 0}{\sim}
\frac{1}{\calS_{\Delta \Delta_{34}}^{\lambda}
\calS_{\tilde \Delta \Delta_{34}}^{\lambda}}
\big( x_{12}^2\big)^{\frac{\Delta-l_1-\Delta_1-\Delta_2}{2}}
\bra \calO_3(x_3) \calO_4(x_4,\bz_{4}) \tilde {\tilde \calO}(x_2, x_{12}, z_{21},  z_{22}) \ket\nonumber\\
&=
\big( x_{12}^2\big)^{\frac{ \Delta-l_1-\Delta_1-\Delta_2}{2}}
\bra \calO_3(x_3) \calO_4(x_4,\bz_{4})  \calO(x_2, x_{12}, z_{21},  z_{22}) \ket\,.
\label{eq:wave_x12_limit}
\end{align}
Doing the limit $x_{34}^a\to 0$ in a similar way leads to
\bea
W_{\Delta,\lambda}^{\lambda_1 \lambda_2 \lambda_3 \lambda_4}
\!
\underset{\substack{x_{12}\to 0 \\  x_{34}\to 0}}{\sim}
\!
\pi_\lambda (x_{12},z_{21},z_{22},\bdel_{\bz_{0}}) 
\frac{
\left(m^{(24)} (z_{01},x_{34})\right)^{l_1} 
\left(m^{(24)}(z_{02}, z_{41})\right)^{l_2} 
\left(m^{(24)}(z_{03}, z_{42})\right)^{l_3}}{
( x_{24}^2)^{\Delta}
( x_{12}^2)^{\frac{1}{2}( \Delta_1+\Delta_2-\Delta+l_1)}
( x_{34}^2)^{\frac{1}{2}( \Delta_3+\Delta_4-\Delta+l_1)}
}\,.
\eea{eq:wave_x12_x34_limit}
In order to extract the normalization it is convenient to consider only the term which is of leading order in the building blocks 
$m^{(24)} (x_{12},x_{34})$, $m^{(24)} (z_{21},x_{41})$ and $m^{(24)} (z_{22},x_{42})$.
This term has a prefactor that can be found by considering the birdtrack diagram in \eqref{eq:young_norm_bt}
and removing all antisymmetrizations, resulting in a factor of
\beq
\frac{n_\lambda}{\prod\limits_{i=1}^{l_1}h_i!}\,.
\eeq
Finally one can use that
\beq
1 - v \underset{\substack{x_{12}\to 0 \\  x_{34}\to 0}}{\sim}
\frac{2 m^{(24)} (x_{12},x_{34})}{x_{24}^2}\,,
\eeq
and find
\begin{align}
W_{\Delta,\lambda}^{\lambda_1 \lambda_2 \lambda_3 \lambda_4}
&\underset{\substack{x_{12}\to 0 \\  x_{34}\to 0}}{\sim}
\frac{n_\lambda}{\prod\limits_{i=1}^{l_1}h_i!}
2^{-l_1}
u^{\frac{1}{2}(\Delta-l_1)}
(1-v)^{l_1}
\frac{
\left(m^{(24)}(z_{21}, z_{41})\right)^{l_2} 
\left(m^{(24)}(z_{22}, z_{42})\right)^{l_3}}{
\big( x_{12}^2\big)^{\frac{\Delta_1+\Delta_2}{2}}
\big( x_{34}^2\big)^{\frac{ \Delta_3+\Delta_4}{2}}
}\nonumber\\
&\qquad+ \calO \left( \left(\frac{1-v}{\sqrt{u}}\right)^{l_1-1} , \left(m^{(24)}(z_{21}, z_{41})\right)^{l_2-1},  \left(m^{(24)}(z_{22}, z_{42})\right)^{l_3-1} \right)\,.
\label{eq:wave_x12_x34_limit_uv}
\end{align}

\section{Spherical tensor harmonics}
\label{sec:spherical_harmonics}

In this appendix the relation between projectors to $SO(d)$ irreps and spherical tensor harmonics on the sphere $S^{d-1}$
is explained. In the case of projectors to traceless symmetric tensors this relation is just the fact that the projectors are
encoded by Gegenbauer polynomials \eqref{eq:symmetric_pi_as_gegenbauer}, which are scalar spherical harmonics.
In the radial coordinates of \cite{Hogervorst:2013sma}, the conformal blocks are naturally written in terms of 
spherical tensor harmonics \cite{Costa:2016other}.

Consider a tensor field on $S^{d-1} \in \mathbb{R}^d$. We shall work in the embedding space $\mathbb{R}^d$ and impose transversality
\beq
x^{a_i} t_{a_1 \dots a_k}(x) =0 \, ,
\label{transversetensor}
\eeq
for all $i=1,\dots,k$ and $x_a x^a=1$.
As shown in subsection \ref{sec:covariant_derivatives} below, covariant derivatives on the sphere are just partial derivatives $\partial_a=\frac{\partial}{\partial x^a}$ on $\mathbb{R}^d$ projected to the sphere (over all indices of the resulting tensor).
Therefore, the laplacian on the sphere can be written as
\beq
\nabla^2  t_{a_1 \dots a_k}=
P^{ab}\partial_a \left(
P_b^c P_{b_1}^{c_1}\dots 
P_{b_k}^{c_k} \partial_c 
t_{c_1 \dots c_k}
\right)
P^{b_1}_{a_1}\dots 
P^{b_k}_{a_k}\, ,
\eeq
where $P_{ab}=\delta_{ab}-x_a x_b$ is a projector onto the unit sphere.
Using (\ref{transversetensor}) on the sphere $x_a x^a=1$, one can simplify this expression to
\beq
\nabla^2  t_{a_1 \dots a_k}=
P^{b_1}_{a_1}\dots 
P^{b_k}_{a_k}
 \left(\partial_a \partial^a -x^a x^b \partial_a \partial_b-(d-1) x^a\partial_a +k
\right)t_{b_1 \dots b_k}\, .
\label{spherelaplacian}
\eeq

Another interesting differential operator to consider is the quadratic Casimir of the symmetry group $SO(d)$, generated by
\beq
J_{ab}=i\left( x_a \partial_b -x_b \partial_a\right) +\Sigma_{ab}\, ,
\eeq
where $\Sigma_{ab}$ rotates the indices of a tensor. More precisely,
\beq
\label{actionSigma}
\Sigma_{ab}t_{a_1 \dots a_k}=
\sum_{i=1}^k \left[ M_{ab}\right]^c_{a_i}
t_{a_1 \dots a_{i-1} c  a_{i+1} \dots a_k}\, ,
\eeq
with $\left[ M^{ab}\right]^c_{e}=i
\left( \delta^a_e \delta^{bc}-
 \delta^b_e \delta^{ac}\right)$.
The quadratic Casimir is then given by
\beq
\frac{1}{2}J_{ab}J^{ab}=
-x^a \partial^b\left( x_a \partial_b -x_b \partial_a\right)
+2ix_a \partial_b \Sigma^{ab}
+ \frac{1}{2}\Sigma_{ab}\Sigma^{ab}\, .
\eeq
Acting on a tensor obeying (\ref{transversetensor}) on the sphere $x_a x^a=1$, it gives 
\beq
\label{CasimirOnTensor}
\frac{1}{2}J_{ab}J^{ab}t_{a_1 \dots a_k}=
P^{b_1}_{a_1}\dots 
P^{b_k}_{a_k}
\left[ -\nabla^2-k
+ \frac{1}{2}\Sigma_{ab}\Sigma^{ab}
\right]t_{b_1 \dots b_k}
\, ,
\eeq
where we used expression (\ref{spherelaplacian}) for the laplacian on the sphere.

Now consider a tensor field defined by the following contraction with a traceless mixed-symmetry tensor
\beq
t_{(b_1 \dots b_{l_2})(c_1 \dots c_{l_3})\dots} = x^{e_1}\dots x^{e_{l_1}}
f_{(e_1 \dots e_{l_1})(b_1 \dots b_{l_2})
(c_1 \dots c_{l_3})\dots}\, .
\eeq
Notice that tracelessness and property (\ref{eq:symmetrization_ex}) of $f$ implies transversality 
 (\ref{transversetensor}) and 
\beq
\nabla^{a_i} t_{a_1 \dots a_k}=0\, ,
\eeq
where $k=\sum_{i=2} l_i$ and $a_i$ can be any of the indices of the tensor.
Using formula (\ref{spherelaplacian}) it is easy to obtain
\beq
\nabla^2 t_{a_1 \dots a_k}=
-\big[l_1(l_1+d-2)-k \big]t_{a_1 \dots a_k}\, .
\eeq
Moreover, 
\beq
\frac{1}{2}\Sigma_{ab}\Sigma^{ab}
t_{(b_1 \dots b_{l_2})(c_1 \dots c_{l_3})\dots}=
\left[ \sum_{i=1}^{h_1-1} l_{i+1}(l_{i+1}+d-2i)\right]
t_{(b_1 \dots b_{l_2})(c_1 \dots c_{l_3})\dots}\, ,
\eeq
because $\Sigma_{ab}$ only rotates the indices, and therefore $\frac{1}{2}\Sigma_{ab}\Sigma^{ab}$ just measures the Casimir of the irreducible tensor. This statement can also be checked explicitly using the definition (\ref{actionSigma}).
One can also check that (\ref{CasimirOnTensor}) leads to the expected result
\beq
\frac{1}{2}J_{ab}J^{ab}
t_{(b_1 \dots b_{l_2})(c_1 \dots c_{l_3})\dots}=
\left[ \sum_{i=1}^{h_1} l_{i}(l_{i}+d-2i)\right]
t_{(b_1 \dots b_{l_2})(c_1 \dots c_{l_3})\dots}\, .
\eeq

Using the projectors of the previous sections, we can construct the following function
\beq
\Omega_{\lambda}^{b_1 \ldots  , b'_1 \ldots  }(x,y) =
c_\lambda\,x_{a_1}\dots x_{a_{l_1}}
\pi_{\lambda}^{(a_1 \ldots a_{l_1}) (b_1 \ldots b_{l_2})\dots, (a'_1 \ldots a'_{l_1}) (b'_1 \ldots b'_{l_2})\dots}
y_{a'_1}\dots y_{a'_{l_1}}\, .
\label{Omegal}
\eeq
The arguments above show that this function is a tensor harmonic at the point $x$ with indices $b_1\dots$. It is also a tensor harmonic at point $y$ with indices $b_1'\dots$.
We shall fix the normalization constant $c_\lambda$ by imposing the following orthogonality relation
\beq
\int_{S^{d-1}} dy\,
\Omega_{\lambda}^{a_1 \ldots  , b_1 \ldots  }(x,y)
\delta_{b_1b'_1}\dots
\Omega_{\lambda'}^{b'_1 \ldots  , c_1 \ldots  }(y,z)
=\delta_{\lambda,\lambda'}\,\Omega_{\lambda}^{a_1 \ldots  , c_1 \ldots  }(x,z)\, .
\label{OmegaOrtho}
\eeq
Using the definition (\ref{Omegal}) the only integral that we need to compute is  
\beq
I^{a_1\dots a_k}=\int_{S^{d-1}} dy\,y^{a_1}\dots y^{a_k}\, .
\eeq  
Rotational and permutation symmetry imply that $I^{a_1\dots a_k}=0$ for odd $k$ and 
\beq
I^{a_1\dots a_{2k}}=q_k\,\delta^{(a_1a_2}\dots 
\delta^{a_{2k-1}a_{2k})}\, ,
\eeq
for some constant $q_k$.
This constant can be determined by computing the full contraction\footnote{The full contraction of $k$ Kronecker-deltas with the total index symmetrization of another set of $k$ Kronecker-deltas can be determined recursively.}
\bea
\delta_{a_1a_2}\dots 
\delta_{a_{2k-1}a_{2k}}I^{a_1\dots a_{2k}}
=q_k 4^k k! \left(\frac{d}{2}\right)_k \hspace{-0.6em} =\text{Vol}(S^{d-1})\,.
\eea{eq:full_contraction}
This is sufficient to verify the orthogonality relation (\ref{OmegaOrtho}) and to determine the normalization constant
\beq
c_\lambda=\frac{1}{l_1!\,q_{l_1}}=\frac{4^{l_1}\left(\frac{d}{2}\right)_{l_1}}{{\rm Vol}(S^{d-1})}\, .
\eeq

Let us now consider again the orthogonality relation (\ref{OmegaOrtho}) in the case of scalar harmonics and let us sum over $l$,
\beq
\int_{S^{d-1}} dy\,
\Omega_{(l')} (x,y) \sum_{l=0}^\infty
\Omega_{(l)} (y,z)
= \Omega_{(l')} (x,z)\, . 
\eeq
This suggests the following completeness relation
\beq
 \sum_{l=0}^\infty
\Omega_{(l)} (y,z)
= \delta (y,z)\, ,
\label{scalarcomplet}
\eeq
where the delta-function is defined with respect to the sphere metric.
This reasoning can be generalized to the case of tensor harmonics. We  illustrate this method in the case of vector harmonics, corresponding to $\lambda=(l_1,1)$.
Summing the orthogonality relation (\ref{OmegaOrtho})  over $l_1$ we find
\beq
\int_{S^{d-1}} dy\,
\Omega_{(l_1',1)}^{a  , b'   }(x,y)
\delta_{b' b } 
\sum_{l_1=1}^\infty 
\Omega_{(l_1,1)}^{b  , c   }(y,z)
= \Omega_{(l_1',1)}^{a  , c  }(x,z)\, ,
\label{eq:OrthoToComple}
\eeq
which leads to
\beq
 \sum_{l_1=1}^\infty 
\Omega_{(l_1,1)}^{b  , c   }(y,z)+
\nabla^b_{y} \nabla^c_{z} Q(y,z)
= \delta^{bc}\delta(x,z)\, ,
\eeq
where $Q(y,z)$ can be any function of $y\cdot z$ because its contribution to (\ref{eq:OrthoToComple}) vanishes by integration by parts. To determine this function we act with $\delta_{b' b } \nabla^{b'}_{y}$ on the last equation,
\beq
 \nabla^c_{z} \nabla^2_{y}  Q(y,z)
=  \nabla^{c}_{y}\delta(x,z)=  -\nabla^{c}_{z}\delta(x,z)\, .
\eeq
Comparing with (\ref{scalarcomplet}) and using the laplacian eigenvalues of the scalar harmonics we conclude that
\beq
 Q(y,z) =  \sum_{l=0}^\infty
 \frac{1}{l(l+d-2)}\,
\Omega_{(l)} (y,z)\, .
\eeq
The discussion here is very similar to the discussion of harmonic functions in AdS in \cite{Costa:2014kfa}.

\subsection{Covariant derivatives}
\label{sec:covariant_derivatives}
We consider a tensor $t_{a_1\dots a_k}$ obeying (\ref{transversetensor}).
Let us denote by $y^\alpha$ a set of $(d-1)$-coordinates parametrizing the unit sphere $S^{d-1}$. In these coordinates, the tensor is given by
\beq
\tilde{t}_{\alpha_1 \dots \alpha_k}=
\frac{\partial x^{a_1}}{\partial y^{\alpha_1}}
\dots
\frac{\partial x^{a_k}}{\partial y^{\alpha_k}}
 t_{a_1\dots a_k}\, ,
\eeq
and its covariant derivative is
\begin{align}
\nabla_\beta \tilde{t}_{\alpha_1 \dots \alpha_k}&=
\frac{\partial }{\partial y^{\beta}}
\tilde{t}_{\alpha_1 \dots \alpha_k}
-\sum_{i=1}^k \Gamma^\gamma_{\beta\alpha_i}
\tilde{t}_{\alpha_1 \dots \alpha_{i-1}\gamma
\alpha_{i+1}
\dots \alpha_k}
\nonumber\\
&=
\frac{\partial x^{b}}{\partial y^{\beta}}
\frac{\partial x^{a_1}}{\partial y^{\alpha_1}}
\dots
\frac{\partial x^{a_k}}{\partial y^{\alpha_k}}
 \frac{\partial }{\partial x^{b}}t_{a_1\dots a_k}
  \label{termthatvanishes}
\\&+\sum_{i=1}^k 
 \left(
 \frac{\partial^2 x^{a_i}}{\partial y^{\beta} \partial y^{\alpha_i}}
 -\Gamma^\gamma_{\beta\alpha_i}\frac{\partial x^{a_i}}{\partial y^{\gamma}}
\right)
\frac{\partial x^{a_1}}{\partial y^{\alpha_1}}
\dots
\frac{\partial x^{a_{i-1}}}{\partial y^{\alpha_{i-1}}}
\frac{\partial x^{a_{i+1}}}{\partial y^{\alpha_{i+1}}}
\dots
\frac{\partial x^{a_k}}{\partial y^{\alpha_k}}
 t_{a_1\dots   a_k}\, .
 \nonumber
\end{align}
We will now show that the last line vanishes and therefore covariant derivatives on the sphere are equal to partial derivatives on $\mathbb{R}^d$ projected to the sphere. 
It is sufficient to show that
\beq
 \frac{\partial^2 x^{a}}{\partial y^{\beta} \partial y^{\alpha}}
 -\Gamma^\gamma_{\beta\alpha}\frac{\partial x^{a}}{\partial y^{\gamma}} \propto x^a\, ,
 \label{eq:christophelidentity1}
\eeq
because the last line in (\ref{termthatvanishes}) vanishes due to the transversality condition (\ref{transversetensor}).
Equation (\ref{eq:christophelidentity1}) is equivalent to
\beq
 \frac{\partial^2 x^{a}}{\partial y^{\beta} \partial y^{\alpha}}
  \frac{\partial x_{a}}{\partial y^{\mu}}
 -\Gamma^\gamma_{\beta\alpha}
 \frac{\partial x^{a}}{\partial y^{\gamma}} \frac{\partial x_{a}}{\partial y^{\mu}}=0\, ,
\eeq
which can be easily verified using the sphere metric
\beq
g_{\alpha\beta}= \frac{\partial x^{a}}{\partial y^{\alpha}} \frac{\partial x_{a}}{\partial y^{\beta}}\, ,
\eeq
and the expression for the Levi-Civita connection
\beq
\Gamma^\gamma_{\beta\alpha}=\frac{1}{2}
g^{\gamma \mu}\left(
g_{\mu\beta,\alpha}+
g_{\mu\alpha,\beta}-
g_{\beta\alpha,\mu}
\right) .
\eeq

\bibliographystyle{JHEP}

\bibliography{projectors}

\providecommand{\href}[2]{#2}\begingroup\raggedright\begin{thebibliography}{10}

\bibitem{Ferrara:1973yt}
S.~Ferrara, A.~Grillo and R.~Gatto, \emph{{Tensor representations of conformal
  algebra and conformally covariant operator product expansion}},
  \href{http://dx.doi.org/10.1016/0003-4916(73)90446-6}{\emph{Annals Phys.}
  {\bf 76} (1973) 161--188}.

\bibitem{Polyakov:1974gs}
A.~Polyakov, \emph{{Nonhamiltonian approach to conformal quantum field
  theory}}, {\emph{Zh.Eksp.Teor.Fiz.} {\bf 66} (1974) 23--42}.

\bibitem{arXiv:0807.0004}
R.~Rattazzi, V.~S. Rychkov, E.~Tonni and A.~Vichi, \emph{{Bounding scalar
  operator dimensions in 4D CFT}},
  \href{http://dx.doi.org/10.1088/1126-6708/2008/12/031}{\emph{JHEP} {\bf 0812}
  (2008) 031}, [\href{http://arxiv.org/abs/0807.0004}{{\tt 0807.0004}}].

\bibitem{arXiv:0905.2211}
V.~S. Rychkov and A.~Vichi, \emph{{Universal Constraints on Conformal Operator
  Dimensions}}, \href{http://dx.doi.org/10.1103/PhysRevD.80.045006}{\emph{Phys.
  Rev.} {\bf D80} (2009) 045006}, [\href{http://arxiv.org/abs/0905.2211}{{\tt
  0905.2211}}].

\bibitem{arXiv:1009.2087}
D.~Poland and D.~Simmons-Duffin, \emph{{Bounds on 4D Conformal and
  Superconformal Field Theories}},
  \href{http://dx.doi.org/10.1007/JHEP05(2011)017}{\emph{JHEP} {\bf 05} (2011)
  017}, [\href{http://arxiv.org/abs/1009.2087}{{\tt 1009.2087}}].

\bibitem{arXiv:1009.2725}
R.~Rattazzi, S.~Rychkov and A.~Vichi, \emph{{Central Charge Bounds in 4D
  Conformal Field Theory}},
  \href{http://dx.doi.org/10.1103/PhysRevD.83.046011}{\emph{Phys. Rev.} {\bf
  D83} (2011) 046011}, [\href{http://arxiv.org/abs/1009.2725}{{\tt
  1009.2725}}].

\bibitem{arXiv:1009.5985}
R.~Rattazzi, S.~Rychkov and A.~Vichi, \emph{{Bounds in 4D Conformal Field
  Theories with Global Symmetry}},
  \href{http://dx.doi.org/10.1088/1751-8113/44/3/035402}{\emph{J. Phys.} {\bf
  A44} (2011) 035402}, [\href{http://arxiv.org/abs/1009.5985}{{\tt
  1009.5985}}].

\bibitem{arXiv:1109.5176}
D.~Poland, D.~Simmons-Duffin and A.~Vichi, \emph{{Carving Out the Space of 4D
  CFTs}}, \href{http://dx.doi.org/10.1007/JHEP05(2012)110}{\emph{JHEP} {\bf 05}
  (2012) 110}, [\href{http://arxiv.org/abs/1109.5176}{{\tt 1109.5176}}].

\bibitem{arXiv:1203.6064}
S.~El-Showk, M.~F. Paulos, D.~Poland, S.~Rychkov, D.~Simmons-Duffin and
  A.~Vichi, \emph{{Solving the 3D Ising Model with the Conformal Bootstrap}},
  \href{http://dx.doi.org/10.1103/PhysRevD.86.025022}{\emph{Phys. Rev.} {\bf
  D86} (2012) 025022}, [\href{http://arxiv.org/abs/1203.6064}{{\tt
  1203.6064}}].

\bibitem{arXiv:1210.4258}
P.~Liendo, L.~Rastelli and B.~C. van Rees, \emph{{The Bootstrap Program for
  Boundary CFT$_d$}},
  \href{http://dx.doi.org/10.1007/JHEP07(2013)113}{\emph{JHEP} {\bf 07} (2013)
  113}, [\href{http://arxiv.org/abs/1210.4258}{{\tt 1210.4258}}].

\bibitem{arXiv:1304.1803}
C.~Beem, L.~Rastelli and B.~C. van Rees, \emph{{The $\mathcal N=4$
  Superconformal Bootstrap}},
  \href{http://dx.doi.org/10.1103/PhysRevLett.111.071601}{\emph{Phys. Rev.
  Lett.} {\bf 111} (2013) 071601}, [\href{http://arxiv.org/abs/1304.1803}{{\tt
  1304.1803}}].

\bibitem{arXiv:1307.3111}
F.~Gliozzi, \emph{{More constraining conformal bootstrap}},
  \href{http://dx.doi.org/10.1103/PhysRevLett.111.161602}{\emph{Phys. Rev.
  Lett.} {\bf 111} (2013) 161602}, [\href{http://arxiv.org/abs/1307.3111}{{\tt
  1307.3111}}].

\bibitem{arXiv:1307.6856}
F.~Kos, D.~Poland and D.~Simmons-Duffin, \emph{{Bootstrapping the $O(N)$ vector
  models}}, \href{http://dx.doi.org/10.1007/JHEP06(2014)091}{\emph{JHEP} {\bf
  06} (2014) 091}, [\href{http://arxiv.org/abs/1307.6856}{{\tt 1307.6856}}].

\bibitem{arXiv:1309.5089}
S.~El-Showk, M.~Paulos, D.~Poland, S.~Rychkov, D.~Simmons-Duffin and A.~Vichi,
  \emph{{Conformal Field Theories in Fractional Dimensions}},
  \href{http://dx.doi.org/10.1103/PhysRevLett.112.141601}{\emph{Phys. Rev.
  Lett.} {\bf 112} (2014) 141601}, [\href{http://arxiv.org/abs/1309.5089}{{\tt
  1309.5089}}].

\bibitem{arXiv:1403.4545}
S.~El-Showk, M.~F. Paulos, D.~Poland, S.~Rychkov, D.~Simmons-Duffin and
  A.~Vichi, \emph{{Solving the 3d Ising Model with the Conformal Bootstrap II.
  c-Minimization and Precise Critical Exponents}},
  \href{http://dx.doi.org/10.1007/s10955-014-1042-7}{\emph{J. Stat. Phys.} {\bf
  157} (2014) 869}, [\href{http://arxiv.org/abs/1403.4545}{{\tt 1403.4545}}].

\bibitem{arXiv:1404.0489}
Y.~Nakayama and T.~Ohtsuki, \emph{{Approaching the conformal window of
  $O(n)\times O(m)$ symmetric Landau-Ginzburg models using the conformal
  bootstrap}}, \href{http://dx.doi.org/10.1103/PhysRevD.89.126009}{\emph{Phys.
  Rev.} {\bf D89} (2014) 126009}, [\href{http://arxiv.org/abs/1404.0489}{{\tt
  1404.0489}}].

\bibitem{arXiv:1406.4814}
S.~M. Chester, J.~Lee, S.~S. Pufu and R.~Yacoby, \emph{{The $ \mathcal{N}=8 $
  superconformal bootstrap in three dimensions}},
  \href{http://dx.doi.org/10.1007/JHEP09(2014)143}{\emph{JHEP} {\bf 09} (2014)
  143}, [\href{http://arxiv.org/abs/1406.4814}{{\tt 1406.4814}}].

\bibitem{arXiv:1406.4858}
F.~Kos, D.~Poland and D.~Simmons-Duffin, \emph{{Bootstrapping Mixed Correlators
  in the 3D Ising Model}},
  \href{http://dx.doi.org/10.1007/JHEP11(2014)109}{\emph{JHEP} {\bf 11} (2014)
  109}, [\href{http://arxiv.org/abs/1406.4858}{{\tt 1406.4858}}].

\bibitem{arXiv:1412.4127}
M.~F. Paulos, \emph{{JuliBootS: a hands-on guide to the conformal bootstrap}},
  \href{http://arxiv.org/abs/1412.4127}{{\tt 1412.4127}}.

\bibitem{arXiv:1412.7541}
C.~Beem, M.~Lemos, P.~Liendo, L.~Rastelli and B.~C. van Rees, \emph{{The
  ${\mathcal N}=2$ superconformal bootstrap}},
  \href{http://arxiv.org/abs/1412.7541}{{\tt 1412.7541}}.

\bibitem{arXiv:1502.02033}
D.~Simmons-Duffin, \emph{{A Semidefinite Program Solver for the Conformal
  Bootstrap}}, \href{http://dx.doi.org/10.1007/JHEP06(2015)174}{\emph{JHEP}
  {\bf 06} (2015) 174}, [\href{http://arxiv.org/abs/1502.02033}{{\tt
  1502.02033}}].

\bibitem{arXiv:1502.07217}
F.~Gliozzi, P.~Liendo, M.~Meineri and A.~Rago, \emph{{Boundary and Interface
  CFTs from the Conformal Bootstrap}},
  \href{http://dx.doi.org/10.1007/JHEP05(2015)036}{\emph{JHEP} {\bf 05} (2015)
  036}, [\href{http://arxiv.org/abs/1502.07217}{{\tt 1502.07217}}].

\bibitem{arXiv:1503.02081}
N.~Bobev, S.~El-Showk, D.~Mazac and M.~F. Paulos, \emph{{Bootstrapping SCFTs
  with Four Supercharges}},
  \href{http://dx.doi.org/10.1007/JHEP08(2015)142}{\emph{JHEP} {\bf 08} (2015)
  142}, [\href{http://arxiv.org/abs/1503.02081}{{\tt 1503.02081}}].

\bibitem{arXiv:1504.07997}
F.~Kos, D.~Poland, D.~Simmons-Duffin and A.~Vichi, \emph{{Bootstrapping the
  O(N) Archipelago}},
  \href{http://dx.doi.org/10.1007/JHEP11(2015)106}{\emph{JHEP} {\bf 11} (2015)
  106}, [\href{http://arxiv.org/abs/1504.07997}{{\tt 1504.07997}}].

\bibitem{arXiv:1507.05637}
C.~Beem, M.~Lemos, L.~Rastelli and B.~C. van Rees, \emph{{The (2, 0)
  superconformal bootstrap}},
  \href{http://dx.doi.org/10.1103/PhysRevD.93.025016}{\emph{Phys. Rev.} {\bf
  D93} (2016) 025016}, [\href{http://arxiv.org/abs/1507.05637}{{\tt
  1507.05637}}].

\bibitem{arXiv:1508.00012}
L.~Iliesiu, F.~Kos, D.~Poland, S.~S. Pufu, D.~Simmons-Duffin and R.~Yacoby,
  \emph{{Bootstrapping 3D Fermions}},
  \href{http://arxiv.org/abs/1508.00012}{{\tt 1508.00012}}.

\bibitem{arXiv:1510.03866}
M.~Lemos and P.~Liendo, \emph{{Bootstrapping $ \mathcal{N}=2 $ chiral
  correlators}}, \href{http://dx.doi.org/10.1007/JHEP01(2016)025}{\emph{JHEP}
  {\bf 01} (2016) 025}, [\href{http://arxiv.org/abs/1510.03866}{{\tt
  1510.03866}}].

\bibitem{arXiv:1511.04065}
Y.-H. Lin, S.-H. Shao, D.~Simmons-Duffin, Y.~Wang and X.~Yin, \emph{{N=4
  Superconformal Bootstrap of the K3 CFT}},
  \href{http://arxiv.org/abs/1511.04065}{{\tt 1511.04065}}.

\bibitem{arXiv:1511.07552}
S.~M. Chester, L.~V. Iliesiu, S.~S. Pufu and R.~Yacoby, \emph{{Bootstrapping
  $O(N)$ Vector Models with Four Supercharges in $3 \leq d \leq4$}},
  \href{http://arxiv.org/abs/1511.07552}{{\tt 1511.07552}}.

\bibitem{arXiv:1511.08025}
D.~Li, D.~Meltzer and D.~Poland, \emph{{Conformal Collider Physics from the
  Lightcone Bootstrap}},
  \href{http://dx.doi.org/10.1007/JHEP02(2016)143}{\emph{JHEP} {\bf 02} (2016)
  143}, [\href{http://arxiv.org/abs/1511.08025}{{\tt 1511.08025}}].

\bibitem{arXiv:1601.03476}
S.~M. Chester and S.~S. Pufu, \emph{{Towards Bootstrapping QED$_3$}},
  \href{http://arxiv.org/abs/1601.03476}{{\tt 1601.03476}}.

\bibitem{arXiv:1602.02810}
C.~Behan, \emph{{PyCFTBoot: A flexible interface for the conformal bootstrap}},
   \href{http://arxiv.org/abs/1602.02810}{{\tt 1602.02810}}.

\bibitem{arXiv:1603.03771}
D.~M. Hofman, D.~Li, D.~Meltzer, D.~Poland and F.~Rejon-Barrera, \emph{{A Proof
  of the Conformal Collider Bounds}},
  \href{http://arxiv.org/abs/1603.03771}{{\tt 1603.03771}}.

\bibitem{arXiv:1603.04436}
F.~Kos, D.~Poland, D.~Simmons-Duffin and A.~Vichi, \emph{{Precision Islands in
  the Ising and $O(N)$ Models}},  \href{http://arxiv.org/abs/1603.04436}{{\tt
  1603.04436}}.

\bibitem{Costa:2011dw}
M.~S. Costa, J.~Penedones, D.~Poland and S.~Rychkov, \emph{{Spinning Conformal
  Blocks}}, \href{http://dx.doi.org/10.1007/JHEP11(2011)154}{\emph{JHEP} {\bf
  1111} (2011) 154}, [\href{http://arxiv.org/abs/1109.6321}{{\tt 1109.6321}}].

\bibitem{Echeverri:2015rwa}
A.~C. Echeverri, E.~Elkhidir, D.~Karateev and M.~Serone, \emph{{Deconstructing
  Conformal Blocks in 4D CFT}},
  \href{http://dx.doi.org/10.1007/JHEP08(2015)101}{\emph{JHEP} {\bf 08} (2015)
  101}, [\href{http://arxiv.org/abs/1505.03750}{{\tt 1505.03750}}].

\bibitem{Ferrara:1972uq}
S.~Ferrara, A.~Grillo, G.~Parisi and R.~Gatto, \emph{{The shadow operator
  formalism for conformal algebra. vacuum expectation values and operator
  products}}, \href{http://dx.doi.org/10.1007/BF02907130}{\emph{Lett.Nuovo
  Cim.} {\bf 4S2} (1972) 115--120}.

\bibitem{Dolan:2011dv}
F.~Dolan and H.~Osborn, \emph{{Conformal Partial Waves: Further Mathematical
  Results}},  \href{http://arxiv.org/abs/1108.6194}{{\tt 1108.6194}}.

\bibitem{SimmonsDuffin:2012uy}
D.~Simmons-Duffin, \emph{{Projectors, Shadows, and Conformal Blocks}},
  \href{http://dx.doi.org/10.1007/JHEP04(2014)146}{\emph{JHEP} {\bf 1404}
  (2014) 146}, [\href{http://arxiv.org/abs/1204.3894}{{\tt 1204.3894}}].

\bibitem{Dolan:2000ut}
F.~Dolan and H.~Osborn, \emph{{Conformal four point functions and the operator
  product expansion}},
  \href{http://dx.doi.org/10.1016/S0550-3213(01)00013-X}{\emph{Nucl.Phys.} {\bf
  B599} (2001) 459--496}, [\href{http://arxiv.org/abs/hep-th/0011040}{{\tt
  hep-th/0011040}}].

\bibitem{Dolan:2003hv}
F.~Dolan and H.~Osborn, \emph{{Conformal partial waves and the operator product
  expansion}},
  \href{http://dx.doi.org/10.1016/j.nuclphysb.2003.11.016}{\emph{Nucl.Phys.}
  {\bf B678} (2004) 491--507}, [\href{http://arxiv.org/abs/hep-th/0309180}{{\tt
  hep-th/0309180}}].

\bibitem{Rejon-Barrera:2015bpa}
F.~Rejon-Barrera and D.~Robbins, \emph{{Scalar-Vector Bootstrap}},
  \href{http://dx.doi.org/10.1007/JHEP01(2016)139}{\emph{JHEP} {\bf 01} (2016)
  139}, [\href{http://arxiv.org/abs/1508.02676}{{\tt 1508.02676}}].

\bibitem{Geyer:2000ig}
B.~Geyer and M.~Lazar, \emph{{Twist decomposition of nonlocal light cone
  operators. 2. General tensors of 2nd rank}},
  \href{http://dx.doi.org/10.1016/S0550-3213(00)00227-3}{\emph{Nucl. Phys.}
  {\bf B581} (2000) 341--390}, [\href{http://arxiv.org/abs/hep-th/0003080}{{\tt
  hep-th/0003080}}].

\bibitem{Eilers:2006kd}
J.~Eilers, \emph{{Geometric twist decomposition off the light-cone for nonlocal
  QCD operators}},  \href{http://arxiv.org/abs/hep-th/0608173}{{\tt
  hep-th/0608173}}.

\bibitem{Costa:2014rya}
M.~S. Costa and T.~Hansen, \emph{{Conformal correlators of mixed-symmetry
  tensors}}, \href{http://dx.doi.org/10.1007/JHEP02(2015)151}{\emph{JHEP} {\bf
  1502} (2015) 151}, [\href{http://arxiv.org/abs/1411.7351}{{\tt 1411.7351}}].

\bibitem{Cvitanovic:2008zz}
P.~Cvitanovic, \emph{{Group theory: Birdtracks, Lie's and exceptional groups}}.
\newblock Princeton University Press, 2008.

\bibitem{Costa:2011mg}
M.~S. Costa, J.~Penedones, D.~Poland and S.~Rychkov, \emph{{Spinning Conformal
  Correlators}}, \href{http://dx.doi.org/10.1007/JHEP11(2011)071}{\emph{JHEP}
  {\bf 1111} (2011) 071}, [\href{http://arxiv.org/abs/1107.3554}{{\tt
  1107.3554}}].

\bibitem{Echeverri:2016dun}
A.~C. Echeverri, E.~Elkhidir, D.~Karateev and M.~Serone, \emph{{Seed Conformal
  Blocks in 4D CFT}},  \href{http://arxiv.org/abs/1601.05325}{{\tt
  1601.05325}}.

\bibitem{Hogervorst:2013sma}
M.~Hogervorst and S.~Rychkov, \emph{{Radial Coordinates for Conformal Blocks}},
  \href{http://dx.doi.org/10.1103/PhysRevD.87.106004}{\emph{Phys.Rev.} {\bf
  D87} (2013) 106004}, [\href{http://arxiv.org/abs/1303.1111}{{\tt
  1303.1111}}].

\bibitem{Penedones:2015aga}
J.~Penedones, E.~Trevisani and M.~Yamazaki, \emph{{Recursion Relations for
  Conformal Blocks}},  \href{http://arxiv.org/abs/1509.00428}{{\tt
  1509.00428}}.

\bibitem{Costa:2016other}
M.~S. Costa, T.~Hansen, J.~Penedones and E.~Trevisani, \emph{{Radial expansion
  for spinning conformal blocks}},  \href{http://arxiv.org/abs/1603.05552}{{\tt
  1603.05552}}.

\bibitem{Dobrev:1975ru}
V.~Dobrev, V.~Petkova, S.~Petrova and I.~Todorov, \emph{{Dynamical Derivation
  of Vacuum Operator Product Expansion in Euclidean Conformal Quantum Field
  Theory}}, \href{http://dx.doi.org/10.1103/PhysRevD.13.887}{\emph{Phys.Rev.}
  {\bf D13} (1976) 887}.

\bibitem{MR552445}
N.~El~Samra and R.~C. King, \emph{Dimensions of irreducible representations of
  the classical {L}ie groups}, {\emph{J. Phys. A} {\bf 12} (1979) 2317--2328}.

\bibitem{Costa:2014kfa}
M.~S. Costa, V.~Gon\c{c}alves and J.~Penedones, \emph{{Spinning AdS
  Propagators}}, \href{http://dx.doi.org/10.1007/JHEP09(2014)064}{\emph{JHEP}
  {\bf 1409} (2014) 064}, [\href{http://arxiv.org/abs/1404.5625}{{\tt
  1404.5625}}].

\end{thebibliography}\endgroup

\end{document}